# Machine Learning Approach for Identifying Anatomical Biomarkers of Early Mild Cognitive Impairment

Alwani Liyana Ahmad[1,2,3], Jose Sanchez-Bornot[4], Roberto C. Sotero[5], Damien Coyle[6], Zamzuri Idris[2,3,7], Ibrahima Faye[1,8,*], for the Alzheimer's Disease Neuroimaging Initiative[©]

[1] Department of Fundamental and Applied Sciences, Faculty of Science and Information Technology, Universiti Teknologi PETRONAS, Perak, Malaysia.
[2] Department of Neurosciences, Hospital Universiti Sains Malaysia, Kelantan, Malaysia.
[3] Brain and Behaviour Cluster, School of Medical Sciences, Universiti Sains Malaysia, Kelantan, Malaysia
[4] Intelligent Systems Research Centre, School of Computing, Engineering and Intelligent Systems, Ulster University, Magee campus, Derry~Londonderry, BT48 7JL, UK.
[5] Department of Radiology and Hotchkiss Brain Institute, University of Calgary, Calgary, AB, Canada.
[6] The Bath Institute for the Augmented Human, University of Bath, Bath, BA2 7AY, UK.
[7] Department of Neurosciences, School of Medical Sciences, Universiti Sains Malaysia, Kelantan, Malaysia
[8] Centre for Intelligent Signal & Imaging Research (CISIR), Universiti Teknologi PETRONAS, Perak, Malaysia.

[©]Data used in preparation of this article were obtained from the Alzheimer's Disease Neuroimaging Initiative (ADNI) database (adni.loni.usc.edu). As such, the investigators within the ADNI contributed to the design and implementation of ADNI and/or provided data but did not participate in analysis or writing of this report. A complete listing of ADNI investigators can be found at: http://adni.loni.usc.edu/wpcontent/uploads/how_to_apply/ADNI_Acknowledgement_List.pdf.

Corresponding Author:
Ibrahima Faye[1,8]
Universiti Teknologi PETRONAS, Perak, 32610, Malaysia
Email address: ibrahima_faye @utp.edu.my

## Abstract

**Background.** Alzheimer's Disease (AD) represents a significant challenge in neurodegenerative disorders, necessitating early detection for effective intervention. Among neuroimaging methods, magnetic resonance imaging (MRI) is widely used because it is easy to apply in clinical practice and cost-effective, making it crucial for studying AD.

**Objective.** This study aims to perform a comprehensive analysis of machine learning (ML) methods used in MRI-based biomarker selection and classification analysis. The goal is to study AD-related early cognitive decline by discriminating between healthy control (HC) participants who stayed stable and those unstable (uHC) who developed mild cognitive impairment (MCI) within five years.

**Methods.** We utilized 3-Tesla (3T) MRI data from the Alzheimer's Disease Neuroinformatic Initiative (ADNI) and the Open Access Series of Imaging Studies 3 (OASIS-3), focusing on HC and uHC. Freesurfer's recon-all, among other tools, was used to extract MRI-based anatomical biomarkers corresponding to semi-automatic segmented subcortical and cortical brain regions. We applied various ML techniques to select features and classify the data. These included methods from preliminary analysis performed in the MATLAB Classification Learner (MCL) app and more sophisticated methods like nested cross-validation and Bayesian optimization implemented in a customized pipeline to enhance classification performance for balanced and imbalanced datasets. Our pipeline was applied to both original imbalanced and randomly balanced datasets within a Monte Carlo analysis. Moreover, we implemented data harmonization approaches based on polynomial regression that enhanced the performance of ML and statistical methods. Complementary performance metrics, such as Accuracy (Acc), area under receiver operating characteristic curve (AROC), F1 score, and Matthew's correlation coefficient (MCC), were used to evaluate the assessed methodologies.

**Results.** In feature selection analyses, consistent outcomes were obtained from ADNI and OASIS-3 datasets: entorhinal, hippocampus, lateral ventricle, and lateral orbitofrontal regions were consistently identified as the most affected areas during early cognitive decline. In classification analyses, outcomes differed between the randomly balanced and imbalanced data analysis, and we also found noticeable differences between analyses involving ADNI and OASIS-3 datasets. Naïve Bayes model, using z-score data harmonization with ReliefF feature selection, performed best for ADNI balanced datasets (Acc = 69.17 ± 6.54 %, AROC = 77.73 ± 7.08 %, F1 = 69.21 ± 7.90 %, MCC′ = 69.28 ± 6.56 %). In contrast, for OASIS-3 balanced analyses, SVM for z-score-corrected data performed better than other methods (Acc = 66.58 ± 2.91 %, AROC = 72.01 ± 2.40 %, MCC′ = 66.78 ± 2.96 %), although the Logistic regression showed best performance according to the F1 score (66.68 ± 1.21%). However, these results differed from those obtained with the imbalanced data analysis. Here, RUSBoost demonstrated the strongest combined performance on ADNI (F1 = 50.60 ± 5.20 %, AROC = 81.54 ± 2.92 %) and OASIS-3 (MCC′ = 63.31 ± 1.43 %), SVM showed the best performance on ADNI according to Acc (82.93 ± 1.59 %) and MCC′ (70.21 ± 3.16 %) metrics, and Naïve Bayes showed the best performance on OASIS-3 according to F1 (42.54 ± 1.71 %) and AROC (70.33 ± 1.00 %).

**Conclusion.** Data harmonization techniques improved the consistency and performance of feature selection and ML classification analyses. Despite the small sample sizes, z-score harmonization produced the best results, especially in ML classification analyses. Our methodology suggests the usefulness of a semi-automatic pipeline for early AD detection using MRI, with prospective integration with other neuroimaging data to enhance the prediction of AD progression.

# Introduction

Alzheimer's Disease (AD) is marked by the gradual accumulation of amyloid-β (amyloid plaques) in the extracellular space and tau proteins (Neurofibrillary tangles – NFT) in the intracellular space of a neuron, leading to cognitive and motor dysfunctions and difficulties in daily activities. The symptomatic onset of AD is gradual, beginning with losses in episodic and semantic memory, progressing to aphasia, apraxia, mood disturbances, and more severe symptoms in the advanced stages [1,2]. Post-mortem examinations reveal patterns of neurodegeneration in brain regions corresponding to these cognitive and behavioral changes, as delineated by Braak's staging [3]. The medial temporal lobe (MTL), including the hippocampus, amygdala, and entorhinal cortex, undergoes significant atrophy, which impacts memory formation and consolidation. Interestingly, early changes are also observed in the limbic system, encompassing the hippocampus, amygdala, cingulate, and parahippocampal gyri, affecting emotion and memory processing. The limbic system is connected to the entorhinal cortex via the subiculum, through which it is hypothesized that AD pathology spreads from one region to adjacent ones [4]. However, Braak & Tredici reported that the very-early AD changes can be observed in the transentorhinal region in stage I when prospective AD patients remain asymptomatic, and from there, it spreads to the entorhinal region and the hippocampal formation in stages II and III, respectively [5]. Therefore, when patients have the first symptoms of AD, they may be already in an irreversible stage.

As AD advances, further anatomical changes include atrophy in association cortical areas and ventricular enlargement [6–8].

The cascade of anatomical changes can be observed in vivo using neuroimaging and clinical data, e.g., using positron emission tomography (PET) and cerebrospinal fluid (CSF) analysis to detect abnormal accumulation of amyloid plaques and tau proteins in the brain [2,9,10]. Additionally, single-photon emission computed tomography (SPECT), utilizing a ligand binding to the dopamine transporter molecule (DaTscan), aids in evaluating Parkinsonian syndrome and distinguishing it and Lewy Body dementia from AD [11–14]. Researchers have also explored combining multiple neuroimaging modalities, including SPECT, PET, MRI, functional MRI (fMRI), and magneto/electro-encephalography (M/EEG) [15,16], and integrating neuroimaging data with cognitive or clinical measurements [17,18]. However, it is essential to recognize that while PET and SPECT provide valuable insights, they are more invasive, costlier, and less globally accessible than MRI scans [12,19]. Essentially, used alone or combined with other neuroimaging data, MRI remains indispensable for evaluating suspected dementia cases, and ruling out alternative causes such as microinfarcts and white matter lesions [12,20,21]. Also, the enhanced resolution of MRI images allows the quantification of regional cerebral atrophy, making it relevant for early dementia assessment despite its limitations [12,20,22–27]

On the other hand, it has been found that pathogenic infections like prions have a significant impact on the neuronal atrophy and disruption of connectivity hubs within the medial temporal lobe [28], leading to the hypothesis that AD could be triggered by the presence of a non-endogenous pathogen [5]. This observation also relates to the AD's disconnection syndrome hypothesis [29,30]. Particularly, Li et al. [31] identified that damage to white and gray matter within these regions disrupts limbic system networks, correlating with memory and behavioral impairments in AD patients. This disruption has been evidenced in neuroimaging studies using diffusion tensor imaging (DTI), MRI, and fMRI data [31–33]. However, minor fluctuations in behavior and emotional states can also be due to changes in diet [34], lifestyle [35] or other less controlled factors, therefore posing a challenge in diagnosing mild cognitive impairment (MCI), a prodromal AD stage, and its progression to AD [36,37]. This has led to a growing focus on developing automated diagnostic tools, primarily leveraging ML methods with neuroimaging data, for cost-effective and objective cognitive assessment [14,15,22–24,36–38].

ML is increasingly utilized in healthcare for early-stage disease diagnosis, including cancer[26,39–41] and AD [27,42–45], reducing the possible subjectivity of diagnostic outcomes. However, AD research often focuses on comparing AD vs. healthy control (HC) participants data or using data from MCI participants who are already in an irreversible or progressive stage, potentially overlooking the early AD stage [27,46–50]. Interestingly, Popuri et al. [51] trained a classifier to discriminate between HC and AD participants using MRI data and posteriorly applied this classifier to predict MCI conversion to AD in 6 months or more, with an area under curve (AROC) outcome of 0.81 for six months conversion and 0.73 for seven years conversion. This study also demonstrated the advantages of using data harmonization, e.g., removing the data variability due to nuisance variables such as age, gender, and intracranial volume (ICV), for increasing the classifier performance. Although not considered in our study, Ma et al. [52] also compared different data harmonization strategies, including three different methods for ICV calculation, and their impact on classification performance. As reported in this study, data harmonization can improve the results as variability in the post-processed data can be more exclusively associated with changes due to AD progression.

Moreover, combining different techniques with classification methods has also helped improve the prediction outcome, as demonstrated by applying graph analysis tools with support vector machine (SVM) for predicting the risk of dementia among MCI patients in an EEG study [53]. Nevertheless, it is critically important to properly evaluate different methodologies to ensure reproducibility and potential implementation for clinical applications. For example, based on a Monte Carlo simulation data analysis, Stamate et al. [54] introduced an ML framework to compare multiple classification models and found that the top-performing methods for predicting dementia and MCI were based on decision trees algorithms and the eXtreme Gradient Boosting model with the ReliefF method applied for feature selection. Significantly, the evaluation and comparison among different classification methods often rely on the performance of the classification accuracy, although this statistic may be biased for analysis involving imbalanced data [55,56]. In the medical field, imbalanced datasets are very common because of the lower number of abnormal cases compared to normal cases. This situation leads to misclassification for cases in the minority group, which may hamper the research on early AD detection [57].

Addressing imbalanced data, various methods have been proposed which mainly combine resampling techniques with cost-sensitive classification approaches [58]. For example, Chawla et al. introduced an oversampling technique known as Synthetic Minority Over-sampling Technique (SMOTE) [59], which was demonstrated in combination with a C4.5 decision tree and Ripper and Naïve Bayes classifiers. In contrast, Rahman et al. explored different under-sampling strategies as alternatives to SMOTE [57]. So far, in the literature on imbalance data classification, RUSBoost is one of the most successful classification methods, combining under-sampling and boosting algorithms [60,61]. However, in general, both under-sampling and over-sampling techniques present advantages and limitations, e.g., whereas over-sampling methods increase the computational time and risk of overfitting due to sample duplication, mainly for the minority class, under-sampling may incur data loss, mainly for the majority class [62].

Our study investigates early MRI-based anatomical changes linked to cognitive decline, ensuring wide applicability, reproducibility, and a comprehensive ML evaluation for balanced and imbalanced datasets. We used the Matthew's correlation coefficient (MCC) and F1 score [63,64], besides accuracy and AROC statistics, to more fittingly evaluate ML classifiers performance. For analyzing the early AD anatomical changes, we assessed the brain regional atrophy using ADNI and OASIS-3 datasets while examining a subset of HC participants who remained stable during these respective studies, in contrast to those participants who converted to MCI in less than 5 years. The analyzed MRI images for both groups were recorded at baseline, where all the participants were healthy. The Freesurfer software [65] was used for the semi-automatic processing of the MRI data. Our approach evaluated the possible advantages of data harmonization while comparing various feature selection and ML classification methods on different dataset cohorts: first, using MATLAB's Classification Learner (MCL) app, and second, using a customized pipeline for a more robust assessment based on Bayesian optimization and nested cross-validation approaches within a Monte Carlo replication analysis. Our main findings showed anatomical changes in MTL brain regions associated with potential cognitive decline, which align well with previous reports and were consistently found across the application of multiple feature selection and ML methods.

## Materials & Methods

Our methodology, illustrated in **Figure 1**, involves five key steps: I) Data selection of participants from ADNI and OASIS-3 datasets who remained healthy during the study (HC) and those progressing to MCI over five years (uHC), producing imbalanced datasets (**Figure 1A**). II) Data processing was optionally used for each data to reduce variability due to gender, age, and ICV, using the HC group as a reference. Two different approaches are evaluated: residual and z-score harmonization (**Figure 1B**). III) The MCL app was used to evaluate different classification and feature selection methods based exclusively on a randomly selected ADNI-balanced cohort to evaluate the most appropriate methods and brain regions for posterior analyses (**Figure 1C**). IV) In parallel, the SPSS statistical software was used to perform analogous feature selection analyses to the MCL app using the ADNI-imbalanced cohort (**Figure 1D**). V) Further validation and evaluation of selected models and features was performed through a customized pipeline, combining nested cross-validation with Bayesian optimization within a Monte Carlo replication analysis (**Figure 1E**). Here, balanced cohorts were created from imbalanced datasets by randomly selecting the same number of samples in the majority as in the minority group.

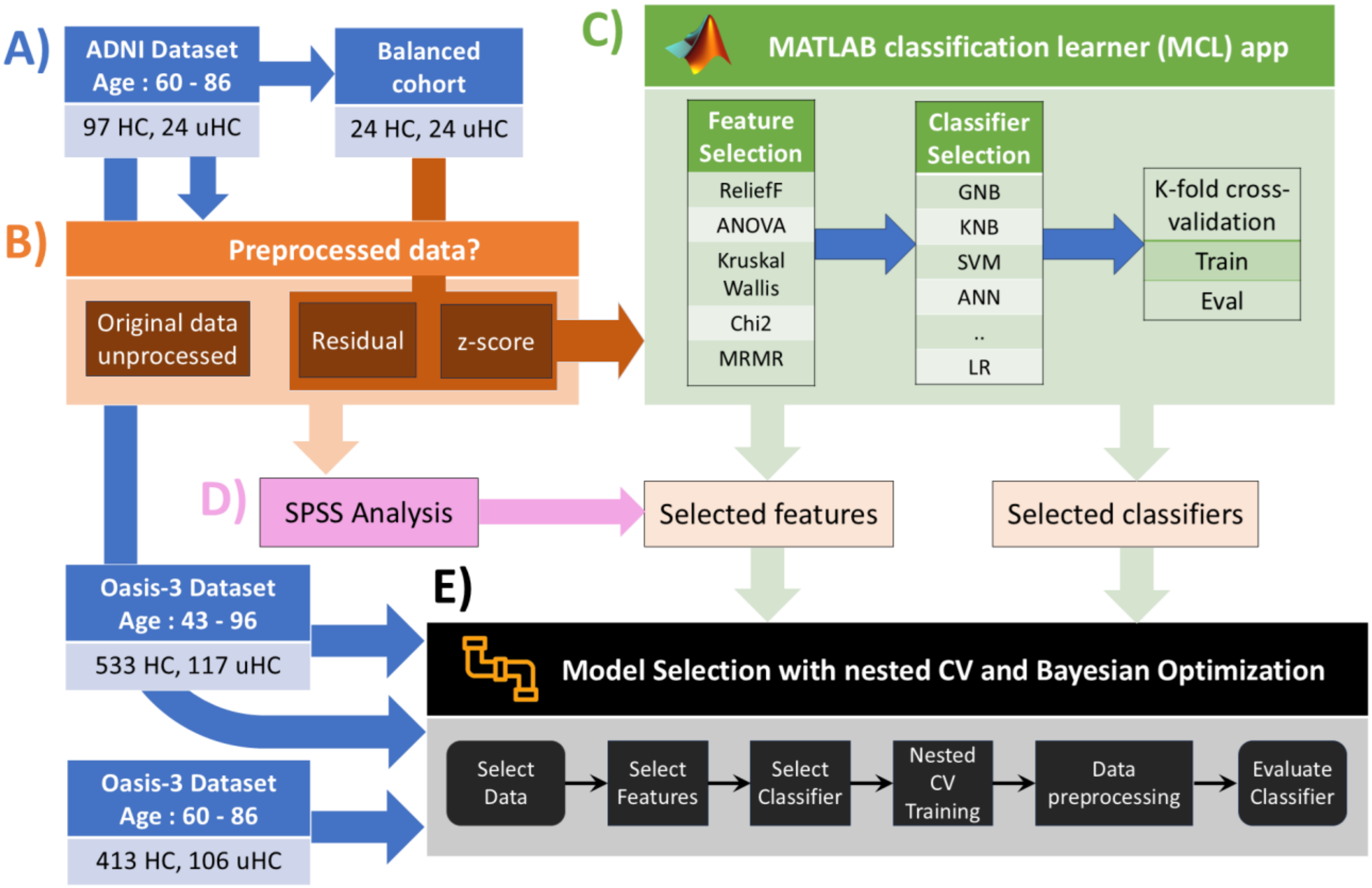

***Figure 1:*** *Workflow illustrating the proposed methodology. A) Selection process of participants data corresponding to healthy controls (HC) and participants who transitioned to MCI (uHC) in a period lower or equal than 5 years for ADNI and OASIS-3 datasets. These are imbalanced datasets as shown by the integer values indicating the number of samples in each group. A manually-balance cohort was extracted from ADNI dataset to the used within MCL app analysis. (B) All data was optionally pre-processed using two different data correction procedures: residual and z-score harmonization. (C) The MATLAB's Classification Learner (MCL) app was utilized for evaluating a wide range of feature selection and classification methods, using an ADNI-balanced cohort. D) In parallel, both unprocessed and processed ADNI data undergone statistical analysis using SPSS software for assessing significant features. Thus, we performed a preliminary selection of "best" classifiers and features from the MCL app and SPSS analysis. (E) Further evaluation of selected features and classification methods was performed through our proposed customized pipeline, implementing nested cross-validation (CV) with Bayesian optimization within a Monte Carlo replication framework. This last analysis was performed for both ADNI and OASIS-3 imbalanced datasets.*

## Participants data

We selected MRI data from two longitudinal studies: the Alzheimer's Disease Neuroimaging Initiative (ADNI) (http://adni.loni.usc.edu/) [66] and the Open Access Series of Imaging Studies 3 (OASIS-3) [67] (www.oasis-brains.org) datasets. The rationale behind using two different datasets is to compare and validate our methods with more heterogeneous data. Even when ADNI is already a multisite project, it follows a much stricter acquisition protocol than other studies. In summary, the ADNI study was launched in 2003 with the primary goal to test whether neuroimaging modalities such as MRI and PET can be analyzed independently or combined with other clinical and neuropsychological data to find Alzheimer's biomarkers and study the progression from HC to AD (https://adni.loni.usc.edu/methods/documents/). The OASIS-3 is a series of neuroimaging studies for which datasets are publicly available, as collected by the Knight Alzheimer Disease Research Center (ADRC) and its affiliated organizations [68]. Similarly to ADNI, OASIS-3 contains longitudinal data involving MRI and PET neuroimaging, as well as clinical, cognitive, and biomarker data from both normal aging and AD participants [67,68].

Subjects with an unavailable 3T MRI image at the baseline were excluded from this study. We specifically chose Magnetization Prepared Rapid Gradient Echo (MP-RANGE) MRI images without repetition. We restricted our analysis to using only 3T MRI images from both datasets to simplify our study's complexity and ensure our results' consistency and reliability. 3T MRI scanners deliver a higher signal-to-noise ratio (SNR) and better spatial resolution than 1.5T scanners, resulting in higher image resolution [69]. Furthermore, we avoided combining data from 1.5T and 3T MRI scanners as it could introduce variability due to differences in image acquisition protocols, and the differential analysis between the results for 3T and 1.5T analysis is beyond our present objectives. Additionally, it has been reported that changes in brain tissue texture detected by 3T MRI can lead to earlier AD diagnosis compared to 1.5T MRI [70]. Moreover, from ADNI and OASIS-3 datasets, we also extracted essential demographic and cognitive information for our analysis, including participants' gender, age, years of education, and Mini-Mental State Examination (MMSE) scores. Specifically, the ADNI participants selected for this study ranged in age from 60 to 86 and were either English or Spanish speakers.

The ADNI dataset is used in this work as the primary data, mainly for feature selection and evaluation of ML classifiers. Here, we selected 97 HC participants who remained stable during the study, as reflected in the ADNIMERGE table downloaded from the ADNI website (http://adni.loni.usc.edu/). Additionally, we selected 24 participants who were diagnosed with HC at baseline and converted to MCI during a 5-year follow-up period after enrolling in the study. Otherwise, from the OASIS-3 dataset, we exclusively focused on MRI images for 533 HC and 117 uHC. Subjects in the OASIS-3 dataset were categorized according to the Clinical Dementia Rating (CDR). Participants with CDR=0 when their MRI image was first acquired and who remained stable during the study were considered HC. In contrast, participants who initially had a CDR of 0 but later showed an increase to a CDR of 0.5 at a subsequent visit were labeled uHC. For both data selected from ADNI and OASIS-3 datasets, the conversion period for uHC participants is 5 years or less from their first visit. We divided the OASIS-3 dataset into two cohorts based on age ranges: 1) the original participants' age range of 43-96 years and 2) a restricted age range of 60-86 years. The purpose of restricting the age range to 60-86 years is to match the ADNI dataset, as structural brain changes depend on age [71]. Critically, MRI data for HC and uHC groups were all selected at baseline, where the participants were regarded as healthy. **Table 1** summarizes the demographic information for selected participants in our study.

***Table 1:*** *Demographic and clinical findings of the subjects. Data are given as mean (SD). (%min) percentage of minority class, referred to as the imbalance dataset for classification analysis.*

| | **ADNI: age 60-86** | | | | **OASIS-3: age 43-96** | | | | **OASIS-3: age 60-86** | | | |
|---|---|---|---|---|---|---|---|---|---|---|---|---|
| | **HC** | **uHC** | **p-value** | **% min class** | **HC** | **uHC** | **p-value** | **% min class** | **HC** | **uHC** | **p-value** | **% min class** |
| **Number of subjects** | 97 | 24 | NA | 24.74 | 533 | 117 | NA | 21.95 | 413 | 106 | NA | 25.67 |
| **Gender (M/F)** | 56/41 | 12/12 | 0.686 | NA | 222/310 | 58/59 | 0.117 | NA | 175/238 | 53/53 | 0.159 | NA |
| **Age (years)** | 72.91 (5.96) | 75.95 (5.79) | 0.026 | NA | 66.71 (8.97) | 76.43 (7.40) | <0.001 | NA | 69.81 (5.11) | 76.08 (5.48) | <0.001 | NA |
| **MMSE** | 29.19 (1.12) | 28.67 (1.47) | 0.056 | NA | 29.20 (1.07) | 28.30 (1.61) | <0.001 | NA | 29.11 (1.11) | 28.31 (1.64) | <0.001 | NA |
| **Years of education** | 16.56 (2.39) | 16.00 (2.72) | 0.318 | NA | 16.38 (2.39) | 15.62 (2.90) | 0.003 | NA | 16.40 (2.37) | 15.65 (2.97) | 0.006 | NA |

## MRI preprocessing pipeline

The MRI images were downloaded in NIFTI format and processed using FreeSurfer software (package version 7.3.2), with the standard cross-sectional pipeline *recon-all*, https://surfer.nmr.mgh.harvard.edu/fswiki/recon-all. In summary, *recon-all* performs operations such as automatic co-registration to the Talairach atlas, image intensity normalization, and removal of non-brain tissue (e.g., skull stripping) by utilizing a hybrid watershed/surface deformation procedure [72], segmentation of grey matter (GM), white matter (WM), cerebrospinal fluid (CSF) tissues, subcortical brain regions automatic segmentation, and cortical automatic parcellation [73,74]. The outcomes of the *recon-all* pipeline were carefully inspected to correct and ameliorate cortical and segmentation defects. Subsequently, the Freesurfer's *asegstats2table* and *aparcstats2table* scripts were run over this output, respectively, to extract the subcortical volume information tables for predefined regions and the different statistics (e.g., volume and cortical thickness) for the cortical brain regions, which were extracted according to the Desikan atlas [75]. The ICV value was also estimated as part of the processing pipeline. Presumably, ICV provides a metric that resists change along aging for adults older than 50 years old, thus serving as a critical measure to control for brain size differences, for example, between female and male populations [52]. Together with demographic information such as age and gender, using ICV can help remove unnecessary variation in the data that is not due to the brain degeneration process occurring in AD. In this study, we used only the brain volume information for the brain subcortical and cortical regions as extracted by the above MRI preprocessing pipeline. Moreover, we calculated total brain volumes by combining the values for the left and right hemispheres. In summary, we analyzed 39 merged brain volumes used as predictors in the ML analysis, in addition to the measurement of the brain segmentation volume without ventricles (BrainSegVolNotVent).

## Data harmonization to eliminate the effects of nuisance factors

The purpose behind employing data correction is to eliminate the uncontrolled effect of nuisance factors on extracted brain regional measures, such as the effects of age, gender, and ICV; therefore, harmonized data would be less dependent on these variables, and thus we can assume that the main source of variability and differences among the HC and uHC harmonized data are due to the AD degenerative process. For example, it has been observed that brain structures vary across the lifespan, even in healthy aging, with non-linear and non-monotonic trajectories, although the trajectories become more linear for adults older than 50 years [71]. Typically, males have a larger average ICV than females, and brain regional volumes are correlated to ICV. Consequently, it may be appreciated that after controlling by ICV, gender-based differences are less noticeable [52].

In general, applying a correction to remove the effect of these variables can increase the performance of statistical and ML analysis [51]. Here, complementarily to previous studies [51,52,76,77], we adopted a multivariate polynomial regression approach for data harmonization, using age, gender, and ICV as covariates and setting the HC group as reference (i.e., using exclusively the HC data to fit the polynomial regression parameters). With the purpose of illustrating the possible advantages of this procedure, we used the whole dataset from HC, MCI, and AD groups available in the ADNIMERGE table and the hippocampus volume as a region of interest, which is one of the central brain regions suffering atrophy due to AD effects.

Two different harmonization approaches are discussed here. The first approach uses the residuals after fitting the polynomial to the HC data, while the second approach relies on the z-score transform, implemented using the following formulations:

$$\hat{\rho}_G = \underset{\rho_G}{\operatorname{argmin}} \left\{ \sum_{i=i}^{N} \left( y_i^{(HC,G)} - \operatorname{fit}\left(\rho, Age_i^{(HC,G)}, ICV_i^{(HC,G)}\right) \right)^2 \right\},$$

$$\hat{\mu}_i, \hat{\sigma}_i = \operatorname{predint}\left(\hat{\rho}_{G_i}, Age_i, ICV_i,\right),$$

$$x_i^{(1)} = y_i - \hat{\mu}_i, \text{ and } x_i^{(2)} = (y_i - \hat{\mu}_i)/\hat{\sigma}_i.$$

Here, the polynomials were fitted separately for the genders, $G = \{Male, Female\}$, using the MATLAB "fit" function, where $\hat{\rho}_G$ represents the best-fitted polynomial model. $y_i^{(HC,G)}$, $Age_i^{(HC,G)}$, and $ICV_i^{(HC,G)}$, represent the $i-th$ measures for each participant in the HC group, considered separately for each gender, corresponding to the involved variables. $\hat{\mu}_i$ and $\hat{\sigma}_i$ are the polynomial interpolation and standard deviation estimates, calculated with the MATLAB "predint" function, for each sample in the dataset, which are required to derive the corrections $x_i^{(1)}$ and $x_i^{(2)}$, corresponding to proposed harmonization procedures called here as residuals-corrected and z-score-corrected harmonized data, respectively.

## Statistical Analysis

We used the IBM SPSS Statistics software, version 28.0.1.1(15), to perform a statistical analysis of all available structural volume features obtained from the preprocessing analysis with FreeSurfer. All paired structures, with left- and right-side volumes, were merged. We conducted a study of covariance (ANCOVA) only for the uncorrected data for each brain feature while using age, gender, years of education, and ICV as covariates [78]. Additionally, we applied both ANOVA and the independent sample non-parametric test of Kruskal Wallis for all uncorrected and harmonized data while controlling for participant gender, age, and ICV variables. We employed the Bonferroni correction to correct for multiple comparisons. For the ADNI dataset, features that exhibited significant differences with p-value $\leq 0.05$ across all three analyses (ANCOVA, ANOVA, and Kruskal Wallis) were selected for further classification analysis. The same analysis was later applied to the OASIS-3 dataset, and consistency among the selected features was evaluated.

## Feature and classification model selection in the MCL app

We utilized the MCL app, a graphical user interface (GUI) that facilitates feature and model selection, through the tuning of predefined classification models based on *K*-fold cross-validation, holdout, or resubstitution validation, for binary and multiclass problems. The utilization of this app in our study is intended to simplify the process of exploring, building, and fine-tuning classification models. Within the MCL app, we explored all the available algorithms, including decision trees, discriminant analysis, logistic regression, naïve Bayes, support vector machines, nearest neighbors, kernel approximation, ensemble methods, and neural networks, combined with the available feature selection techniques. We evaluated all these methods using the default-defined architectures and hyperparameter values. For example, the MCL app includes predefined Bilayered Neural Network (BNN) and Wide Neural Network (WNN) architectures for the neural networks' family. The classifiers showing better preformance were saved as MATLAB scripts and further refined to be used within our customized ML pipeline for a more comprehensive analysis based on nested cross-validation combined with Bayesian optimization.

In addition, we applied all the available combinations between feature selection and classification methods within the MCL app. The feature selection procedure not only aids in reducing overfitting [79], but also facilitates faster training and decreases model complexity, making interpretation easier. Due to scale differences, the scores are converted into percentages to make feature selection more clear-cut. Particularly, the available feature selection methods in the MCL app are (https://uk.mathworks.com/help/stats/feature-selection-and-feature-transformation.html):

- *Minimum Redundancy Maximum Relevance (MRMR)*

MRMR algorithm identifies the importance of predictor variables, selecting highly relevant features concerning the target variable while ensuring low redundancy among the chosen features.

- *Chi-square (Chi2)*

Ranks the features based on the p-value derived from the chi-square test. The potential independence between each predictor variable and the response variable was assessed through a separate chi-square test for each variable. The scores are represented as -log(p)

- *ReliefF*

This method is particularly effective for evaluating the significance of features in distance-based supervised models, which rely on pairwise distances between observations to make predictions about the response variable.

- *ANOVA*

Conducts individual one-way analysis of variance for each predictor variable, categorized by class, and subsequently prioritizes features ranking based on the p-value. The score is represented as -log(p).

- *Kruskal Wallis*

Ranks the features based on the p-values derived from the Kruskal-Wallis's test. The scores are represented as -log(p).

Using the MCL app's GUI options, we split the data into train (80%) and test (20%) subsets with $K = 10$ fold cross-validation to train and evaluate each classifier using the app's feature selection criteria, separately in successive runs. This process was repeated 10 times with different random partitions to average the results and ensure more stable outcomes. For each process, we recorded the classifiers that achieved the highest accuracy. This process is aimed at identifying the best models and features for subsequent classification analysis. Ultimately, we exported the best-performing classifiers (those that appeared most frequently as top performers) to corresponding implementations in MATLAB functions. This allowed for further performance evaluation using balanced and imbalanced data analysis with our customized ML pipeline.

Moreover, combining the SPSS statistical analyses in the previous section with the feature selection analyses in the MCL app, we ultimately proposed the following four selection criteria (selected features under these criteria are referred to as subset A-D features later in our analyses, denoting each subset with the corresponding letter in the below list):

A) Average score percentage from MCL app

We combine the four scores calculated with the MCL app (Chi-square, ANOVA, Kruskal Wallis, and ReliefF) to create an average score. The selected features are those with score at or above the median value.

B) ReliefF

We selected only the features with positive scores from the ReliefF feature selection method in the MCL app, as negative scores indicate features of lesser importance [80].

C) Frequent feature appearances from all feature ranking analysis

We select the features that consistently appeared across all the explored feature selection approaches, among those selected from MCL app and SPSS analyses.

D) Features selection according to SPSS analysis

We select the features with significant differences in the HC vs. uHC statistical analysis performed in SPSS (e.g., combining ANOVA, ANCOVA, and Kruskal Wallis outcome).

Finally, we exclusively used and ADNI-balanced cohort dataset for this preliminary analysis (**Figure 1A**), since available MATLAB's classifiers are primarily optimized for balanced data analysis. Note that ADNI dataset adheres to a much stricter acquisition protocol and has been extensively used in numerous previous studies [71,76,81–83], offering a more reliable basis for comparison than the OASIS-3 dataset. At the same time, our research emulates the case when the outcome of one study is attempted to be replicated in other studies by using different datasets. Thus, we evaluated this preliminary selection in a posterior analysis which involves the application of our customized ML pipeline to the imbalanced ADNI and OASIS-3 datasets.

## Classification performance metrics

To evaluate the performance in binary classification problems, we calculated several statistical scores for the different techniques in our study, such as accuracy ($Acc$), $F1$, and Matthew's correlation coefficient ($MCC$), also known as the Yule's phi coefficient. The $F1$ and $MCC$ scores are essentially recommended for imbalanced classification problems. However, the $MCC$ score has been reported as superior to accuracy and $F1$ in general binary classification problems [63,64]. For clarity and self-content reasons, we present these metrics as follows, based on the variables represented in **Table 2**:

$$Acc = \frac{TP+TN}{Total},\ F1 = \frac{2}{\frac{1}{PPV}+\frac{1}{TPR}} = \frac{2*PPV*TPR}{PPV+TPR},$$

$$MCC = \sqrt{TPR*TNR*PPV*NPV} - \sqrt{(1-TPR)*(1-TNR)*(1-PPV)*(1-NPV)},$$

$$= (TP*TN - FP*FN)/\sqrt{CE*CA*EP*EN}$$

$$MCC' = 0.5*(1+MCC),$$

where $TPR$ and $TNR$ represent the true positive and negative rate, also known as sensitivity (recall) and specificity, respectively ($TPR = TP/CE$ and $TNR = TN/CA$). $PPV$ and $NPV$ represent the positive and negative predictive values, respectively ($PPV = TP/EP$ and $NPV = TN/EN$). The $PPV$ is also commonly known as precision.

The $Acc$ and $F1$ scores are defined in the range $[0,1]$, where a value near to 1 refers to an excellent performance. Otherwise, $MCC$ is defined in the range $[-1,1]$, reaching 1 for perfect classification, when $TP = CE = EP$, and $TN = CA = EN$, and reaching $-1$ for a completely wrong classification when $FN = CE = EN$ and $FP = CA = EP$. However, we prefer to use the modified $MCC$ ($MCC'$) score as it is equivalent to the original but defined in the range $[0,1]$, which eases the visual comparison with the $Acc$ and $F1$ scores.

***Table 2:*** *The contingency table represents the number of cases with an existing/absent condition (CE/CA) evaluated using a generic test procedure, resulting in positive/negative examination (EP/EN) cases. Combining the Condition and Examination labels, the cases can be partitioned as true/false positive (TP/FP) and true/false negative (TN/FN).*

| | **Examination** | | |
|---|---|---|---|
| **Condition** | **Positive** | **Negative** | $Total = CE + CA$ |
| Existing | $TP$ | $FN$ | $CE$ |
| Absent | $FP$ | $TN$ | $CA$ |
| $Total = EP + EN$ | $EP$ | $EN$ | |

## Further validation with a customized ML pipeline

After selecting the feature and classification approaches using the MCL app and SPSS tools for each data harmonization approach, we evaluated each method combination further with nested cross-validation and Bayesian optimization within a Monte Carlo replication analysis. To implement nested cross-validation, in the external holdout loop, for each $k = 1, \ldots, K$ ($K = 10$), 10% of samples are left out as the holdout subset. Then, the optimal hyperparameters were selected for each corresponding model using a MATLAB-based Bayesian optimization procedure, automatically implementing an internal $K - 1$ fold cross-validation. Here, the partitions were created using MATLAB's "cvpartition" function, taking into consideration the sample group information (HC or uHC). This guarantees that each partition has similar proportions in each group (stratified partitions), which is critical to increasing robustness in imbalanced data analysis. Within a Monte Carlo replication analysis, we repeated this procedure 20 times to obtain individual measurements for the above performance metrics, enabling a statistical comparison analysis to assess the better methodological combination.

Moreover, for the Bayesian optimization approach, we used 200 iterations to enable the algorithm to find the "optimal" configuration of hyperparameters for each corresponding classifier. Several optimizable options were selected among the available ones as follows:

1. *Naïve Bayes (*https://uk.mathworks.com/help/stats/fitcnb.html*)*
   Data distribution assumption: "normal" or "kernel".
   Kernel smoother type: "box", "epanechnikov", "normal", or "triangle".
   Kernel smoothing window width: unbounded positive real number.

2. *K nearest neighbors (*https://uk.mathworks.com/help/stats/fitcknn.html*)*
   Number of neighbors: integer number, restricted for values in the range [5, 30].
   Distance metric: "cityblock", "chebychev", "correlation", "cosine", "euclidean", "hamming", "jaccard", "mahalanobis", "minkowski", "seuclidean", or "spearman".

3. *SVM (*https://uk.mathworks.com/help/stats/fitcsvm.html*)*
   Kernel function: "gaussian", "rbf", "linear", or "polynomial".
   Kernel scale parameter: positive real value constrained in the range [$10^{-1}$, 10].
   Box constraint: positive real value constrained in the range [$10^{-1}$, 10].

4. *Logistic regression (*https://uk.mathworks.com/help/stats/fitclinear.html*)*
   Lambda (logistic regression implemented with Lasso regularization): positive real value evaluated in the range [$10^{-3}$, 10]. Score transformation: "none", "logit", "invlogit", or "doublelogit".

5. *RUSBoost (*https://uk.mathworks.com/help/stats/fitcensemble.html
   Ensemble aggregation method: "RUSBoost".
   Number of ensemble learning cycles: positive integer (unbounded).
   Learning rate for shrinkage: positive real number defined in the range (0, 1].
   Maximal number of decision splits: positive integer number (unbounded).

The above procedure was implemented in our customized ML pipeline. The pipeline was directly applied to the imbalanced ADNI and OASIS-3 datasets, with the latter having an age range of 43–96 years, and to the imbalanced OASIS-3 dataset restricted to the same age range as ADNI for imbalance data analyses. As a comparison, a similar application was performed for the same method combination but for balanced datasets, which were randomly generated within each Monte Carlo replication step, i.e., by randomly undersampling the larger group to match the same number of samples as in the smaller group before the evaluation of each method combination.

Finally, we performed a statistical analysis involving N-way ANOVA and pairwise comparisons, to assess the influence of the different options in our analyses, including the selection of harmonization, feature selection and classification combination. For the control of spurious outcomes due to multiple comparisons, we applied both the Bonferroni correction and the Benjamini-Hochberg method, which controls the false discovery rate (FDR). We also used the post hoc Tukey Honestly Significant Difference (HSD) test, assuming a significance threshold of p-value $\leq 0.05$ to identify statistically significant differences. For Benjamini-Hochberg method correction, we applied a 5% FDR correction.

# Results

## Data correction to eliminate the nuisance factors.

**Figure 2** illustrates the data harmonization procedure using polynomial regression of hippocampal volumes for data extracted from the ADNIMERGE table for HC, MCI, and AD participants (see **Materials and Methods**). The effect of harmonization is illustrated for the various subgroups, obtained from the combination of the diagnostic (HC, MCI, or AD), participants' gender (M – male, F – female), and three artificial subdivisions of the participants according to their ICV size (group id = 0 for smaller ICV, id=1 for medium ICV, and id=2 for larger ICV), as identified in the legend inset (**Figure 2C**). After harmonization, linear models were fitted for the corrected volumes for each subgroup as a function of participants' age to uncover the general data trends during the aging process.

As expected, the negative trend in uncorrected hippocampal volume is observed even for aging in healthy conditions. **Figure 2A** shows that hippocampal volume data points for participants with larger ICV are primarily localized on the top ("+" marker). In contrast, hippocampal measures for smaller ICV are mainly localized on the bottom ("×" marker), which exposes the positive correlation between ICV and hippocampal volume. It is also clear that the graphs for the linear fit of female hippocampal volume are lower than for male data for each diagnostic subgroup, reflecting that females have lower hippocampal volume on average. Moreover, the linear fit slopes are more similar except for the AD participants, where the slope is less negative for females than males (darker/brighter intensity for each color corresponds to the male/female data).

**Figure 2B** shows the differences among the combined subgroups for the harmonized data derived with the residual-data correction approach, equivalent to using the residuals from fitting the polynomial models for each gender, separately (**Figure 2D**). Similarly, **Figure 2C** illustrates the changes observed from the second proposed harmonization procedure, called z-score-data correction, which uses the estimated mean and standard deviation at each interpolation point to calculate the z-scores (see **Materials and Methods**). Data harmonization was utilized to remove the effects of gender, ICV, and age over the harmonized data. For both corrections, we observed that the slopes of the a-posteriori fitted linear models are near zero for each combined subgroup. At the same time, the differences between the genders are smaller within each diagnostic subgroup. Noticeably, we can more easily appreciate that female AD participants at older ages have relatively larger hippocampal volumes than males after data harmonization. For male participants, the differences are more stable between diagnostic subgroups, i.e., the slopes nearly remain the same regardless of the group. **Figure 2D** illustrates the polynomial surface interpolation separately per gender, which is mostly linear except in the borders, where interpolation errors may increase due to scarcer points (more female/male data points for smaller/larger ICV and fewer points for the age range extremes). However, it can also be appreciated that there are apparent nonlinear local changes in the surfaces and slightly more curvature for males than females for the hippocampal volumes (**Figures 2D-E**).

As shown next, the calculated harmonized data (**Figures 2B-C**) can be used for statistical or classification analysis. For example, data harmonization may help to increase the statistical power necessary for variable selection to reduce dimensions before the classification analysis. Moreover, using higher-order polynomial regression may be advantageous in better fitting the nonlinearity in the data. However, this may be an advantage only for larger datasets. In scenarios with a small amount of data, it is advisable to use linear interpolation, especially as the data fitting can be biased at the borders. In our case, we used polynomial fit ('poly22') only for illustrative purposes (based on hippocampal volume data), but in the analyses that follow we used linear interpolation ('poly11') for both residual data and z-score-data correction procedures. The following analyses applied the proposed harmonization procedures to all the features extracted from the Freesurfer's pipeline.

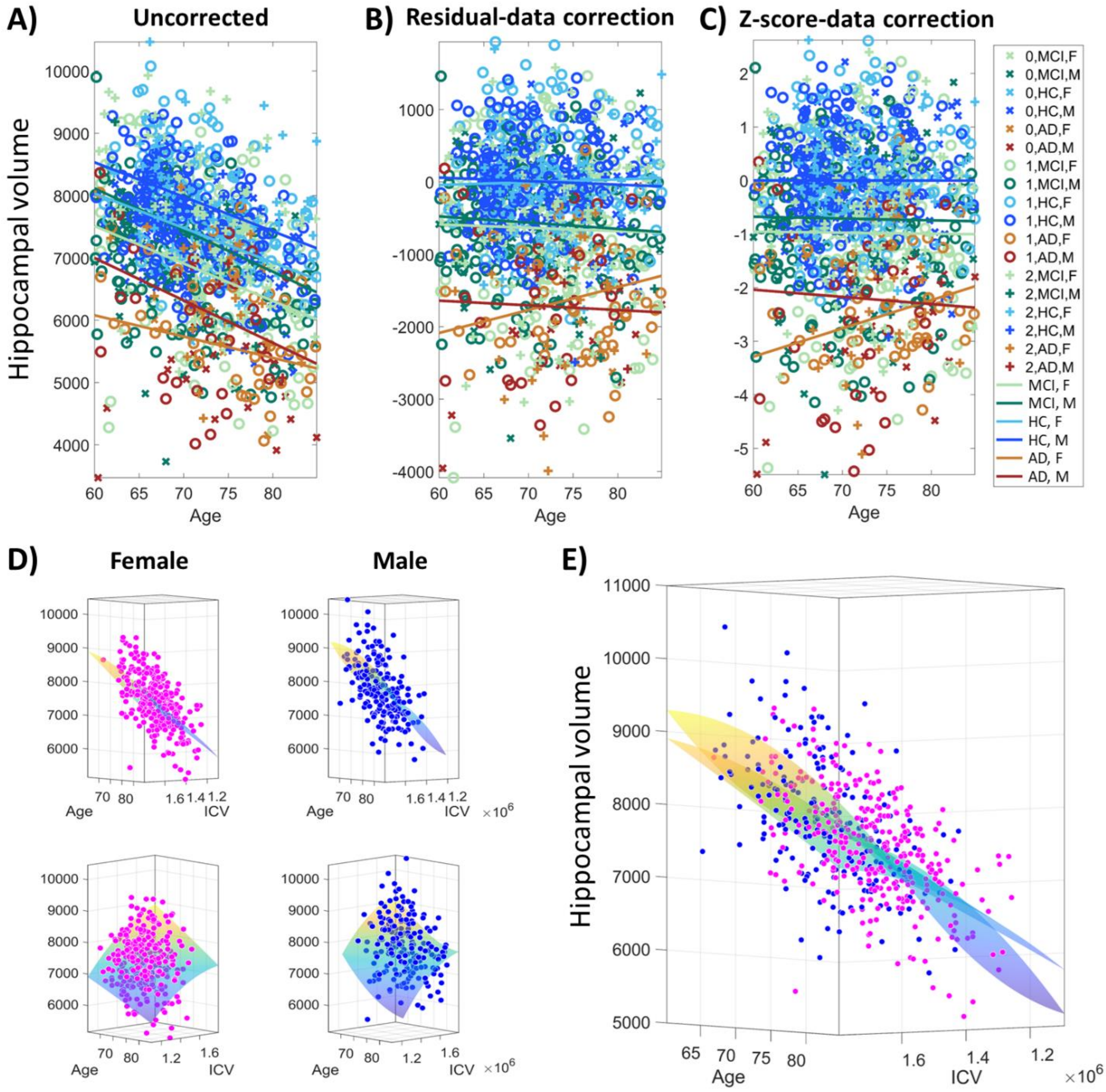

***Figure 2:*** *Data harmonization procedure illustrated for hippocampal volume variable in ADNIMERGE dataset. Healthy Control (HC): females = 306, males = 213. Mild cognitive impairment (MCI): females = 219, males = 284. Alzheimer's disease (AD), females = 56, males = 76. A) Original/uncorrected volume data as a function of age. B) Trend correction using residual -based (linear regression) fit with HC data as reference, calculated separately for female/male subgroups using age and intracranial volume (ICV) as covariates. C) z-score correction using polynomial fit of degree (3,3) for interactions between age and ICV covariates, calculated separately for HC female/male subgroups. The mean and standard deviation of the polynomial fit in every point of the age-ICV subspace is used to calculate the z-score. D) Illustration of the polynomial fitting procedure, separately for HC female/male subgroups. E) Illustration of the polynomial fitting procedure for hippocampal volume in HC female and male subgroups.*

## Statistical analysis

Initially, we investigated early anatomical changes of AD based on the volumes of the Freesurfer-extracted brain regions for the uncorrected data, using ANOVA, Kruskal Wallis, and ANCOVA tests in the SPSS statistical software. Whereas the ANCOVA analysis was performed for the uncorrected data for each brain feature, using age, gender, years of education, and ICV as nuisance variables, ANOVA and Kruskal Wallis were directly applied to all the uncorrected and harmonized data. **Table 3** shows that results are more significant for the harmonized data than the uncorrected data. From these analyses, eight features were consistently found to significantly differ between the HC and uHC groups for the ADNI imbalanced dataset. These eight features were selected for posterior analyses. In contrast, highlighted here only for comparison purposes, sixteen and twelve features were significantly different for the analyses involving the imbalanced OASIS-3 dataset, original and age-matched participants, respectively. As expected, more significant results were obtained as the OASIS-3 cohorts have a larger sample size (**Table 1**). Interestingly, the results demonstrate consistency across the datasets as the eight features found significant with the ADNI data analysis also showed significant results for the OASIS-3 cohorts. Overall, these analyses revealed some advantages of using data harmonization, as the corresponding outcome revealed more significant differences.

***Table 3:*** *Mean volume values for the features that showed significant differences while controlling for multiple comparison using the Bonferroni's correction (p ≤ 0.05), for ADNI and OASIS-3 imbalanced datasets. Values are reported as mean (SD) calculated for the uncorrected and harmonized data. The original uncorrected values for the regional measures, followed by the results for the ANOVA, Kruskal Wallis, and ANCOVA tests are presented across columns. For ANCOVA test for the uncorrected data, the covariates were age, gender, years of education and ICV. Symbol "*" denotes that p-value is not significant. The main important outcomes are for the ADNI data as the corresponding selected brain regions will be used in posterior analysis. The results for OASIS-3 data cohorts are only illustrated for comparison purposes.*

| **Features** | **Uncorrected data** | | | | | **corrected by linear regression** | | **Z-scores** | |
|---|---|---|---|---|---|---|---|---|---|
| | **HC (original volumes, mm$^3$) (SD)** | **uHC (original volumes, mm$^3$) (SD)** | **ANOVA p-value** | **Kruskal Wallis p-value** | **ANCOVA p-value** | **ANOVA p-value** | **Kruskal Wallis p-value** | **ANOVA p-value** | **Kruskal Wallis p-value** |
| **ADNI: age 60-86** | | | | | | | | | |
| **Lateral Ventricle** | 32077.74 (15998.18) | 50914.57 (26158.63) | < 0.001 | < 0.001 | 0.001 | < 0.001 | 0.003 | < 0.001 | 0.002 |
| **Inf-Lat-Vent** | 1165.48 (672.08) | 2163.08 (1471.34) | < 0.001 | < 0.001 | < 0.001 | < 0.001 | 0.002 | < 0.001 | 0.002 |
| **Hippocampus** | 7592.62 (837.74) | 7213.36 (829.68) | *0.051 | *0.092 | 0.010 | 0.005 | 0.015 | 0.005 | 0.014 |
| **Accumbens-area** | 900.29 (165.10) | 761.66 (186.26) | < 0.001 | 0.005 | 0.004 | 0.003 | 0.011 | 0.002 | 0.011 |
| **Entorhinal** | 3582.20 (648.10) | 3378.46 (719.29) | *0.191 | *0.246 | 0.010 | 0.007 | 0.010 | 0.007 | 0.011 |
| **Lateral orbitofrontal** | 13717.41 (1373.75) | 13239.04 (1379.935) | *0.140 | *0.123 | 0.002 | 0.002 | 0.003 | 0.002 | 0.002 |
| **Middle temporal** | 20807.65 (2386.15) | 20222.25 (2563.62) | *0.303 | *0.349 | 0.015 | 0.012 | 0.027 | 0.011 | 0.021 |
| **BrainSegVolNotVent** | 1025466.82 (103015.64) | 1011076.79 (94271.69) | *0.567 | *0.626 | 0.002 | < 0.001 | 0.002 | < 0.001 | 0.002 |
| **OASIS-3: age 43-96** | | | | | | | | | |
| **Lateral Ventricle** | 27870.93 (16230.48) | 43413.22 (23753.60) | < 0.001 | < 0.001 | 0.006 | < 0.001 | 0.032 | < 0.001 | 0.032 |
| **Inf-Lat-Vent** | 1103.52 (665.42) | 2062.28 (1464.24) | < 0.001 | < 0.001 | < 0.001 | < 0.001 | < 0.001 | < 0.001 | < 0.001 |
| **Hippocampus** | 7763.43 (868.24) | 6996.65 (879.65) | < 0.001 | < 0.001 | < 0.001 | < 0.001 | < 0.001 | < 0.001 | < 0.001 |
| **Amygdala** | 3166.92 (464.77) | 2821.13 (529.46) | < 0.001 | < 0.001 | < 0.001 | < 0.001 | < 0.001 | < 0.001 | < 0.001 |
| **Accumbens-area** | 965.69 (192.28) | 806.29 (192.31) | < 0.001 | < 0.001 | < 0.001 | < 0.001 | < 0.001 | < 0.001 | < 0.001 |
| **Entorhinal** | 3691.79 (689.42) | 3401.94 (781.46) | < 0.001 | < 0.001 | < 0.001 | < 0.001 | < 0.001 | < 0.001 | < 0.001 |
| **Fusiform** | 17639.88 (2282.77) | 16793.58 (2541.72) | < 0.001 | < 0.001 | < 0.001 | < 0.001 | < 0.001 | < 0.001 | < 0.001 |
| **Inferior temporal** | 19586.25 (2783.38) | 18433.00 (2929.76) | < 0.001 | < 0.001 | < 0.001 | < 0.001 | < 0.001 | < 0.001 | < 0.001 |
| **Isthmus cingulate** | 4657.37 (687.63) | 4570.60 (685.20) | *0.217 | *0.180 | 0.049 | 0.014 | 0.020 | 0.013 | 0.018 |
| **Lateral orbitofrontal** | 13572.41 (1565.69) | 13221.11 (1591.28) | 0.029 | 0.026 | 0.040 | 0.005 | 0.008 | 0.006 | 0.009 |
| **Medial orbitofrontal** | 10136.71 (1151.98) | 9957.09 (1243.07) | *0.133 | *0.098 | 0.004 | < 0.001 | 0.001 | < 0.001 | 0.002 |
| **Middle temporal** | 20397.11 (2793.42) | 19323.31 (2741.79) | < 0.001 | < 0.001 | < 0.001 | < 0.001 | < 0.001 | < 0.001 | < 0.001 |
| **Para hippocampal** | 2830.92 (526.19) | 3628.37 (556.32) | < 0.001 | < 0.001 | 0.004 | < 0.001 | < 0.001 | < 0.001 | < 0.001 |
| **Superior temporal** | 21659.54 (2508.39) | 20609.39 (2840.67) | < 0.001 | < 0.001 | 0.007 | < 0.001 | < 0.001 | < 0.001 | < 0.001 |
| **Insula** | 13042.72 (1545.59) | 12915.63 (1673.89) | *0.428 | *0.583 | 0.020 | 0.004 | 0.006 | 0.004 | 0.007 |
| **BrainSegVol NotVent** | 1042137.77 (109896.64) | 1006024.23 (109950.49) | 0.001 | 0.003 | < 0.001 | < 0.001 | < 0.001 | < 0.001 | < 0.001 |
| **OASIS-3: age 60-86** | | | | | | | | | |
| **Inf-Lat-Vent** | 1181.78 (681.07) | 2062.96 (1463.92) | < 0.001 | < 0.001 | < 0.001 | < 0.001 | < 0.001 | < 0.001 | < 0.001 |
| **Hippocampus** | 7639.15 (801.46) | 7013.75 (825.28) | < 0.001 | < 0.001 | < 0.001 | < 0.001 | < 0.001 | < 0.001 | < 0.001 |
| **Amygdala** | 3122.47 (430.58) | 2825.05 (480.07) | < 0.001 | < 0.001 | < 0.001 | < 0.001 | < 0.001 | < 0.001 | < 0.001 |
| **Accumbens-area** | 936.00 (173.92) | 805.50 (185.47) | < 0.001 | < 0.001 | 0.003 | < 0.001 | < 0.001 | < 0.001 | < 0.001 |
| **Entorhinal** | 3673.07 (689.16) | 3425.49 (785.00) | 0.001 | < 0.001 | < 0.001 | < 0.001 | < 0.001 | < 0.001 | < 0.001 |
| **Fusiform** | 17451.08 (2175.02) | 16908.48 (2423.70) | 0.026 | 0.032 | 0.017 | < 0.001 | < 0.001 | < 0.001 | < 0.001 |
| **Inferior temporal** | 19404.56 (2752.38) | 18550.90 (2845.48) | 0.005 | 0.008 | < 0.001 | < 0.001 | < 0.001 | < 0.001 | < 0.001 |
| **Medial orbitofrontal** | 10102.50 (1158.77) | 10012.27 (1198.65) | 0.478* | 0.453* | 0.019 | < 0.001 | 0.004 | < 0.001 | 0.004 |
| **Middle temporal** | 20153.47 (2723.91) | 19491.38 (2551.67) | 0.024 | 0.033 | 0.013 | < 0.001 | < 0.001 | < 0.001 | < 0.001 |
| **Para hippocampal** | 3791.19 (516.83) | 3639.69 (545.66) | 0.008 | 0.004 | 0.027 | 0.001 | 0.001 | 0.002 | 0.002 |
| **Superior temporal** | 21337.81 (2289.95) | 20739.03 (2575.37) | 0.020 | 0.014 | 0.051 | 0.002 | 0.003 | 0.003 | 0.004 |
| **BrainSegVol NotVent** | 1031148.72 (103676.92) | 1011836.22 (104411.90) | 0.088* | 0.153* | < 0.001 | < 0.001 | < 0.001 | < 0.001 | < 0.001 |

## Comparison between data harmonization approaches using the MCL app

Here, we performed a preliminary analysis to assess which harmonization procedure could offer superior performance for classification analysis using the MCL app for the ADNI-balanced cohort (see **Figure 1C**). **Table 4** and **Figure 3** display the average performance of the different data harmonization procedures for top-performance classification methods. The best results were achieved for the residual harmonization procedure, with Kernel Naïve Bayes achieving an accuracy of 76.95% and AROC of 84.0% with comparable superior sensitivity and specificity results to other methods. Similarly, the results for the other classification methods were superior for this harmonization procedure except for Coarse Tree. Different classification methods, including Support Vector Machine (SVM) and Logistic Regression (LR), were also evaluated but their results were inferior. This analysis produced better results for residual-corrected data. This may be because, for z-score harmonization, a smaller sample size may negatively impact the calculation of the z-scores, particularly at the borders of the data space. Since the same analysis for the uncorrected data produced the worst results (not shown), we may conclude that data harmonization is important to increase classification performance.

***Table 4:*** *Performance comparison between corrected data centered by linear regression and z-score. Acc=accuracy, Sen=sensitivity, Spec=specificity, AUC= area under the curve.*

| Model | Centered by linear regression | | | | Z-score | | | |
|---|---|---|---|---|---|---|---|---|
| | Acc (%) | Sen (%) | Spec (%) | AUC (%) | Acc (%) | Sen (%) | Spec (%) | AUC (%) |
| **Kernel Naïve Bayes** | 76.95 | 76.84 | 77.11 | 84.0 | 71.80 | 67.43 | 76.05 | 76.0 |
| **Cosine KNN** | 75.23 | 70.17 | 80.00 | 78.0 | 64.10 | 47.37 | 80.00 | 65.0 |
| **Coarse Tree** | 74.40 | 75.00 | 73.68 | 74.0 | 76.90 | 80.00 | 73.68 | 77.0 |
| **Ensemble: Bagged Trees** | 76.90 | 73.68 | 80.00 | 77.0 | 66.65 | 63.16 | 70.00 | 67.0 |

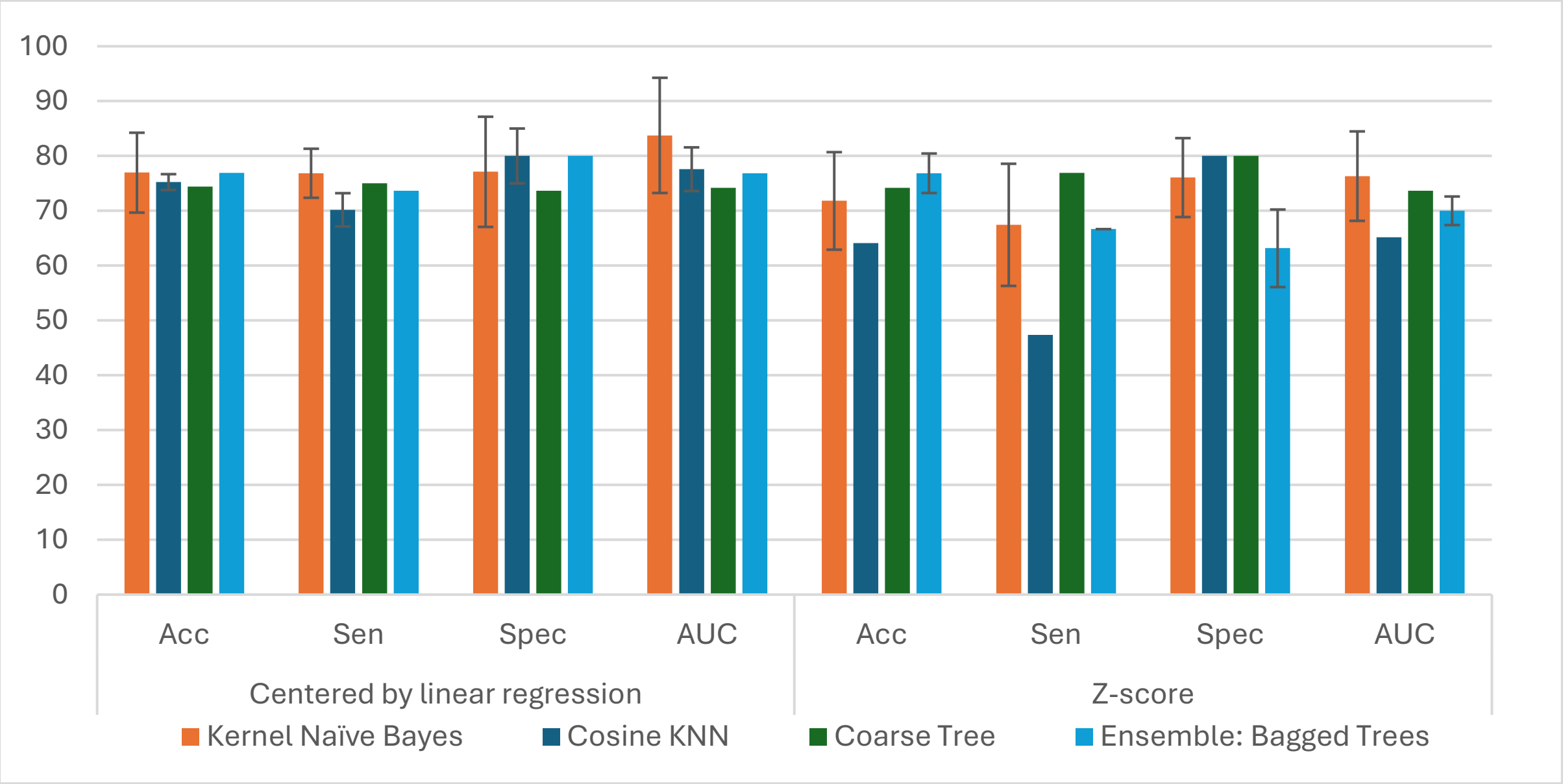

***Figure 3:*** *Comparison in classification between data harmonization procedures using linear regression based centered data and polynomial regression based calculated z-scores. The values in the figure are the same as in Table 3 but the error bars are shown as supplementary information. The columns present the results, in this order, for the four presented methods and the different calculated statistics (Acc – accuracy, Sen – sensitivity, Spec – specificity, AUC – area under curve of the ROC graph).*

## Features and classification methods evaluation using the MCL App

As a complement to the previous analysis, we also performed an exhaustive analysis to select the "best" feature selection and classification methods as provided in the MCL app, using the ADNI-balanced cohort (see **Figure 1C**). Using the feature selection methods available in the app, we calculated the percentages for the 40 features. We ignored the MRMR outcome, as only one feature (Inf-Lat-Vent volume) exhibited a score greater than 0. **Table 5** reveals that the chi-square provides lower scores when compared to ANOVA and Kruskal Wallis. Conversely, ANOVA and Kruskal Wallis exhibited minimal discrepancies in their scores. This is the primary reason we converted the scores into percentages to enhance visual comparison and selection of the most relevant features. Then, we ranked the features based on how frequently the feature selection methods selected them. As shown in **Table 6**, we found five features selected by all selection criteria. These features are BrainSegVolNotVent, inferior lateral ventricle, entorhinal, lateral orbitofrontal, and lateral ventricle. Then, with a slight rank, the parahippocampal and hippocampus regions were selected by 5/6 of the selection criteria.

***Table 5:*** *Score rating by classification learner application with percentage.*

| Features | Chi-square | | ANOVA | | Kruskal Wallis | | ReliefF | | Average Score (%) Median= 1.79 |
|---|---|---|---|---|---|---|---|---|---|
| | Score | % Median= 1.94 | Score | % Median= 2.13 | Score | % Median= 1.94 | Score | % | |
| **entorhinal** | 2.28 | 3.80 | 5.48 | 6.07 | 5.48 | 6.35 | 0.04 | 13.50 | 7.43 |
| **fusiform** | 1.44 | 2.40 | 1.22 | 1.35 | 1.68 | 1.94 | 0.03 | 10.51 | 4.05 |
| **Inf-Lat-Vent** | 2.83 | 4.72 | 4.13 | 4.58 | 4.49 | 5.20 | 0.03 | 9.88 | 6.09 |
| **temporalpole** | 3.48 | 5.80 | 0.44 | 0.48 | 2.06 | 2.39 | 0.03 | 8.89 | 4.39 |
| **posteriorcingulate** | 1.44 | 2.40 | 0.68 | 0.75 | 0.92 | 1.07 | 0.02 | 7.11 | 2.83 |
| **isthmuscingulate** | 1.54 | 2.57 | 1.30 | 1.44 | 2.27 | 2.63 | 0.02 | 7.02 | 3.42 |
| **Hippocampus** | 1.14 | 1.90 | 4.29 | 4.75 | 4.14 | 4.80 | 0.02 | 6.70 | 4.54 |
| **parahippocampal** | 4.09 | 6.82 | 3.27 | 3.62 | 2.71 | 3.14 | 0.02 | 5.84 | 4.86 |
| **parsopercularis** | 1.34 | 2.23 | 1.79 | 1.99 | 1.75 | 2.03 | 0.02 | 5.62 | 2.97 |
| **Lateral-Ventricle** | 2.83 | 4.72 | 4.57 | 5.07 | 5.42 | 6.28 | 0.02 | 4.92 | 5.25 |
| **transversetemporal** | 0.63 | 1.05 | 0.02 | 0.02 | 0.26 | 0.30 | 0.01 | 4.70 | 1.52 |
| **precentral** | 0.70 | 1.17 | 3.81 | 4.22 | 3.81 | 4.42 | 0.01 | 4.45 | 3.56 |
| **insula** | 3.55 | 5.91 | 2.04 | 2.26 | 2.14 | 2.49 | 0.01 | 4.16 | 3.70 |
| **frontalpole** | 0.20 | 0.34 | 1.49 | 1.65 | 0.73 | 0.84 | 0.01 | 2.70 | 1.38 |
| **superiortemporal** | 1.39 | 2.32 | 0.28 | 0.31 | 0.36 | 0.42 | 0.00 | 1.52 | 1.14 |
| **Amygdala** | 2.77 | 4.61 | 0.87 | 0.95 | 1.16 | 1.35 | 0.00 | 1.17 | 2.02 |
| **lateralorbitofrontal** | 2.22 | 3.70 | 4.93 | 5.46 | 4.79 | 5.55 | 0.00 | 0.60 | 3.83 |
| **lingual** | 0.42 | 0.70 | 0.37 | 0.41 | 0.12 | 0.14 | 0.00 | 0.57 | 0.46 |
| **BrainSegVolNotVent** | 5.45 | 9.08 | 9.80 | 10.86 | 9.76 | 11.30 | 0.00 | 0.13 | 7.84 |
| **parstriangularis** | 0.48 | 0.81 | 0.22 | 0.25 | 0.30 | 0.35 | -0.00 | | 0.35 |
| **middletemporal** | 1.14 | 1.90 | 4.01 | 4.44 | 3.19 | 3.70 | -0.00 | | 2.51 |
| **paracentral** | 1.34 | 2.23 | 2.48 | 2.75 | 1.60 | 1.87 | -0.01 | | 1.71 |
| **precuneus** | 0.83 | 1.38 | 2.21 | 2.45 | 1.50 | 1.73 | -0.01 | | 1.39 |
| **Accumbens-area** | 1.44 | 2.40 | 4.12 | 4.56 | 3.65 | 4.23 | -0.01 | | 2.80 |
| **medialorbitofrontal** | 0.16 | 0.27 | 0.86 | 0.95 | 0.81 | 0.94 | -0.01 | | 0.54 |
| **inferiorparietal** | 0.42 | 0.70 | 1.61 | 1.78 | 1.20 | 1.39 | -0.01 | | 0.97 |
| **superiorfrontal** | 1.05 | 1.75 | 1.93 | 2.14 | 1.68 | 1.94 | -0.02 | | 1.46 |
| **postcentral** | 0.27 | 0.46 | 1.77 | 1.96 | 1.26 | 1.46 | -0.02 | | 0.97 |
| **caudalmiddlefrontal** | 0.48 | 0.81 | 1.97 | 2.18 | 1.98 | 2.30 | -0.02 | | 1.32 |
| **lateraloccipital** | 1.60 | 2.66 | 1.92 | 2.13 | 1.68 | 1.94 | -0.02 | | 1.68 |
| **cuneus** | 0.05 | 0.08 | 0.10 | 0.11 | 0.26 | 0.30 | -0.21 | | 0.12 |
| **bankssts** | 0.70 | 1.17 | 2.34 | 2.59 | 2.36 | 2.73 | -0.02 | | 1.62 |
| **caudalanteriorcingulate** | 1.19 | 1.98 | 1.17 | 1.30 | 1.01 | 1.17 | -0.02 | | 1.11 |
| **parsorbitalis** | 1.99 | 3.31 | 2.02 | 2.23 | 1.64 | 1.90 | -0.03 | | 1.86 |
| **superiorparietal** | 0.20 | 0.34 | 0.29 | 0.32 | 0.34 | 0.40 | -0.03 | | 0.26 |
| **inferiortemporal** | 1.00 | 1.67 | 2.49 | 2.76 | 1.64 | 1.90 | -0.03 | | 1.58 |
| **rostralmiddlefrontal** | 2.77 | 4.61 | 3.16 | 3.49 | 2.40 | 2.78 | -0.03 | | 2.72 |
| **rostralanteriorcingulate** | 1.05 | 1.75 | 3.43 | 3.80 | 2.86 | 3.30 | -0.03 | | 2.21 |
| **pericalcarine** | 1.14 | 1.90 | 0.35 | 0.39 | 0.22 | 0.25 | -0.03 | | 0.64 |
| **supramarginal** | 0.96 | 1.60 | 1.08 | 1.20 | 0.67 | 0.78 | -0.04 | | 0.89 |

***Table 6:*** *Feature selection according to different selection criteria.*

| **Features** | **Chi-square** | **ReliefF** | **ANOVA** | **Kruskal Wallis** | **Average score (MCL app): chi-square, ReliefF, ANOVA & Kruskal Wallis** | **Statistical analysis (SPSS): ANOVA, ANCOVA & Kruskal Wallis** | **Total** |
|---|---|---|---|---|---|---|---|
| **BrainSegVolNotVent** | / | / | / | / | / | / | 6 |
| **Inf-Lat-Vent** | / | / | / | / | / | / | 6 |
| **Entorhinal** | / | / | / | / | / | / | 6 |
| **Lateral orbitofrontal** | / | / | / | / | / | / | 6 |
| **Lateral Ventricle** | / | / | / | / | / | / | 6 |
| **Parahippocampal** | / | / | / | / | / | | 5 |
| **Hippocampus** | | / | / | / | / | / | 5 |
| **Accumbens-area** | | | / | / | / | / | 4 |
| **Middle temporal** | | | / | / | / | / | 4 |
| **Precentral** | | / | / | / | / | | 4 |
| **Insula** | / | / | | | / | | 3 |
| **Temporal pole** | / | / | | | / | | 3 |
| **Rostral middle frontal** | / | | / | | / | | 3 |
| **Rostral anterior cingulate** | | | / | / | / | | 3 |
| **Amygdala** | / | / | | | / | | 2 |
| **Fusiform** | | / | | | / | | 2 |
| **Posterior cingulate** | | / | | | / | | 2 |
| **Isthmuscingulate** | | / | | | / | | 2 |
| **Parsopercularis** | | / | | | / | | 2 |
| **Parsorbitalis** | / | | | | / | | 2 |
| **Transverse temporal** | | / | | | | | 1 |
| **Frontal pole** | | / | | | | | 1 |
| **Superior temporal** | | / | | | | | 1 |
| **Lingual** | | / | | | | | 1 |

Subsequently, we calculated the average for classification accuracy, sensitivity, specificity and AROC statistics for each classifier and feature selection method for ten random replications. **Table 7** reveals that Kernel Naïve Bayes was selected 34% of the time as the best-performance classifier, and its average accuracy was 77.3% by pooling together all the corresponding outcomes from the feature selection methods. Gaussian Naïve Bayes and Cosine KNN were tied up in second place, selected 7% of the time as the top performer, and with average accuracy performance of 73.05% and 71.5%, respectively. For the other classifier, the LR achieved an average accuracy of 75.65% with an AROC of 0.7592 when using the ReliefF method. However, it did not perform well for the other feature selection criteria. Regarding the best feature selection methods in the MCL app, ReliefF outperformed the other methods (**Figure 4**). It is important to emphasize that the outcome from **Tables 3-7** and **Figures 3-4** were derived from evaluation exclusively on the ADNI dataset. In the next section, we evaluate the generalization of these results using both ADNI and OASIS-3 derived cohorts with our customized pipeline.

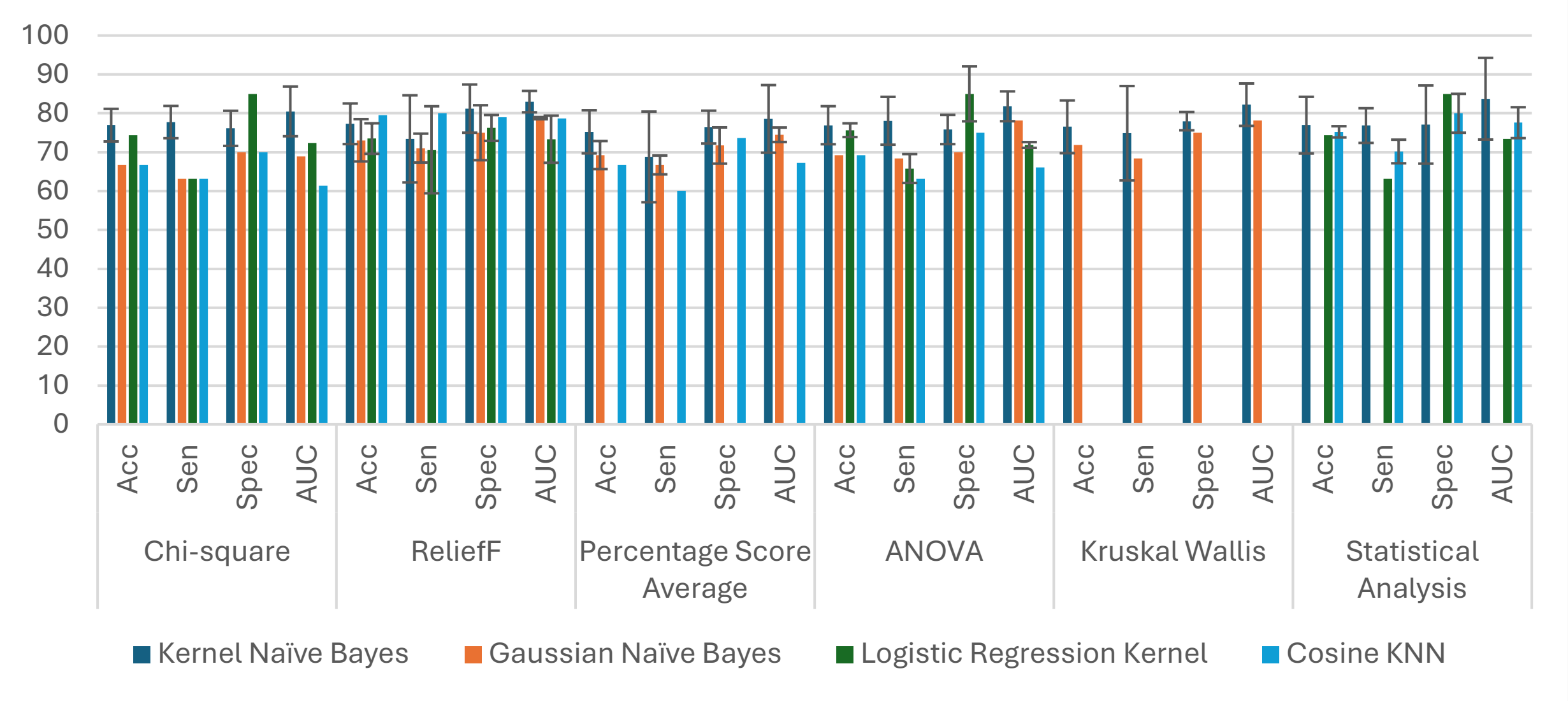

***Figure 4:*** *Average classification in model selection analysis. Some graphs do not display the standard error since the model only appears once throughout the procedure. Acc=accuracy, Sen=sensitivity, Spec=specificity, AUC= area under the curve.*

***Table 7:*** *Selection frequency as top performer for each classification method under different feature selection criteria. The results for the classification methods are presented across the rows, whereas the results for the different feature selection strategies.*

| **Models** | **Chi-square** | **ReliefF** | **ANOVA** | **Kruskal Wallis** | **Average score (MCL app): chi-square, ReliefF, ANOVA & Kruskal Wallis** | **Statistical analysis (SPSS): ANOVA, ANCOVA & Kruskal Wallis** | **Total appearances** | **Total (%)** |
|---|---|---|---|---|---|---|---|---|
| **Kernel Naïve Bayes** | 7 | 7 | 6 | 6 | 6 | 2 | 34 | 34 |
| **Gaussian Naïve Bayes** | 1 | 2 | 1 | 1 | 2 | | 7 | 7 |
| **Cosine KNN** | 1 | 1 | 1 | | 1 | 3 | 7 | 7 |
| **Logistic Regression Kernel** | 1 | 2 | 2 | | | 1 | 6 | 6 |
| **Weighted KNN** | | | 1 | | | 3 | 4 | 4 |
| **Subspace Discriminant** | 3 | | 1 | | | | 4 | 4 |
| **Subspace KNN** | 1 | | 1 | 1 | | | 3 | 3 |
| **Medium KNN** | | | 1 | 1 | 1 | | 3 | 3 |
| **SVM Kernel** | 1 | 1 | | 1 | | | 3 | 3 |
| **Linear SVM** | 1 | 1 | | | | 1 | 3 | 3 |
| **Fine tree** | 1 | | | | 1 | 1 | 3 | 3 |
| **Medium tree** | 1 | | | | 1 | 1 | 3 | 3 |
| **Coarse tree** | 1 | | | | 1 | 1 | 3 | 3 |
| **Linear Discriminant** | 2 | | | | 1 | | 3 | 3 |
| **Quadratic SVM** | | 1 | 1 | | | | 2 | 2 |
| **Trilayered Neural Network** | | 1 | | | 1 | | 2 | 2 |
| **Cubic KNN** | | | | | | 2 | 2 | 2 |
| **Fine KNN** | | 1 | | | | 1 | 2 | 2 |
| **Logistic Regression** | | 1 | | | | | 1 | 1 |
| **Ensemble: Subspace Discriminant** | | | | | 1 | | 1 | 1 |
| **Cubic SVM** | | 1 | | | | | 1 | 1 |
| **Medium Gaussian SVM** | | 1 | | | | | 1 | 1 |
| **Bagged Trees** | 1 | | | | | | 1 | 1 |
| **Ensemble: Bagged Trees** | | | | | | 1 | 1 | 1 |

## Balanced data analysis with customized pipeline

In the present and next section, we further evaluate the selected "best" combination for classification methods, selected features, and data harmonization procedures with our customized pipeline, for balanced and imbalance datasets respectively. Five classifiers are compared in these analyses: Naïve Bayes, KNN, SVM, LR, and RUSBoost. The purpose is mainly to further evaluate the "best" candidates selected from the above MCL app analysis, compared against RUSBoost which is expected to show superior performance for imbalanced datasets. In contrast to the MCL app and feature selection analyses above, which could have had some bias due to the MCL app analysis been restricted to a single balanced ADNI cohort, the current analyses are extended to include both the ADNI and OASIS-3 imbalanced cohorts, as well as the OASIS-3 age-matched cohort, to evaluate the selected method combinations. However, in this section we performed randomly undersampling to balance these datasets within a Monte Carlo replication analysis which subsequently runs our customized pipeline for each of the combined choices, among four different groups of selected features (subsets **A-D** as shown in **Table 8**), five classifiers, three datasets, and two harmonization procedures. This analysis may favor classifiers that perform better in balanced data scenarios, which can be compared against the following section results, where the same analysis will be applied without undersampling, i.e., for the original imbalanced datasets. Evaluations are based on the following performance metrics: Acc, AROC, F1, and MCC′. Interestingly, for balanced data analysis, metrics such as Acc and AROC serve as standard to evaluate the "best" classification performances, but this may not be the case for imbalanced data analysis, where F1 and MCC′ are recommended (**Materials and Methods**).

**Figure 5** shows the results for the balanced data analysis. The best outcome for ADNI balanced cohorts was achieved using a Naïve Bayes classifier using the ReliefF feature selection (subset B) and *z*-score data harmonization, achieving Acc = 69.17 ± 6.54 %, AROC = 77.73 ± 7.08 %, F1 = 69.21 ± 7.90 %, and MCC′ = 69.28 ± 6.56 % (FDR-adjusted p-value $p_{FDR} < 0.05$ for all multiple comparisons of Naïve Bayes vs other classifiers for each metric). However, this result was not replicated for the OASIS-3 cohorts, possibly revealing a selection bias as an individual balanced ADNI cohort was utilized in the previous MCL analysis for feature selection. For OASIS-3 age-matched dataset, the best performance was obtained for the SVM classifier using the features selected in subset D and z-score data harmonization, with Acc = 66.58 ± 2.91 %, AROC = 72.01 ± 2.40 %, and MCC′ = 66.78 ± 2.96 %. Logistic regression showed the best performance according to the F1 score of 66.68 ± 1.21% for ReliefF features and residual harmonization.

When pooling together measures calculated for the four feature subsets for the F1 score, ANOVA with multiple comparison analysis revealed that LR was the best classifier, significantly superior from all the other classifiers for all the three datasets using the residual harmonization approach (see **Supplementary Figure 7** and **Supplementary Table 9**). Similar analysis for the MCC′ score revealed that Naïve Bayes using z-score harmonization for ADNI dataset was superior to the other approaches except for SVM for all the three datasets and z-score harmonization (see **Supplementary Figure 8** and **Supplementary Table 10**).

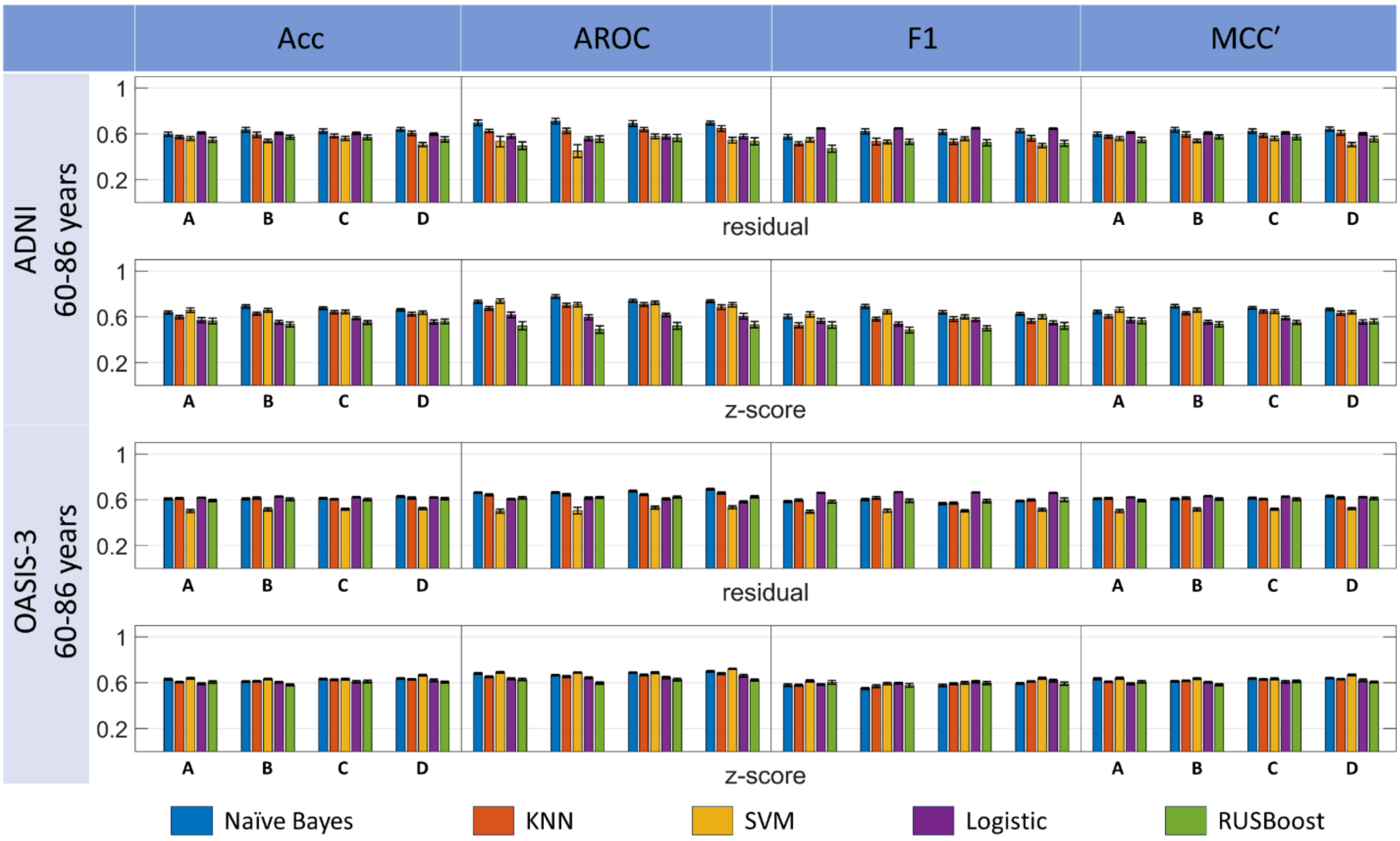

***Figure 5:*** *Comparison among multiple classification pipeline options, involving five classifiers, four feature selection and two harmonization techniques. Performance is measured for randomly balanced cohorts extracted from ADNI and OASIS-3 imbalanced datasets within a Monte Carlo replication analysis. Results are presented for Naïve Bayes, KNN, SVM, Logistic and RUSBoost, residual and z-score harmonization procedures, as represented in the x-axis and legend labels. Bar groups denoted by letters **A-D** indicate the outcomes corresponding to the different features selected in the MCL analysis: (**A**) Features selected using the average scores (**Table 8A** lists the feature labels); (**B**) Features selected based on the ReliefF criterion (**Table 8B** lists the feature labels); (**C**) Features selected according to the combination of all evaluated feature selection algorithms (**Table 8C** lists the feature labels); (**D**) Features selected within the SPSS statistical analysis (**Table 8D** lists the feature lables). Column labels: Acc=accuracy. AROC=Area under Receiver Operating Curve. F1= F1 score, MCC=Matthew's correlation coefficient. Performance metrics are normalized in the range [0, 1] and plotted together to enhance visual comparison.*

## Imbalanced data analysis with customized pipeline

Analogous to the above analysis, **Figure 6** shows the results for the imbalanced data analysis. Here, it is clear the divergence among performance metrics. Although the accuracy indicates that SVM may be the best classifier, F1 and MCC′ significantly favor RUSBoost at least for the ADNI data analysis. With a detailed inspection, we may realize that the accuracy could be biased in this case as the SVM tends to favor the majority group (HC) at expense of poor classification for the minority group (uHC). Apart from being reflected by the corresponding F1 and MCC scores, this is more clearly visible by inspecting the corresponding true positive rate (TPR) and positive predictive value (PPV) scores which highlights an overall instability of the SVM classifier in imbalanced data analysis (see **Supp. Materials' Figure 2 and Tables 3-4**).

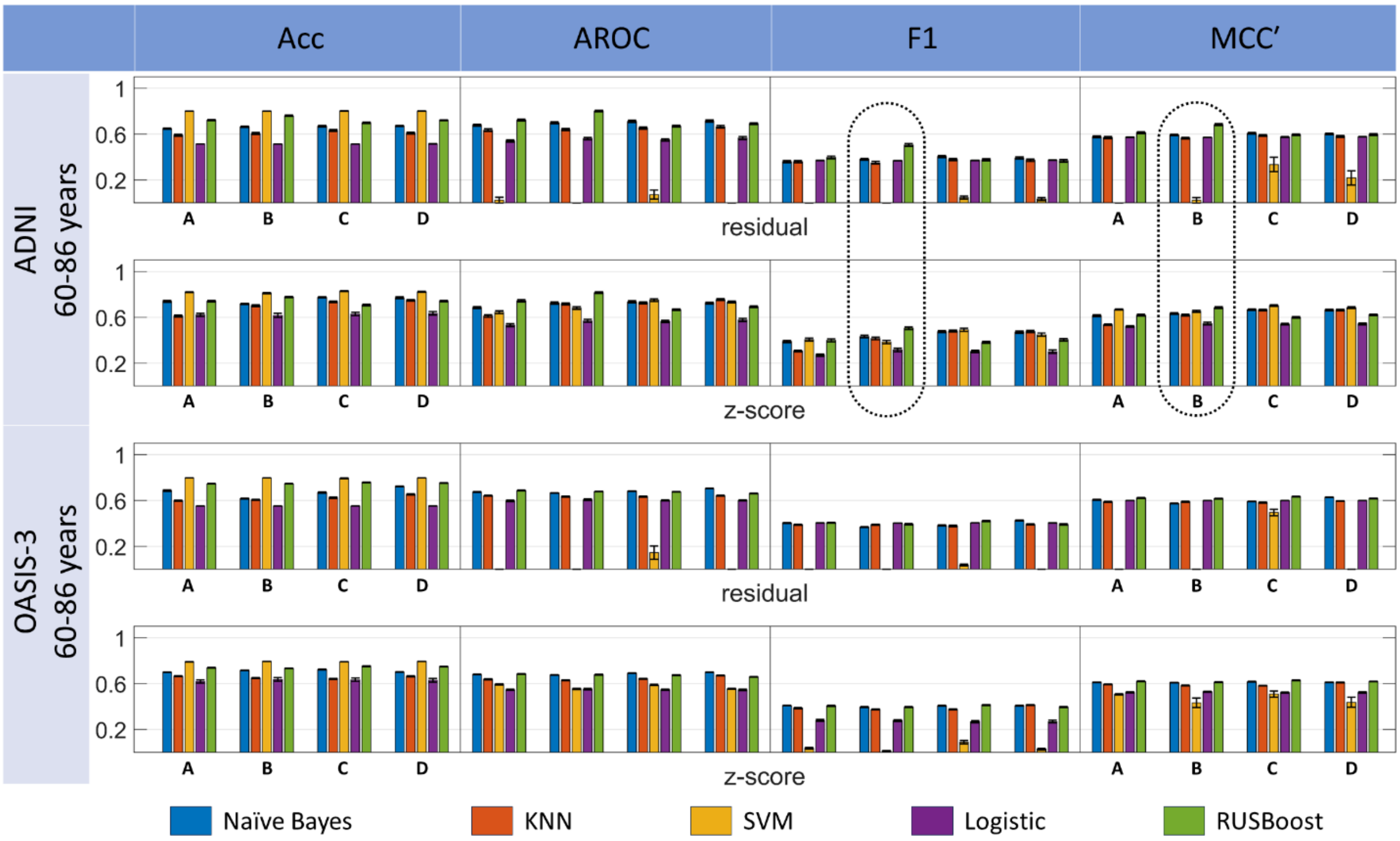

***Figure 6:*** *Comparison among multiple classification pipeline options, involving five classifiers, four feature selection, and two harmonization techniques. Performance is measured directly for ADNI and OASIS-3 imbalanced datasets within a Monte Carlo replication analysis. See* ***Figure 4*** *for complementary information regarding balanced data analysis and figure caption.*

For the ADNI imbalanced cohort, RUSBoost achieved the best performance according to two metrics: F1 = 50.60 ± 5.20 % (based on ReliefF features and residual harmonization) and AROC = 81.54 ± 2.92 % (based on ReliefF features and z-score harmonization). Whereas SVM showed the best performance according to the other metrics for subset C features and z-score harmonization in both cases: Acc = 82.93 ± 1.59 % and MCC′ = 70.21 ± 3.16 %. For the OASIS-3 age-matched dataset, Naïve Bayes showed the best performance according to F1 (42.54 ± 1.71 %, $p_{FDR} < 0.05$) and AROC (70.33 ± 1.00 %; $p_{FDR} < 0.05$), for subset D features and residual harmonization in both cases, with RUSBoost dominating for the MCC′ score (63.31 ± 1.43 %), for subset C features and residual harmonization. Here, the accuracy performance was dominated by SVM (79.58 ± 0.00), but this result is invalid since it matches the percentage of the majority class for this dataset, i.e., 413 ÷ (413 + 106) × 100 % (**Table 1**).

Interestingly, although in the balanced data analysis above we employed the default cost error matrix (i.e., $[0\ 1; 1\ 0]$ in order following MATLAB notation), for the imbalanced data analysis we compensated the other classifiers (except RUSBoost) with a customized cost matrix penalizing more the error committed for classifying a sample in the majority class when the true class is the critical one: $[0\ 1; \delta\ 0]$ where $\delta$ is the ratio between the cardinalities of the majority and minority groups. When this correction is ignored, the other methods show very poor results. This correction is not needed for RUSBoost as confirmed with our evaluations due to RUSBoost implementation directly based on random undersampling (RUS).

For the imbalanced data analysis, when pooling together measures calculated for the four feature subsets for the F1 score, ANOVA with multiple comparison analysis revealed that Naïve Bayes using z-score harmonization for ADNI dataset was superior to the other approaches except for SVM for the same combination (see **Supplementary Figure 9** and **Supplementary Table 11**). In contrast, for the MCC′ score, the roles were reversed with SVM followed by Naïve Bayes as superior to the rest, also for z-score harmonization of ADNI data, (see **Supplementary Figure 10** and **Supplementary Table 12**). Remarkably, results for imbalance data analysis were significantly worse than the corresponding ones for balanced analysis, and results were also inferior for the OASIS-3 with respect to ADNI dataset.

# Discussion

In this paper, our primary objective was to develop an MRI-based methodology for early AD prediction, motivated by the fact that MRI is a well-established and widely used technique, providing detailed images for assessing brain regional integrity. This approach enables tracking anatomical changes in the brain during healthy aging and disease progression. In summary, we target the detection of brain changes associated with early cognitive decline by comparing MRI images of elders that remained healthy (HC group) to the images of other initially healthy elders, who were later diagnosed as mild cognitive impairment (MCI) in a period of 5 years (uHC group), with data provided by ADNI and OASIS-3 longitudinal studies. We presented a machine learning (ML) approach to evaluate multiple feature selection and classification methods. Particularly, combining feature selection and statistical analysis methods, we found that six out of eight significantly detected brain regions in our analyses are consistently reported in the literature as related to early AD anatomical changes: entorhinal, hippocampus, lateral ventricle, lateral orbitofrontal, accumbens area, and middle temporal (see **Tables 3**, **5-6**).

These regions are central to the limbic system's functioning and pivotal in regulating emotions, memory, executive functions, and behavior [84]. Therefore, anatomical and functional alterations observed in these regions could be critically associated with the early progression of neurodegenerative disorders [85,86]. Notably, the hippocampus and entorhinal cortex [87,88] are frequently observed to be affected to a significant degree during the early MCI stage [89]. Additionally, changes in the lateral ventricle's size or shape can indicate certain neurological conditions, including neurodegenerative disease, as previously observed in AD [8]. Moreover, the orbitofrontal cortex is critical in decision-making, impulse control, and evaluating reward and punishment stimuli. Damage or dysfunction in this area can lead to impairment in these functions and changes in behavior, which are the most common observed symptoms in AD patients, but even in individuals who may receive an MCI diagnosis, as demonstrated by a post-mortem analysis [90]. Moreover, consistent with our results, previous studies have found that the entorhinal cortex is one of the earliest brain regions affected by AD, leading to gradual memory deficits [87–89]. The entorhinal cortex is closely connected to the hippocampus via the subiculum, and it is a critical brain region involved in memory formation, spatial navigation, and the processing of associations between different pieces of information.

Our proposed methodology also evaluated the performance of different ML approaches for balanced and imbalanced data analyses using ADNI and OASIS-3 datasets. First, mainly using the MATLAB Classification Leaner (MCL) app for an accelerated exploration and discover of "best" method candidates for further analysis, among a vast number of available techniques. This preliminary analysis enabled the evaluation of the consistency of selected features and classifiers' performance in different conditions. In this analysis, we found that using ReliefF [80] consistently outperformed other techniques, although the same methods were not observed as superior through different scenarios (**Supplementary Material Tables 1-4**). Interestingly, the stable selection of the same group of brain regions by the different techniques highlighted the importance of the regions mentioned above and our methodology to uncover early AD-linked brain changes.

Although, our main goal is to discover MRI-based biomarkers associated with AD, this must go through the exploration and analysis using a wide range of available techniques, as some may be more appropriate than others. In this sense, the MCL app helped to considerably reduce our preselection efforts. This app includes many popular algorithms such as decision trees, discriminant analysis, logistic regression, naïve Bayes, support vector machines, nearest neighbors, kernel approximation, ensemble methods, and neural networks, within predefined templates. Particularly, it predefines Bilayered Neural Networks (BNNs) and Wide Neural Network (WNNs) architecture models from the neural networks' family. MCL app's BNNs consist of two hidden layers with 10 neurons in each layer. In contrast, WNNs are predefined with a single hidden layer with 100 neurons by default in the MCL app. The activation function is ReLu in both cases by default. Although a more flexible option exists within the MCL app graphical interface to run these models using Bayesian optimization, we preferred to run and evaluate all the predefined models in the MCL app, including the above predefined BNN and WNN models, since these calculations are computationally intensive. In general, using the "Optimizable" model options in the MCL app enables selecting between different hyperparameter options automatically through Bayesian optimization. For example, in the case of neural networks, these options are for the number of layers, number of neurons per layer, and activation function type, among others. However, apart from the intensive computational reason, we decided to evaluate "only" all the predefined models, since we also assess in our study a more advanced ML pipeline implementing nested cross-validation with Bayesian optimization for selected models, evaluated within a Monte Carlo replication framework.

Note also that above analyses in MCL may be limited as we exclusively used a single randomly balanced cohort, extracted from the ADNI dataset, for this analysis, since MCL app's algorithms are optimized for balanced data analysis. This is compensated in our study as our customized ML pipeline was applied after this preliminary analysis using the original imbalanced ADNI and OASIS-3 datasets, for a more in-depth evaluation of selected features and classification models. This helped us to obtain a more solid evaluation of five popular techniques (including Naïve Bayes, SVM, and RUSBoost), based on the implementation of nested cross-validation with Bayesian optimization in our pipeline, evaluated within a Monte Carlo replication analysis designed to produce more stable results. The same pipeline was used for both balanced and imbalanced data analyses for the same ADNI and OASIS-3 datasets. The unique difference is the implementation of random rebalancing (generate random subsets from the original HC and uHC groups with same number of samples) within the Monte Carlo analysis, before the evaluation of classification models with the balanced cohorts. We included RUSBoost in our study since it has been reported as one of the best ML algorithms for imbalanced data analysis [60,61]. We also implemented a rich set of evaluation metrics, including the F1 score, and Matthew's correlation coefficient (MCC), in addition to the traditional accuracy (Acc) and area under receiver operating characteristic curve (AROC) measures, because they are deemed more appropriate for imbalanced classification analysis [63,64]. Overall, our study suggests that popular algorithms such as Naïve Bayes and Logistic regression could be very competitive even for imbalanced data analysis when the algorithm's cost matrix is set conveniently as in our case, or when random undersampling is considered as in our pipeline implementation (see **Figures 5-6** and discussion therein).

As a warning for future research in this area, all the performance metrics in our study (e.g., Acc, F1 and MCC) showed overfitting (see **Supplementary Figures 3-6**, in contrast with **Supplementary Figures 7-10**), and incorrectly addressing this issue could have negatively impacted our observations. In our case, the use of nested cross-validation (nested-CV) helped us to address the overfitting effects and achieve more robust outcomes, as noted in the literature [91,92] On the negative side, RUSBoost outcome seems to be more affected by overfitting as it shows very superior performance for imbalanced data analysis based on direct CV measurements (**Supplementary Figures 5 and 6**, for F1 and MCC′ scores, respectively) but this advantage completely disappeared when using the nested-CV (holdout) measurements (see **Supplementary Figures 9 and 10**, respectively, for corresponding score comparisons). This may question the validity of previous results based on RUSBoost for imbalanced data analysis, if these studies did not implement a more cautious strategy, like in our case, based on nested cross-validation.

Not least relevant, when comparing directly the balanced analysis vs their imbalanced counterparts for nested-CV measurements (see **Supplementary Figures 7 vs. 9**, for visual comparison based on the F1 score, and **Supplementary Figures 8 vs. 10**, for visual comparison based on the MCC score), it is thought-provoking that balanced analysis produced significantly superior performance results than imbalanced (note the difference in x-axis tick range). This observation takes into consideration that both analyses used the same nested-CV pipeline for the same datasets, with the only difference being that balanced analysis applied the nested-CV pipeline for randomly balanced cohorts, extracted from the same imbalanced datasets within a Monte Carlo replication analysis (see **Material and Methods** for more information; see also **Supplementary Materials Table 13-14** for more detail). This suggests that the rebalancing approach, evaluated here with our customized pipeline, could serve to improve imbalanced data analysis.

## Limitations

It is essential to acknowledge the limitations of our study. Primarily, for the balanced data analysis using the ADNI dataset, a smaller sample size can be used to argue for the possible unreliability of the presented results. This small number was mainly due to our consideration of data acquired only with the 3T MRI technique, but it also may be attributed to the challenge of studying very early AD-linked brain changes. Notice that ADNI also provides data for 1.5T acquired for the very early participants in this study, which we did not consider avoiding this additional confounding factor. However, we can increase the sample size in future studies by considering this and robustly controlling for the possible heterogeneity between 1.5T and 3T MRI images. Additionally, as ADNI is a still ongoing longitudinal study, we can access more data in the future or possibly use different search criteria and methodology to increase the sample size. Another explicit limitation is that the balanced ADNI cohort used in the MCL app analysis may have been selected arbitrarily. We performed a random selection to exclude subjective bias, but once selected a unique ADNI-balanced cohort was used for the feature and classifier selection process. However, this limitation appears in many studies limited by small sample size data. In our case, this was compensated later with more robust analyses, based on the original extracted imbalanced ADNI and OASIS-3 datasets, using our customized ML pipeline which enables the implementation of both balanced and imbalanced data analysis. Another clear limitation is restricting our research to MRI-based AD biomarkers, which we are currently addressing in an ongoing study which also includes magneto-electroencephalographic (MEG/EEG) features. However, the use of our pipeline, as well as the findings reported in our study, could be valid for more complex analyses involving multimodal neuroimaging features. Lastly, the current research is still far from the goal of developing a quasi-automatic procedure to evaluate early Alzheimer's disease cases.

## Conclusions

This study comprehensively compared multiple strategies to identify the most effective predictors and optimize classifier models for early cognitive decline prediction from MRI data. We identified the predictors through a comprehensive statistical analysis conducted on uncorrected and corrected/harmonized data using three different analytical approaches: one-way ANOVA, ANCOVA, and Kruskal Wallis. Moreover, using the MCL app, we also analyzed four feature ranking methods (Chi-square, ANOVA, Kruskal Wallis, and ReliefF) and multiple classification methods to reduce the number of selected features and classification models for posterior analyses. Subsequently, we used a customized pipeline implementing nested cross-validation pipeline and Bayesian optimization to evaluate the selected features and classification models further within a Monte Carlo replication framework. We enhanced our assessment of the "best" features and models by analyzing this pipeline's outcome using N-way ANOVA and multiple comparison methods for assessed performance metrics (e.g., accuracy, F1, and MCC). To ensure the robustness and reproducibility of our results, we validated our methodology by using both ADNI and OASIS-3 datasets. Overall, we corroborated that using harmonization approaches improves the evaluation and selection of biomarkers and classification algorithms, and that imbalanced data analysis could be improved with ideas such as random rebalancing and using nested cross-validation, as implemented together with our customized pipeline. Extending our pipeline for use with other multimodal neuroimaging and improves its automatization could be critical for early detection of Alzheimer's disease and related brain disorders.

## Ethics statement of possible datasets to be used in this study

No data was collected during the implementation of this study. The different needed brain datasets are available in online repositories and were collected following different protocols and ethical standards, as presented on the different organization websites. Access to the different datasets was granted for each application, and our study followed the compromise assumed in the application.

## Declaration of Competing Interest

The authors declare that they do not possess any discernible competing financial interests or personal affiliations that might have conceivably impacted the research presented in this manuscript.

## Acknowledgments

Data collection and dissemination for this project were funded by the Alzheimer's Disease Neuroimaging Initiative (ADNI): the National Institutes of Health (grant number U01 AG024904) and the Department of Defense (award number W81XWH-12-2-0012). ADNI is funded by the National Institute of Aging and the National Institute of Biomedical Imaging and Bioengineering as well as through generous contributions from the following organizations: AbbVie, Alzheimer's Association, Alzheimer's Drug Discovery Foundation, Araclon Biotech, BioClinica Inc., Biogen, Bristol-Myers Squibb Company, CereSpir Inc., Eisai Inc., Elan Pharmaceuticals Inc., Eli Lilly and Company, EuroImmun, F. Hoffmann-La Roche Ltd. and its affiliated company Genentech Inc., Fujirebio, GE Healthcare, IXICO Ltd., Janssen Alzheimer Immunotherapy Research & Development LLC., Johnson & Johnson Pharmaceutical Research & Development LLC., Lumosity, Lundbeck, Merck & Co. Inc., Meso Scale Diagnostics LLC., NeuroRx Research, Neurotrack Technologies, Novartis Pharmaceuticals Corporation, Pfizer Inc., Piramal Imaging, Servier, Takeda Pharmaceutical Company, and Transition Therapeutics. The Canadian Institutes of Health Research are providing funds to support ADNI clinical sites in Canada. Private sector contributions are facilitated by the Foundation for the National Institutes of Health (http://www.fnih.org/). The grantee organization is the Northern California Institute for Research and Education, and the study is coordinated by the Alzheimer's Therapeutic Research Institute at the University of Southern California. ADNI data are disseminated by the Laboratory for Neuro Imaging at the University of Southern California. The authors are also grateful for access to the Tier 2 High-Performance Computing resources provided by the Northern Ireland High-Performance Computing (NI-HPC) facility funded by the UK Engineering and Physical Sciences Research Council (EPSRC), Grant No. EP/T022175/1. ALA would like to thank Universiti Sains Malaysia and Ministry of Higher Education (MOHE) Malaysia (Scholarship *Hadiah Latihan Persekutuan (HLP)*) for sponsoring and Universiti Teknologi PETRONAS for supporting this study. DC is supported by a UKRI Turing AI Fellowship 2021-2025, funded by the EPSRC, Grant No, EP/V025724/1. RCS was partially supported by grant RGPIN-2022-03042 from Canada's Natural Sciences and Engineering Research Council. JASK's research is supported by the FAU Foundation.

## References

1. Frisoni GB, Weiner MW. Alzheimer's disease neuroimaging initiative special issue. *Neurobiol Aging*. 2010;31(8):1259-1262. doi:10.1016/j.neurobiolaging.2010.05.006
2. Petrella JR, Coleman RE, Doraiswamy PM. State of the Art Radiology Neuroimaging and Early Diagnosis of Alzheimer Disease : A Look to the Future 1. *Radiology*. 2003;226(2):315-336. http://radiology.rsna.org/content/226/2/315.short

3. Braak H, Alafuzoff I, Arzberger T, Kretzschmar H, Tredici K. Staging of Alzheimer disease-associated neurofibrillary pathology using paraffin sections and immunocytochemistry. *Acta Neuropathol*. 2006;112(4):389-404. doi:10.1007/s00401-006-0127-z

4. Didic, M., Barbeau, E. J., Felician, O., Tramoni, E., Guedj, E., Poncet, M., & Ceccaldi M. Which memory system is impaired first in Alzheimer's disease?. Journal of Alzheimer's Disease, 27(1), 11-22. *J Alzheimers Dis*. 2011;27(1):11-22.

5. Braak H, Del Tredici K. The preclinical phase of the pathological process underlying sporadic Alzheimer's disease. *Brain*. 2015;138(10):2814-2833. doi:10.1093/brain/awv236

6. Thompson PM, Hayashi KM, De Zubicaray G, et al. Dynamics of gray matter loss in Alzheimer's disease. *J Neurosci*. 2003;23(3):994-1005. doi:10.1523/jneurosci.23-03-00994.2003

7. Apostolova LG, Steiner CA, Akopyan GG, et al. Three-dimensional gray matter atrophy mapping in mild cognitive impairment and mild Alzheimer disease. *Arch Neurol*. 2007;64(10):1489-1495. doi:10.1001/archneur.64.10.1489

8. Sean M. Nestor, Raul Rupsingh, Michael Borrie, Matthew Smith, Vittorio Accomazzi, Jennie L. Wells, Jennifer Fogarty, Robert Bartha the ADNI. Ventricular enlargement as a possible measure of Alzheimer's disease progression validated using the Alzheimer's disease neuroimaging initiative database. *Brain*. 2008;131(9):2443-2454. doi:10.1093/brain/awn146

9. Faull M, Ching SYL, Jarmolowicz AI, Beilby J, Panegyres PK. Comparison of two methods for the analysis of CSF Aβ and tau in the diagnosis of Alzheimer's disease. *Am J Neurodegener Dis*. 2014;3(3):143-151.

10. Apostolova LG. Alzheimer Disease. *Contin Lifelong Learn Neurol*. 2016;22(2, Dementia):419-434. doi:10.1212/CON.0000000000000307

11. de la Fuente-Fernández R. Role of DaTSCAN and clinical diagnosis in Parkinson disease. *Neurology*. 2012;78(10):696-701.

12. Sullivan, V., Majumdar, B., Richman, A., & Vinjamuri S. To scan or not to scan: neuroimaging in mild cognitive impairment and dementia. *Adv Psychiatr Treat*. 2012;18(6):457-466.

13. Papathanasiou, N. D., Boutsiadis, A., Dickson, J., & Bomanji JB. Diagnostic accuracy of 123I-FP-CIT (DaTSCAN) in dementia with Lewy bodies: a meta-analysis of published studies. *Parkinsonism Relat Disord*. 2012;18(3):225-229.

14. Magesh, P. R., Myloth, R. D., & Tom RJ. An explainable machine learning model for early detection of Parkinson's disease using LIME on DaTSCAN imagery. *Comput Biol Med*. 2020;126:104041.

15. Liu S, Cai W, Liu S, Zhang F, Fulham M, Feng D, Pujol S KR. Multimodal neuroimaging computing: a review of the applications in neuropsychiatric disorders. *Brain Informatics*. 2015;2(3):167-180. doi:10.1007/s40708-015-0019-x

16. Liu S, Cai W, Liu S, Zhang F. Multimodal neuroimaging computing : the workflows , methods , and platforms. *Brain Informatics*. 2015;2(3):181-195. doi:10.1007/s40708-015-0020-4

17. Mofrad SA, Lundervold AJ, Vik A, Lundervold AS. Cognitive and MRI trajectories

for prediction of Alzheimer's disease. *Sci Rep*. 2021;11(1):1-10. doi:10.1038/s41598-020-78095-7

18. Liu, S., Cao, Y., Liu, J., Ding, X., & Coyle D. A Novelty Detection Approach to Effectively Predict Conversion from Mild Cognitive Impairment to Alzheimer's Disease. *Int J Mach Learn Cybern*. 2022;14:213-228. doi:https://doi.org/10.1007/s13042-022-01570-2

19. Wernickand, M.N.; Aarsvold JN. *Emission Tomography: The Fundamentals of PET and SPECT*. Elsevier: New York, NY, USA; 2004.

20. Chouliaras, L., & O'Brien JT. The use of neuroimaging techniques in the early and differential diagnosis of dementia. *Mol Psychiatry*. Published online 2023:1-14.

21. Harper, L., Barkhof, F., Scheltens, P., Schott, J. M., & Fox NC. An algorithmic approach to structural imaging in dementia. *J Neurol Neurosurg Psychiatry*. Published online 2013.

22. Beltrán JF, Wahba BM, Hose N, Shasha D, Kline RP. Inexpensive, non-invasive biomarkers predict Alzheimer transition using machine learning analysis of the Alzheimer's Disease Neuroimaging (ADNI) database. *PLoS One*. 2020;15(7 July):1-26. doi:10.1371/journal.pone.0235663

23. Salvatore C, Cerasa A, Battista P, Gilardi MC, Quattrone A, Castiglioni I. Magnetic resonance imaging biomarkers for the early diagnosis of Alzheimer's disease: A machine learning approach. *Front Neurosci*. 2015;9(SEP):1-13. doi:10.3389/fnins.2015.00307

24. Harper, Lorna, Frederik Barkhof, Nick C. Fox and JMS. Using visual rating to diagnose dementia: a critical evaluation of MRI atrophy scales. *J Neurol Neurosurg Psychiatry*. Published online 2015.

25. Harper L, Fumagalli GG, Barkhof F, Scheltens P, O'Brien JT, Bouwman F, Burton EJ, Rohrer JD, Fox NC, Ridgway GR SJ. MRI visual rating scales in the diagnosis of dementia: evaluation in 184 post-mortem confirmed cases. *Brain*. 2016;139(4):1211-1225. doi:10.1093/brain/aww005

26. Yue W, Wang Z, Chen H, Payne A LX. Machine Learning with Applications in Breast Cancer Diagnosis and Prognosis. *Designs*. Published online 2018:1-17. doi:10.3390/designs2020013

27. Baseline MRI Predictors of Conversion from MCI to Probable AD in the ADNI Cohort. Published online 2009:347-361.

28. Rábano, Alberto, Carmen Guerrero Márquez, Ramón A. Juste, María V. Geijo and MC. Medial Temporal Lobe Involvement in Human Prion Diseases: Implications for the Study of Focal Non Prion Neurodegenerative Pathology." 11, no. 3 (2021): 413. *Biomolecules*. 2021;11(3):413.

29. Smailovic U, Koenig T, Savitcheva I, et al. Regional disconnection in alzheimer dementia and amyloid-positive mild cognitive impairment: Association between eeg functional connectivity and brain glucose metabolism. *Brain Connect*. 2020;10(10):555-565. doi:10.1089/brain.2020.0785

30. Delbeuck X, Collette F, Van der Linden M. Is Alzheimer's disease a disconnection syndrome?. Evidence from a crossmodal audio-visual illusory experiment. *Neuropsychologia*. 2007;45(14):3315-3323.

doi:10.1016/j.neuropsychologia.2007.05.001

31. Xiaoshu Li, Haibao Wang, Yanghua Tian, Shanshan Zhou, Xiaohu Li KW and YY. Impaired white matter connections of the limbic system networks associated with impaired emotional memory in alzheimer's diseas. *Front Aging Neurosci*. 2016;8(October):1-14. doi:10.3389/fnagi.2016.00250

32. Talwar P, Kushwaha S, Chaturvedi M, Mahajan V. Systematic Review of Different Neuroimaging Correlates in Mild Cognitive Impairment and Alzheimer's Disease. *Clin Neuroradiol*. 2021;31(4):953-967. doi:10.1007/s00062-021-01057-7

33. Kehoe, Elizabeth G., Dervla Farrell, Claudia Metzler-Baddeley, Brian A. Lawlor, Rose Anne Kenny, Declan Lyons, Jonathan P. McNulty, Paul G. Mullins, Damien Coyle and ALB. Fornix white matter is correlated with resting-state functional connectivity of the thalamus and hippocampus in healthy aging but not in mild cognitive impairment–a preliminary study. *Front Aging Neurosci*. 2015;7(10).

34. Mohatar-barba M, Fern E. Mediterranean Diet and the Emotional Well-Being of Students of the Campus of Melilla (University of Granada). Published online 2020:1-12.

35. Tang J, Society IC, Zhang Y, Sun J, Rao J. Quantitative Study of Individual Emotional States in Social Networks. 2012;3(2):132-144. doi:10.1109/T-AFFC.2011.23

36. Moradi E, Pepe A, Gaser C, Huttunen H, Tohka J. Machine learning framework for early MRI-based Alzheimer's conversion prediction in MCI subjects. *Neuroimage*. 2015;104:398-412. doi:10.1016/j.neuroimage.2014.10.002

37. McCombe N, Bamrah J, Sanchez-Bornot JM, Finn DP, McClean PL, Wong-Lin KF. Alzheimer's disease classification using cluster-based labelling for graph neural network on heterogeneous data. *Healthc Technol Lett*. 2022;9(6):102-109. doi:10.1049/htl2.12037

38. Klöppel S, Abdulkadir A, Jack CR, Koutsouleris N, Mourão-Miranda J, Vemuri P. Diagnostic neuroimaging across diseases. *Neuroimage*. 2012;61(2):457-463. doi:10.1016/j.neuroimage.2011.11.002

39. Kourou K, Exarchos TP, Exarchos KP, Karamouzis M V, Fotiadis DI. Machine learning applications in cancer prognosis and prediction. *CSBJ*. 2015;13:8-17. doi:10.1016/j.csbj.2014.11.005

40. Cruz JA, Wishart DS. Applications of Machine Learning in Cancer Prediction and Prognosis. Published online 2006:59-77.

41. Amrane M. Breast cancer classification using machine learning. *2018 Electr Electron Comput Sci Biomed Eng Meet*.:1-4. doi:10.1109/EBBT.2018.8391453

42. Lebedeva AK, Westman E, Borza T, Beyer MK, Engedal K, Aarsland D, Selbaek G HA. MRI-based classification models in prediction of mild cognitive impairment and dementia in late-life depression. *Front Aging Neurosci*. 2017;9(FEB):1-11. doi:10.3389/fnagi.2017.00013

43. Delshad Vaghari, Ricardo Bruna, Laura E. Hughes, David Nesbitt, Roni Tibon, James B. Rowe, Fernando Maestu RNH. A multi-site, multi-participant magnetoencephalography resting-state dataset to study dementia: The BioFIND dataset. *Neuroimage*. 2022;258(May):119344. doi:10.1016/j.neuroimage.2022.119344

44. Islam J, Zhang Y. Early diagnosis of alzheimer's disease: A neuroimaging study with

deep learning architectures. *IEEE Comput Soc Conf Comput Vis Pattern Recognit Work*. 2018;2018-June:1962-1964. doi:10.1109/CVPRW.2018.00247

45. Li Y, Zhang L, Bozoki A, Zhu DC, Choi J, Maiti T. Early prediction of Alzheimer's disease using longitudinal volumetric MRI data from ADNI. *Heal Serv Outcomes Res Methodol*. 2020;20(1):13-39. doi:10.1007/s10742-019-00206-3

46. Echávarri C, Aalten P, Uylings HB, Jacobs HI, Visser PJ, Gronenschild EH, Verhey FR BS. Atrophy in the parahippocampal gyrus as an early biomarker of Alzheimer ' s disease. Published online 2011:265-271. doi:10.1007/s00429-010-0283-8

47. Sun Z, Qiao Y, Lelieveldt BPF, Staring M. Integrating spatial-anatomical regularization and structure sparsity into SVM: Improving interpretation of Alzheimer's disease classification. *Neuroimage*. 2018;178(February 2018):445-460. doi:10.1016/j.neuroimage.2018.05.051

48. Albert M, Zhu Y, Moghekar A, Mori S, Miller MI, Soldan A, Pettigrew C, Selnes O, Li S WM. Predicting progression from normal cognition to mild cognitive impairment for individuals at 5 years. *Brain*. 2018;141(3):877-887. doi:10.1093/brain/awx365

49. Alderson, T., Kehoe, E., Maguire, L., Farrell, D., Lawlor, B., Kenny, R. A., Lyons, D., Bokde, A. L. W., & Coyle D. Disrupted Thalamus White Matter Anatomy and Posterior Default Mode Network Effective Connectivity in Amnestic Mild Cognitive Impairment. *Front Aging Neurosci*. 2017;9:1-15.

50. Garg, G., Prasad, G., Grag, L., Miyakoshi, M., Nakai, T., & Coyle D. Regional optimum frequency analysis of resting-state fMRI data for early detection of Alzheimer's disease biomarkers. *Multimed Tools Appl*. 2022;81:41953-41977. https://doi.org/10.1007/s11042-022-13523-6

51. Popuri K, Ma D, Wang L, Beg MF. Using machine learning to quantify structural MRI neurodegeneration patterns of Alzheimer's disease into dementia score: Independent validation on 8,834 images from ADNI, AIBL, OASIS, and MIRIAD databases. *Hum Brain Mapp*. 2020;41(14):4127-4147. doi:10.1002/hbm.25115

52. Ma D, Popuri K, Bhalla M, Sangha O, Lu D, Cao J, Jacova C, Wang L BMADNI. Quantitative assessment of field strength, total intracranial volume, sex, and age effects on the goodness of harmonization for volumetric analysis on the ADNI database. *Hum Brain Mapp*. 2019;40(5):1507-1527. doi:10.1002/hbm.24463

53. Rossini PM, Miraglia F, Vecchio F. Early dementia diagnosis, MCI-to-dementia risk prediction, and the role of machine learning methods for feature extraction from integrated biomarkers, in particular for EEG signal analysis. *Alzheimer's Dement*. 2022;18(12):2699-2706. doi:10.1002/alz.12645

54. Stamate D, Alghambdi W, Ogg J, Hoile R, Murtagh F. A Machine Learning Framework for Predicting Dementia and Mild Cognitive Impairment. *Proc - 17th IEEE Int Conf Mach Learn Appl ICMLA 2018*. Published online 2019:671-678. doi:10.1109/ICMLA.2018.00107

55. Douzas G, Bacao F, Last F. Improving imbalanced learning through a heuristic oversampling method based on k-means and SMOTE. *Inf Sci (Ny)*. 2018;465:1-20. doi:10.1016/j.ins.2018.06.056

56. Chawla N V., Japkowicz N, Kotcz A. Editorial. *Special issue on learning from imbalanced data sets, ACM SIGKDD Explorations Newsletter*. 2004:1-6.

57. Rahman MM, Davis DN. Addressing the Class Imbalance Problem in Medical Datasets. *Int J Mach Learn Comput*. 2013;3(2):224-228. doi:10.7763/ijmlc.2013.v3.307

58. Ling CX, Sheng VS. Cost-Sensitive Learning and the Class Imbalance Problem. *Encycl Mach Learn*. Published online 2008:231-235. http://www.springer.com/computer/ai/book/978-0-387-30768-8%5Cnhttp://citeseerx.ist.psu.edu/viewdoc/download?doi=10.1.1.164.4418&rep=rep1&type=pdf

59. Chawla N V., Bowyer KW, Hall LO, Kegelmeyer WP. SMOTE: Synthetic Minority Over-sampling Technique. *J Artif Intell Res*. 2002;16:321-357. doi:10.1613/jair.953

60. Seiffert C, Khoshgoftaar TM, Van Hulse J, Napolitano A. RUSBoost: A hybrid approach to alleviating class imbalance. *IEEE Trans Syst Man, Cybern Part ASystems Humans*. 2010;40(1):185-197. doi:10.1109/TSMCA.2009.2029559

61. Van Hulse J, Khoshgoftaar TM, Napolitano A. Experimental perspectives on learning from imbalanced data. *ACM Int Conf Proceeding Ser*. 2007;227:935-942. doi:10.1145/1273496.1273614

62. Drummond C, Holte RC. Class Imbalance, and Cost Sensitivity: Why Under-Sampling beats Over-Sampling. *Phys Rev Lett*. 2003;91(3).

63. Boughorbel S, Jarray F, El-Anbari M. Optimal classifier for imbalanced data using Matthews Correlation Coefficient metric. *PLoS One*. 2017;12(6):1-17. doi:10.1371/journal.pone.0177678

64. Chicco D, Jurman G. The advantages of the Matthews correlation coefficient (MCC) over F1 score and accuracy in binary classification evaluation. *BMC Genomics*. 2020;21(1):1-13. doi:10.1186/s12864-019-6413-7

65. Fischl B. FreeSurfer. *Neuroimage*. 2012;62(2):774-781. doi:10.1016/j.neuroimage.2012.01.021

66. Jack CR Jr, Bernstein MA, Fox NC, Thompson P, Alexander G, Harvey D, Borowski B, Britson PJ, L Whitwell J, Ward C, Dale AM, Felmlee JP, Gunter JL, Hill DL, Killiany R, Schuff N, Fox-Bosetti S, Lin C, Studholme C, DeCarli CS, Krueger G, Ward HA, Metzger GJ WM. The Alzheimer ' s Disease Neuroimaging Initiative ( ADNI ): MRI Methods. 2008;691(February):685-691. doi:10.1002/jmri.21049

67. Pamela J. LaMontagne, Tammie LS. Benzinger, John C. Morris, Sarah Keefe, Russ Hornbeck, Chengjie Xiong, Elizabeth Grant, Jason Hassenstab, Krista Moulder, Andrei G. Vlassenko, Marcus E. Raichle, Carlos Cruchaga DM. OASIS-3: Longitudinal Neuroimaging, Clinical, and Cognitive Dataset for Normal Aging and Alzheimer Disease. *medRxiv*. Published online 2019. doi:https://doi.org/10.1101/2019.12.13.19014902

68. Marcus DS, Fotenos AF, Csernansky JG, Morris JC, Buckner RL. Open access series of imaging studies: Longitudinal MRI data in nondemented and demented older adults. *J Cogn Neurosci*. 2010;22(12):2677-2684. doi:10.1162/jocn.2009.21407

69. Graves MJ. 3 T: the good, the bad and the ugly. *Br J Radiol*. 2022;95(1130). doi:10.1259/bjr.20210708

70. Leandrou S, Lamnisos D, Kyriacou PA, Constanti S, Pattichis CS. Comparison of 1.5 T and 3 T MRI hippocampus texture features in the assessment of Alzheimer's

disease. *Biomed Signal Process Control*. 2020;62(July):3-8. doi:10.1016/j.bspc.2020.102098

71. R. A. I. Bethlehem, J. Seidlitz, S. R. White, J. W. Vogel, K. M. Anderson, C. Adamson, S. Adler, G. S. Alexopoulos, E. Anagnostou, A. Areces-Gonzalez, D. E. Astle, B. Auyeung, M. Ayub, J. Bae, G. Ball, S. Baron-Cohen, R. Beare, S. A. Bedford, V. Benegal, ETB& AFAB. Brain charts for the human lifespan. *Nature*. 2022;604(February). doi:10.1038/s41586-022-04554-y

72. Ségonne F, Dale AM, Busa E, et al. A hybrid approach to the skull stripping problem in MRI. *Neuroimage*. 2004;22(3):1060-1075. doi:10.1016/j.neuroimage.2004.03.032

73. Bruce Fischl, David H. Salat, André J.W. van der Kouwe, Nikos Makris, Florent Ségonne, Brian T. Quinn AMD. Sequence-independent segmentation of magnetic resonance images. *Neuroimage*. 2004;23(SUPPL. 1):69-84. doi:10.1016/j.neuroimage.2004.07.016

74. Fischl B, Salat DH, Busa E, Albert M, Dieterich M, Haselgrove C, van der Kouwe A, Killiany R, Kennedy D, Klaveness S, Montillo A, Makris N, Rosen B DA. Whole brain segmentation: Automated labeling of neuroanatomical structures in the human brain. *Neuron*. 2002;33(3):341-355. doi:10.1016/S0896-6273(02)00569-X

75. Desikan RS, Ségonne F, Fischl B, Quinn BT, Dickerson BC, Blacker D, Buckner RL, Dale AM, Maguire RP, Hyman BT, Albert MS KR. An automated labeling system for subdividing the human cerebral cortex on MRI scans into gyral based regions of interest. *Neuroimage*. 2006;31(3):968-980. doi:10.1016/j.neuroimage.2006.01.021

76. Ledig C, Schuh A, Guerrero R, Heckemann RA, Rueckert D. Structural brain imaging in Alzheimer's disease and mild cognitive impairment: biomarker analysis and shared morphometry database. *Sci Rep*. 2018;8(1):1-16. doi:10.1038/s41598-018-29295-9

77. Koikkalainen J, Pölönen H, Mattila J, van Gils M, Soininen H, Lötjönen J. Improved classification of alzheimer's disease data via removal of nuisance variability. *PLoS One*. 2012;7(2). doi:10.1371/journal.pone.0031112

78. Sarica A, Vasta R, Novellino F, Vaccaro MG, Cerasa A, Quattrone A. MRI asymmetry index of hippocampal subfields increases through the continuum from the mild cognitive impairment to the alzheimer's disease. *Front Neurosci*. 2018;12(AUG):1-12. doi:10.3389/fnins.2018.00576

79. Johnson MK and K. *Applied Predictive Modeling*. Springer New, York; 2013.

80. Robnik-Šikonja, M., Kononenko I. Theoretical and Empirical Analysis of ReliefF and RReliefF. *Mach Learn*. 2003;53:23-69. http://lkm.fri.uni-lj.si/xaigor/slo/clanki/MLJ2003-FinalPaper.pdf

81. Feng J, Zhang S wu, Chen L, Neuroimaging D. Identification of Alzheimer ' s disease based on wavelet transformation energy feature of the structural MRI image and NN classifier. 2020;108(August). doi:10.1016/j.artmed.2020.101940

82. Grueso S, Viejo-sobera R. Machine learning methods for predicting progression from mild cognitive impairment to Alzheimer ' s disease dementia : a systematic review. Published online 2021.

83. Pellegrini E, Ballerini L, Hernandez MDCV, Chappell FM, González-Castro V, Anblagan D, Danso S, Muñoz-Maniega S, Job D, Pernet C, Mair G, MacGillivray TJ, Trucco E WJ. Machine learning of neuroimaging for assisted diagnosis of cognitive impairment and dementia: A systematic review. *Alzheimer's Dement Diagnosis, Assess Dis Monit*. 2018;10:519-535. doi:10.1016/j.dadm.2018.07.004

84. Patestas, M. A., and Gartner LP. *Limbic System," in A Textbook of Neuroanatomy.* Oxford, England: Blackwell; 2006.

85. RajMohan V, Mohandas E. The limbic system. *Indian J Psychiatry*. 2007;49(2):132. doi:10.4103/0019-5545.33264

86. Mori S, Aggarwal M. In vivo magnetic resonance imaging of the human limbic white matter. *Front Aging Neurosci*. 2014;6(NOV):1-6. doi:10.3389/fnagi.2014.00321

87. deToledo-Morrell L, Stoub TR, Bulgakova M, Wilson RS, Bennett DA, Leurgans S, Wuu J T DA. MRI-derived entorhinal volume is a good predictor of conversion from MCI to AD. 2004;i:1197-1203. doi:10.1016/j.neurobiolaging.2003.12.007

88. Pennanen C, Kivipelto M, Tuomainen S, Hartikainen P, Hänninen T, Laakso MP, Hallikainen M, Vanhanen M, Nissinen A, Helkala EL, Vainio P, Vanninen R, Partanen K SH. Hippocampus and entorhinal cortex in mild cognitive impairment and early AD. *Neurobiol Aging*. 2004;25(3):303-310. doi:10.1016/S0197-4580(03)00084-8

89. Tero Tapiola, Corina Pennanen, Mia Tapiola, Susanna Tervo, Miia Kivipelto, Tuomo Hänninen, Maija Pihlajamäki, Mikko P. Laakso, Merja Hallikainen, Anne Hämäläinen, Matti Vanhanen, Eeva-Liisa Helkala, Ritva Vanninen, Aulikki Nissinen, Roberta Rossi, Giovann HS. MRI of hippocampus and entorhinal cortex in mild cognitive impairment : A follow-up study. 2008;29:31-38. doi:10.1016/j.neurobiolaging.2006.09.007

90. Van Hoesen GW, Parvizi J, Chu CC. Orbitofrontal cortex pathology in Alzheimer's disease. *Cereb Cortex*. 2000;10(3):243-251. doi:10.1093/cercor/10.3.243

91. Varoquaux G, Raamana PR, Engemann DA, Hoyos-Idrobo A, Schwartz Y, Thirion B. Assessing and tuning brain decoders: Cross-validation, caveats, and guidelines. *Neuroimage*. 2017;145(August 2015):166-179. doi:10.1016/j.neuroimage.2016.10.038

92. Varoquaux G. Cross-validation failure: Small sample sizes lead to large error bars. *Neuroimage*. 2018;180(June 2017):68-77. doi:10.1016/j.neuroimage.2017.06.061

# Supplementary Material

# A Machine Learning Approach for Identifying Anatomical Biomarkers of Early Mild Cognitive Impairment

Alwani Liyana Ahmad[1,2,3], Jose Sanchez-Bornot[4], Roberto C. Sotero[5], Damien Coyle[6], Zamzuri Idris[2,3,7], Ibrahima Faye[1,8,*], for the Alzheimer's Disease Neuroimaging Initiative[©]

[1] Department of Fundamental and Applied Sciences, Faculty of Science and Information Technology, Universiti Teknologi PETRONAS, Perak, Malaysia.
[2] Department of Neurosciences, Hospital Universiti Sains Malaysia, Kelantan, Malaysia.
[3] Brain and Behaviour Cluster, School of Medical Sciences, Universiti Sains Malaysia, Kelantan, Malaysia
[4] Intelligent Systems Research Centre, School of Computing, Engineering and Intelligent Systems, Ulster University, Magee campus, Derry~Londonderry, BT48 7JL, UK.
[5] Department of Radiology and Hotchkiss Brain Institute, University of Calgary, Calgary, AB, Canada.
[6] The Bath Institute for the Augmented Human, University of Bath, Bath, BA2 7AY, UK.
[7] Department of Neurosciences, School of Medical Sciences, Universiti Sains Malaysia, Kelantan, Malaysia
[8] Centre for Intelligent Signal & Imaging Research (CISIR), Universiti Teknologi PETRONAS, Perak, Malaysia.

[©]Data used in preparation of this article were obtained from the Alzheimer's Disease Neuroimaging Initiative (ADNI) database (adni.loni.usc.edu). As such, the investigators within the ADNI contributed to the design and implementation of ADNI and/or provided data but did not participate in analysis or writing of this report. A complete listing of ADNI investigators can be found at: http://adni.loni.usc.edu/wpcontent/uploads/how_to_apply/ADNI_Acknowledgement_List.pdf.

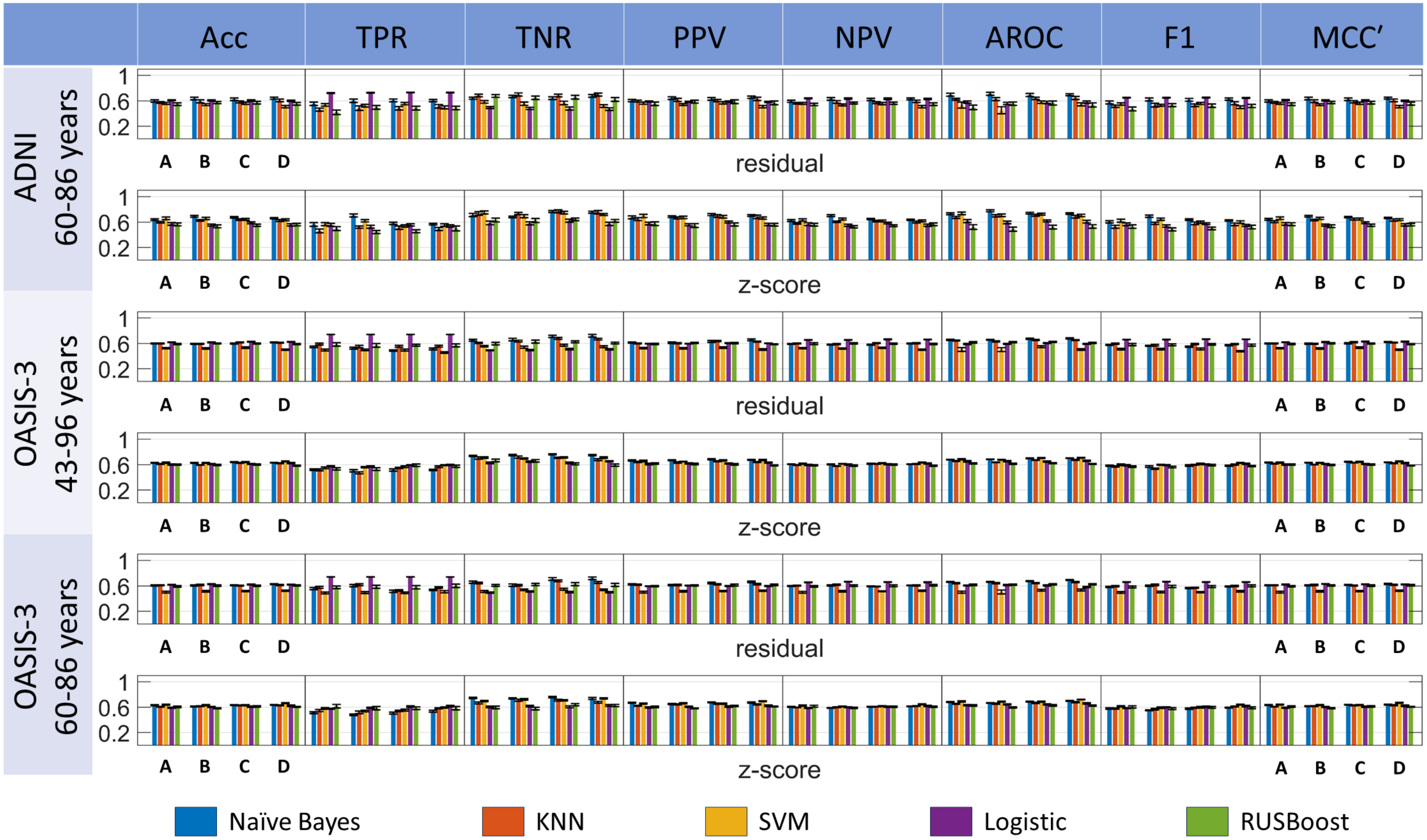

**Supplementary Figure 1:** Comparison among multiple classification pipeline options, involving five classifiers, four feature selection and two harmonization techniques. Performance is measured for randomly balanced cohorts extracted from ADNI and OASIS-3 imbalanced datasets within a Monte Carlo replication analysis. Results are presented for Naïve Bayes, KNN, SVM, Logistic and RUSBoost, residual and z-score harmonization procedures, as represented in the x-axis and legend labels. Bar groups denoted by letters **A-D** indicate the outcomes corresponding to the different features selected in the MCL analysis: (**A**) Features selected using the average scores (**Table *8*A** lists the feature labels); (**B**) Features selected based on the ReliefF criterion (**Table *8*B** lists the feature labels); (**C**) Features selected according to the combination of all evaluated feature selection algorithms (**Table *8*C** lists the feature labels); (**D**) Features selected within the SPSS statistical analysis (**Table *8*D** lists the feature lables). Column labels: **Acc**=accuracy. **TPR** = True Positive Rate. **TNR** = True Negative Rate. **PPV** = Positive Predictive Value. **NPV** = Negative Predictive Value. **AROC**=Area under Receiver Operating Curve. **F1**= F1 score, **MCC**=Matthew's correlation coefficient. Performance metrics are normalized in the range [0, 1] and plotted together to enhance visual comparison.

**Supplementary Table 1:** Comparative analysis of classifier models and feature selection algorithms using different normalization techniques (see **Fig. 4** for complementary information) for the balanced analysis for ADNI dataset.

****classifier model pairwise comparison: p value post hoc test (Tukey's HSD)
**** Pairwise comparison: p-value: Adjustment for multiple comparison: Bonferroni
**** NS = Not significant
***** Features selection methods:
A = Average score percentage (MCL app): chi-square, ReliefF, ANOVA & Kruskal Wallis
**B** = ReliefF (MCL app)
**C** = Frequent feature appearances from all feature ranking analysis
**D** = features extracted from statistical analysis (SPSS software): ANOVA, ANCOVA & Kruskal Wallis

| Dataset | Classification | Features selection group | Classifier model | Harmonization | Mean (%) | Standard deviation | Adjustment for multiple comparison: Bonferroni | | | Adjustment for multiple comparison: false discovery rate (FDR) Benjamini-Hochberg ($\alpha$ = 0.05) | | | |
|---|---|---|---|---|---|---|---|---|---|---|---|---|---|
| | | | | | | | p-value (**Displaying only statistically significant p-values**) | 95% confidence interval | Pairwise comparison | Hypothesis ID | Raw p-value | adj. p-value | Pairwise comparison |
| ADNI | Accuracy | A | Naïve Bayes | Residual | 59.6875 | 7.7738 | | | NS for feature selection group | | | | NS for feature selection group |
| | | | | z-score | 63.7500 | 5.9080 | < 0.001 | [0.0171, 0.0631] | Naïve Bayes > KNN | H4 | < 0.001 | 0.005 | Naïve Bayes > RUSBoost |
| | | | KNN | Residual | 57.2917 | 5.7154 | < 0.001 | [0.0272, 0.0731] | Naïve Bayes > SVM | H3 | < 0.001 | 0.01 | Naïve Bayes > Logistic |
| | | | | z-score | 59.7917 | 6.1245 | < 0.001 | [0.0367, 0.0826] | Naïve Bayes > Logistic | H2 | < 0.001 | 0.015 | Naïve Bayes > SVM |
| | | | SVM | Residual | 56.0417 | 7.4013 | < 0.001 | 0.0660, 0.1119] | Naïve Bayes > RUSBoost | H7 | < 0.001 | 0.02 | KNN > RUSBoost |
| | | | | z-score | 65.8333 | 8.8749 | < 0.001 | [0.0259, 0.0718] | KNN > RUSBoost | H1 | < 0.001 | 0.025 | Naïve Bayes > KNN |
| | | | Logistic | Residual | 60.8333 | 3.4199 | < 0.001 | [0.0158, 0.0618] | SVM > RUSBoost | H9 | < 0.001 | 0.03 | SVM > RUSBoost |
| | | | | z-score | 57.1875 | 9.4744 | 0.005 | [0.0063, 0.0523] | Logistic > RUSBoost | H10 | < 0.001 | 0.035 | Logistic > RUSBoost |
| | | | RUSBoost | Residual | 54.6875 | 9.4599 | < 0.001 | [-0.038, -0.017] | Residual < Z-score | H6 | 0.02 | 0.04 | KNN > Logistic |
| | | | | z-score | 56.4583 | 10.4035 | | | | H1 | < 0.001 | 0.05 | Residual < Z-score |
| | | B | Naïve Bayes | Residual | 63.5417 | 9.2801 | | | | | | | |
| | | | | z-score | 69.1667 | 6.5394 | | | | | | | |
| | | | KNN | Residual | 59.1667 | 9.9890 | | | | | | | |
| | | | | z-score | 62.7083 | 5.6106 | | | | | | | |
| | | | SVM | Residual | 53.8542 | 6.4929 | | | | | | | |
| | | | | z-score | 65.7292 | 7.6614 | | | | | | | |
| | | | Logistic | Residual | 60.4167 | 4.1115 | | | | | | | |
| | | | | z-score | 55.2083 | 7.2010 | | | | | | | |
| | | | RUSBoost | Residual | 57.1875 | 7.8090 | | | | | | | |
| | | | | z-score | 53.3333 | 9.8276 | | | | | | | |
| | | C | Naïve Bayes | Residual | 62.3958 | 8.5090 | | | | | | | |
| | | | | z-score | 67.3958 | 5.1599 | | | | | | | |
| | | | KNN | Residual | 58.2292 | 7.1365 | | | | | | | |
| | | | | z-score | 63.9583 | 6.0494 | | | | | | | |
| | | | SVM | Residual | 56.0417 | 7.8506 | | | | | | | |
| | | | | z-score | 64.2708 | 7.1621 | | | | | | | |
| | | | Logistic | Residual | 60.4167 | 4.7315 | | | | | | | |
| | | | | z-score | 58.9583 | 6.2719 | | | | | | | |
| | | | RUSBoost | Residual | 56.9792 | 9.1506 | | | | | | | |
| | | | | z-score | 55.0000 | 7.4474 | | | | | | | |
| | | D | Naïve Bayes | Residual | 64.0625 | 7.4005 | | | | | | | |
| | | | | z-score | 66.1458 | 4.3740 | | | | | | | |
| | | | KNN | Residual | 60.5208 | 8.9282 | | | | | | | |
| | | | | z-score | 62.3958 | 7.1044 | | | | | | | |
| | | | SVM | Residual | 50.6250 | 8.0005 | | | | | | | |
| | | | | z-score | 63.6458 | 6.7075 | | | | | | | |
| | | | Logistic | Residual | 59.8958 | 4.3214 | | | | | | | |
| | | | | z-score | 55.5208 | 8.2306 | | | | | | | |
| | | | RUSBoost | Residual | 55.3125 | 10.0136 | | | | | | | |
| | | | | z-score | 56.0417 | 8.6004 | | | | | | | |
| | F1 | A | Naïve Bayes | Residual | 57.4317 | 9.8654 | | | NS for feature selection group | H1 | 0.029 | 0.029 | A < B |
| | | | | z-score | 60.2462 | 8.0501 | < 0.001 | [0.0485, 0.1035] | Naïve Bayes > KNN | H4 | < 0.001 | 0.005 | Naïve Bayes > RUSBoost |
| | | | KNN | Residual | 51.5039 | 7.6424 | < 0.001 | [0.0219, 0.0770] | Naïve Bayes > SVM | H10 | < 0.001 | 0.01 | Logistic > RUSBoost |
| | | | | z-score | 52.6057 | 9.8138 | < 0.001 | [0.0870, 0.1420] | Naïve Bayes > RUSBoost | H1 | < 0.001 | 0.015 | Naïve Bayes > KNN |
| | | | SVM | Residual | 54.8608 | 8.1817 | < 0.001 | [-0.0809, -0.0259] | KNN < Logistic | H9 | < 0.001 | 0.02 | SVM > RUSBoost |
| | | | | z-score | 62.2155 | 10.1956 | 0.001 | [0.0110, 0.0660] | KNN > RUSBoost | H6 | < 0.001 | 0.025 | KNN < Logistic |
| | | | Logistic | Residual | 64.9093 | 2.0702 | < 0.001 | [0.0375, 0.0926] | SVM > RUSBoost | H2 | < 0.001 | 0.03 | Naïve Bayes > SVM |
| | | | | z-score | 56.4569 | 9.3587 | < 0.001 | [0.0644, 0.1194] | Logistic > RUSBoost | H7 | < 0.001 | 0.035 | KNN > RUSBoost |
| | | | RUSBoost | Residual | 47.1448 | 13.7933 | | | NS for harmonization | H8 | 0.008 | 0.04 | SVM < Logistic |
| | | | | z-score | 52.8764 | 12.5829 | | | | H5 | 0.008 | 0.045 | KNN < SVM |
| | | B | Naïve Bayes | Residual | 62.0382 | 10.5443 | | | | H3 | 0.025 | 0.05 | Naïve Bayes > Logistic |
| | | | | z-score | 69.2068 | 7.9010 | | | | | | | NS for harmonization |
| | | | KNN | Residual | 53.3203 | 13.5570 | | | | | | | |
| | | | | z-score | 57.9945 | 6.9649 | | | | | | | |
| | | | SVM | Residual | 52.9191 | 7.1089 | | | | | | | |
| | | | | z-score | 64.3700 | 7.9005 | | | | | | | |
| | | | Logistic | Residual | 64.8133 | 2.5936 | | | | | | | |
| | | | | z-score | 53.6583 | 8.2601 | | | | | | | |
| | | | RUSBoost | Residual | 53.1524 | 10.1856 | | | | | | | |
| | | | | z-score | 48.4634 | 11.2111 | | | | | | | |
| | | C | Naïve Bayes | Residual | 61.5740 | 9.4366 | | | | | | | |
| | | | | z-score | 63.9038 | 6.2569 | | | | | | | |
| | | | KNN | Residual | 53.0155 | 9.8951 | | | | | | | |
| | | | | z-score | 57.9984 | 9.5041 | | | | | | | |
| | | | SVM | Residual | 55.7485 | 7.2894 | | | | | | | |
| | | | | z-score | 59.9889 | 8.1232 | | | | | | | |
| | | | Logistic | Residual | 64.9878 | 2.3530 | | | | | | | |
| | | | | z-score | 57.4366 | 6.4739 | | | | | | | |
| | | | RUSBoost | Residual | 52.3661 | 12.1037 | | | | | | | |
| | | | | z-score | 50.1003 | 9.2563 | | | | | | | |
| | | D | Naïve Bayes | Residual | 62.6978 | 7.3766 | | | | | | | |
| | | | | z-score | 62.5414 | 4.9028 | | | | | | | |
| | | | KNN | Residual | 56.0604 | 10.9740 | | | | | | | |
| | | | | z-score | 56.3436 | 8.7631 | | | | | | | |
| | | | SVM | Residual | 49.9094 | 8.7740 | | | | | | | |
| | | | | z-score | 60.0735 | 8.1566 | | | | | | | |
| | | | Logistic | Residual | 64.5919 | 2.6035 | | | | | | | |
| | | | | z-score | 54.7249 | 7.8028 | | | | | | | |
| | | | RUSBoost | Residual | 51.7989 | 11.1857 | | | | | | | |
| | | | | z-score | 52.1366 | 12.7316 | | | | | | | |
| | MCC′ | A | Naïve Bayes | Residual | 59.7425 | 7.8510 | | | NS for feature selection group | | | | NS for feature selection group |
| | | | | z-score | 64.3034 | 6.2119 | < 0.001 | [0.0142, 0.0610] | Naïve Bayes > KNN | H4 | < 0.001 | 0.005 | Naïve Bayes > RUSBoost |
| | | | KNN | Residual | 57.5802 | 5.7863 | < 0.001 | [0.0273, 0.0742] | Naïve Bayes > SVM | H3 | < 0.001 | 0.01 | Naïve Bayes > Logistic |
| | | | | z-score | 60.4929 | 6.5552 | < 0.001 | [0.0366, 0.0835] | Naïve Bayes > Logistic | H7 | < 0.001 | 0.015 | KNN > RUSBoost |
| | | | SVM | Residual | 56.0582 | 7.4115 | < 0.001 | [0.0663, 0.1132] | Naïve Bayes > RUSBoost | H2 | < 0.001 | 0.02 | Naïve Bayes > SVM |
| | | | | z-score | 66.2192 | 9.0403 | < 0.001 | [0.0287, 0.0756] | KNN > RUSBoost | H9 | < 0.001 | 0.025 | SVM > RUSBoost |
| | | | Logistic | Residual | 61.1193 | 3.3306 | < 0.001 | [0.0155, 0.0624] | SVM > RUSBoost | H1 | < 0.001 | 0.03 | Naïve Bayes > KNN |
| | | | | z-score | 57.2190 | 9.4996 | 0.005 | [0.0062, 0.0531] | Logistic > RUSBoost | H10 | < 0.001 | 0.035 | Logistic > RUSBoost |
| | | | RUSBoost | Residual | 54.7348 | 9.8026 | < 0.001 | [-0.040, -0.018] | Residual < Z-score | H6 | 0.009 | 0.04 | KNN > Logistic |
| | | | | z-score | 56.5636 | 10.5851 | | | | H1 | < 0.001 | 0.05 | Residual < Z-score |
| | | B | Naïve Bayes | Residual | 63.5963 | 9.3091 | | | | | | | |
| | | | | z-score | 69.2764 | 6.5585 | | | | | | | |
| | | | KNN | Residual | 59.4508 | 10.0564 | | | | | | | |
| | | | | z-score | 63.1079 | 5.7277 | | | | | | | |
| | | | SVM | Residual | 53.8823 | 6.5564 | | | | | | | |
| | | | | z-score | 65.8920 | 7.7067 | | | | | | | |
| | | | Logistic | Residual | 60.7032 | 4.1150 | | | | | | | |
| | | | | z-score | 55.2654 | 7.2695 | | | | | | | |
| | | | RUSBoost | Residual | 57.3907 | 8.0443 | | | | | | | |
| | | | | z-score | 53.4779 | 10.2273 | | | | | | | |
| | | C | Naïve Bayes | Residual | 62.4446 | 8.5585 | | | | | | | |
| | | | | z-score | 67.8335 | 5.3061 | | | | | | | |
| | | | KNN | Residual | 58.4681 | 7.2421 | | | | | | | |
| | | | | z-score | 64.6844 | 6.3114 | | | | | | | |
| | | | SVM | Residual | 56.0591 | 7.9551 | | | | | | | |
| | | | | z-score | 64.6948 | 7.3937 | | | | | | | |
| | | | Logistic | Residual | 60.7016 | 4.6615 | | | | | | | |
| | | | | z-score | 59.0714 | 6.3156 | | | | | | | |
| | | | RUSBoost | Residual | 57.1610 | 9.6013 | | | | | | | |
| | | | | z-score | 55.1329 | 7.7137 | | | | | | | |
| | | D | Naïve Bayes | Residual | 64.1713 | 7.4711 | | | | | | | |
| | | | | z-score | 66.5337 | 4.4947 | | | | | | | |
| | | | KNN | Residual | 60.8509 | 9.1927 | | | | | | | |
| | | | | z-score | 63.1865 | 7.6431 | | | | | | | |
| | | | SVM | Residual | 50.6314 | 8.0519 | | | | | | | |
| | | | | z-score | 63.8691 | 6.6921 | | | | | | | |
| | | | Logistic | Residual | 60.2011 | 4.2792 | | | | | | | |
| | | | | z-score | 55.5883 | 8.2901 | | | | | | | |
| | | | RUSBoost | Residual | 55.5329 | 10.3346 | | | | | | | |
| | | | | z-score | 56.1325 | 8.7837 | | | | | | | |
| | AROC | A | Naïve Bayes | Residual | 69.6042 | 10.7847 | | | NS for feature selection group | | | | NS for feature selection group |
| | | | | z-score | 73.1458 | 6.3096 | < 0.001 | [0.0260, 0.0932] | Naïve Bayes > KNN | H4 | < 0.001 | 0.005 | Naïve Bayes > RUSBoost |
| | | | KNN | Residual | 62.4792 | 6.3356 | < 0.001 | [0.0663, 13.3574] | Naïve Bayes > SVM | H7 | < 0.001 | 0.01 | KNN > RUSBoost |
| | | | | z-score | 67.3125 | 7.3963 | < 0.001 | [0.0978, 0.1650] | Naïve Bayes > Logistic | H3 | < 0.001 | 0.015 | Naïve Bayes > Logistic |
| | | | SVM | Residual | 53.3333 | 20.6651 | < 0.001 | [0.1619, 0.2291] | Naïve Bayes > RUSBoost | H2 | < 0.001 | 0.02 | Naïve Bayes > SVM |
| | | | | z-score | 73.7083 | 8.5647 | 0.009 | [0.0067, 0.0740] | KNN > SVM | H9 | < 0.001 | 0.025 | SVM > RUSBoost |
| | | | Logistic | Residual | 57.9167 | 8.0273 | < 0.001 | [0.0382, 0.1054] | KNN > Logistic | H6 | < 0.001 | 0.03 | KNN > Logistic |
| | | | | z-score | 61.6875 | 10.5846 | < 0.001 | [0.1023, 0.1695] | KNN > RUSBoost | H10 | < 0.001 | 0.035 | Logistic > RUSBoost |
| | | | RUSBoost | Residual | 49.5625 | 15.2845 | < 0.001 | [0.0619, 0.1292] | SVM > RUSBoost | H1 | < 0.001 | 0.04 | Naïve Bayes > KNN |
| | | | | z-score | 52.2500 | 16.0029 | < 0.001 | [0.0305, 0.0977] | Logistic > RUSBoost | H5 | 0.001 | 0.045 | KNN > SVM |
| | | B | Naïve Bayes | Residual | 71.1042 | 10.7955 | < 0.001 | [-0.078, -0.047] | Residual < Z-score | H8 | 0.011 | 0.05 | SVM > Logistic |
| | | | | z-score | 77.7292 | 7.0756 | | | | H1 | < 0.001 | 0.05 | Residual < Z-score |
| | | | KNN | Residual | 62.6667 | 10.2637 | | | | | | | |
| | | | | z-score | 70.0625 | 7.9060 | | | | | | | |
| | | | SVM | Residual | 45.0417 | 24.8349 | | | | | | | |
| | | | | z-score | 70.6667 | 85.0095 | | | | | | | |
| | | | Logistic | Residual | 55.8333 | 7.9839 | | | | | | | |
| | | | | z-score | 59.6250 | 9.6596 | | | | | | | |
| | | | RUSBoost | Residual | 55.5208 | 12.0579 | | | | | | | |
| | | | | z-score | 48.8750 | 14.2932 | | | | | | | |
| | | C | Naïve Bayes | Residual | 69.1042 | 11.2441 | | | | | | | |
| | | | | z-score | 74.0833 | 6.3459 | | | | | | | |
| | | | KNN | Residual | 63.7708 | 7.8132 | | | | | | | |
| | | | | z-score | 70.8542 | 6.9030 | | | | | | | |
| | | | SVM | Residual | 57.9167 | 8.8791 | | | | | | | |
| | | | | z-score | 72.3333 | 6.5427 | | | | | | | |
| | | | Logistic | Residual | 57.4792 | 8.9971 | | | | | | | |
| | | | | z-score | 61.7500 | 6.5834 | | | | | | | |
| | | | RUSBoost | Residual | 56.5417 | 13.6938 | | | | | | | |
| | | | | z-score | 52.1250 | 13.3982 | | | | | | | |
| | | D | Naïve Bayes | Residual | 69.5625 | 7.3790 | | | | | | | |
| | | | | z-score | 73.5417 | 5.2904 | | | | | | | |
| | | | KNN | Residual | 64.6250 | 10.8560 | | | | | | | |
| | | | | z-score | 68.4375 | 8.9261 | | | | | | | |
| | | | SVM | Residual | 54.4167 | 11.2556 | | | | | | | |
| | | | | z-score | 70.5000 | 9.0620 | | | | | | | |
| | | | Logistic | Residual | 57.83333 | 8.3603 | | | | | | | |
| | | | | z-score | 60.6458 | 10.3629 | | | | | | | |
| | | | RUSBoost | Residual | 53.4375 | 13.7316 | | | | | | | |
| | | | | z-score | 53.1667 | 12.1954 | | | | | | | |

**Supplementary Table 2:** Comparative analysis of classifier models and feature selection algorithms using different normalization techniques (see **Fig. 4** for complementary information) for a balanced analysis for OASIS-3 (age 60-86) dataset

| Dataset | Classification | Features selection group | Classifier model | Harmonization | Mean (%) | Standard deviation | Adjustment for multiple comparison: Bonferroni | | | Adjustment for multiple comparison: Benjamini-Hochberg | | | |
|---|---|---|---|---|---|---|---|---|---|---|---|---|---|
| | | | | | | | p-value (**Displaying only statistically significant p-values**) | 95% confidence interval | Pairwise comparison | Hypothesis ID | Raw p-value | adj. p-value | Pairwise comparison |
| OASIS-3 (age 60-86) | Accuracy | A | Naïve Bayes | Residual | 60.7311 | 3.1284 | < 0.001 | [-0.02460, -0.0058] | A < D | H3 | < 0.001 | 0.0083 | A < D |
| | | | | z-score | 62.9953 | 3.2911 | < 0.001 | [-0.0236, -0.0048] | B < D | H5 | < 0.001 | 0.0167 | B < D |
| | | | KNN | Residual | 61.0613 | 3.0008 | < 0.001 | [0.0310, 0.0533] | Naïve Bayes > SVM | H6 | 0.019 | 0.025 | C < D |
| | | | | z-score | 60.5660 | 2.3380 | < 0.001 | [0.0079, 0.0302] | Naïve Bayes > RUSBoost | H2 | < 0.001 | 0.005 | Naïve Bayes > SVM |
| | | | SVM | Residual | 50.0943 | 5.9382 | < 0.001 | [0.0259, 0.0482] | KNN > SVM | H4 | < 0.001 | 0.01 | Naïve Baye s > RUSBoost |
| | | | | z-score | 63.8443 | 3.2402 | 0.006 | [0.0028, 0.0251] | KNN > RUSBoost | H5 | < 0.001 | 0.015 | KNN > SVM |
| | | | Logistic | Residual | 61.8160 | 2.0842 | < 0.001 | [-0.0470, -0.0247] | SVM < Logistic | H7 | < 0.001 | 0.02 | KNN > |
| | | | | | | | | | | | | | RUSBoost |
| | | | | z-score | 59.1038 | 3.3322 | < 0.001 | [-0.0343, -0.0120] | SVM < RUSBoost | H8 | < 0.001 | 0.025 | SVM < Logistic |
| | | | RUSBoost | Residual | 59.2217 | 4.4604 | 0.016 | [0.0016, 0.0239] | Logistic > RUSBoost | H9 | < 0.001 | 0.03 | SVM < RUSBoost |
| | | | | z-score | 60.5660 | 5.0210 | < 0.001 | [-0.031, -0.021] | Residual < Z-score | H10 | 0.002 | 0.035 | Logistic > RUSBoost |
| | | B | Naïve Bayes | Residual | 60.7075 | 3.4497 | | | | H1 | < 0.001 | 0.05 | Residual < Z-score |
| | | | | z-score | 60.9906 | 2.5704 | | | | | | | |
| | | | KNN | Residual | 61.5802 | 4.3783 | | | | | | | |
| | | | | z-score | 61.3915 | 2.7087 | | | | | | | |
| | | | SVM | Residual | 51.5802 | 5.4070 | | | | | | | |
| | | | | z-score | 63.1368 | 3.2075 | | | | | | | |
| | | | Logistic | Residual | 62.8302 | 1.9333 | | | | | | | |
| | | | | z-score | 60.2594 | 3.0641 | | | | | | | |
| | | | RUSBoost | Residual | 60.4009 | 4.4918 | | | | | | | |
| | | | | z-score | 58.1604 | 3.6316 | | | | | | | |
| | | C | Naïve Bayes | Residual | 61.1321 | 2.9013 | | | | | | | |
| | | | | z-score | 63.2311 | 2.5948 | | | | | | | |
| | | | KNN | Residual | 60.4481 | 2.9129 | | | | | | | |
| | | | | z-score | 62.5943 | 2.8347 | | | | | | | |
| | | | SVM | Residual | 51.7689 | 3.7138 | | | | | | | |
| | | | | z-score | 63.0896 | 3.3917 | | | | | | | |
| | | | Logistic | Residual | 62.2642 | 2.4149 | | | | | | | |
| | | | | z-score | 60.7783 | 4.8062 | | | | | | | |
| | | | RUSBoost | Residual | 60.2594 | 4.8198 | | | | | | | |
| | | | | z-score | 61.0849 | 4.6833 | | | | | | | |

| | | | | | | | | | | | | |
|---|---|---|---|---|---|---|---|---|---|---|---|---|
| | | D | Naïve Bayes | Residual | 62.8066 | 3.1597 | | | | | | |
| | | | | z-score | 63.6085 | 2.2714 | | | | | | |
| | | | KNN | Residual | 61.5094 | 3.8879 | | | | | | |
| | | | | z-score | 62.9717 | 2.7653 | | | | | | |
| | | | SVM | Residual | 52.3585 | 4.1800 | | | | | | |
| | | | | z-score | 66.5802 | 2.9129 | | | | | | |
| | | | Logistic | Residual | 62.0283 | 2.5859 | | | | | | |
| | | | | z-score | 62.0755 | 5.0974 | | | | | | |
| | | | RUSBoost | Residual | 60.8019 | 4.5986 | | | | | | |
| | | | | z-score | 60.4717 | 3.1720 | | | | | | |
| | F1 | A | Naïve Bayes | Residual | 58.5075 | 3.9593 | 0.020 | [-0.0259, -0.0015] | A < D | H6 | 0.001 | 0.0083 | C < D |
| | | | | z-score | 57.9142 | 5.1847 | 0.010 | [-.0270, -0.0026] | B < D | H5 | 0.002 | 0.0167 | B < D |
| | | | KNN | Residual | 59.4643 | 4.4422 | 0.007 | [-0.0275, -0.0031] | C < D | H3 | 0.004 | 0.025 | A < D |
| | | | | z-score | 57.8484 | 4.0731 | < 0.001 | [0.0085, 0.0374] | Naïve Bayes > SVM | H2 | < 0.001 | 0.005 | Naïve Bayes > SVM |
| | | | SVM | Residual | 49.5125 | 5.7867 | < 0.001 | [-0.0665, -0.0376] | Naïve Bayes < Logistic | H3 | < 0.001 | 0.01 | Naïve Bayes < Logistic |
| | | | | z-score | 61.5109 | 4.2382 | < 0.001 | [0.0191, 0.0480] | KNN > SVM | H5 | < 0.001 | 0.015 | KNN > SVM |
| | | | Logistic | Residual | 66.0576 | 1.2930 | < 0.001 | [-0.0559, -0.0270] | KNN < Logistic | H6 | < 0.001 | 0.02 | KNN < Logistic |
| | | | | z-score | 58.4754 | 3.1268 | < 0.001 | [-0.0895, -0.0605] | SVM < Logistic | H8 | < 0.001 | 0.025 | SVM < |
| | | | | | | | | | | | | | Logistic |
| | | | RUSBoost | Residual | 58.2931 | 6.2170 | < 0.001 | [-0.0484, -0.0195] | SVM < RUSBoost | H9 | < 0.001 | 0.03 | SVM < Logistic |
| | | | | z-score | 60.4379 | 7.3150 | < 0.001 | [0.0266, 0.0555] | Logistic > RUSBoost | H10 | < 0.001 | 0.035 | Logistic > RUSBoost |
| | | B | Naïve Bayes | Residual | 60.3393 | 4.2908 | | | NS for harmonization | H4 | 0.038 | 0.04 | Naïve Bayes < RUSBoost |
| | | | | z-score | 55.0534 | 3.7307 | | | | H1 | 0.046 | 0.045 | Naïve Bayes < KNN |
| | | | KNN | Residual | 61.6495 | 4.9947 | | | | | | | NS for harmonization |
| | | | | z-score | 56.9109 | 5.4564 | | | | | | | |
| | | | SVM | Residual | 50.4654 | 6.1437 | | | | | | | |
| | | | | z-score | 59.1945 | 4.400 | | | | | | | |
| | | | Logistic | Residual | 66.6847 | 1.2132 | | | | | | | |
| | | | | z-score | 59.7058 | 3.2938 | | | | | | | |
| | | | RUSBoost | Residual | 59.1552 | 7.0273 | | | | | | | |
| | | | | z-score | 57.7608 | 6.7986 | | | | | | | |
| | | C | Naïve Bayes | Residual | 56.6650 | 3.9187 | | | | | | | |
| | | | | z-score | 57.6528 | 4.9979 | | | | | | | |
| | | | KNN | Residual | 57.1254 | 3.2750 | | | | | | | |
| | | | | z-score | 59.0665 | 3.4155 | | | | | | | |
| | | | SVM | Residual | 50.2765 | 4.0691 | | | | | | | |
| | | | | z-score | 59.8483 | 4.7931 | | | | | | | |
| | | | Logistic | Residual | 66.3434 | 1.5135 | | | | | | | |
| | | | | z-score | 60.8942 | 4.5101 | | | | | | | |
| | | | RUSBoost | Residual | 58.8272 | 6.6679 | | | | | | | |
| | | | | z-score | 59.7105 | 5.8883 | | | | | | | |
| | | D | Naïve Bayes | Residual | 59.0026 | 2.9076 | | | | | | | |
| | | | | z-score | 59.3751 | 3.4392 | | | | | | | |
| | | | KNN | Residual | 59.8176 | 4.0491 | | | | | | | |
| | | | | z-score | 61.0869 | 2.6516 | | | | | | | |
| | | | SVM | Residual | 51.3963 | 5.6613 | | | | | | | |
| | | | | z-score | 63.9394 | 3.4068 | | | | | | | |
| | | | Logistic | Residual | 66.1368 | 1.5895 | | | | | | | |
| | | | | z-score | 61.8414 | 2.0448 | | | | | | | |
| | | | RUSBoost | Residual | 59.9989 | 6.9815 | | | | | | | |
| | | | | z-score | 59.1212 | 6.3616 | | | | | | | |
| | MCC′ | A | Naïve Bayes | Residual | 60.9037 | 3.2194 | < 0.001 | [-0.0251, -0.0061] | A < D | H3 | < 0.001 | 0.0083 | A < D |
| | | | | z-score | 63.3963 | 3.2959 | < 0.001 | [-0.0237, -0.0047] | B < D | H5 | < 0.001 | 0.0167 | B < D |
| | | | KNN | Residual | 61.1478 | 2.9901 | < 0.001 | [0.0335, 0.0560] | Naïve Bayes > SVM | H6 | 0.025 | 0.025 | C < D |
| | | | | z-score | 60.7269 | 2.3645 | < 0.001 | [0.0096, 0.0321] | Naïve Bayes > RUSBoost | H2 | 0.048 | 0.033 | A < C |
| | | | SVM | Residual | 50.0969 | 5.9472 | < 0.001 | [0.0264, 0.0490] | KNN > SVM | H2 | < 0.001 | 0.005 | Naïve Bayes > SVM |
| | | | | z-score | 63.9673 | 3.2327 | 0.008 | [0.0025, 0.0251] | KNN > RUSBoost | H4 | < 0.001 | 0.01 | Naïve Bayes > RUSBoost |
| | | | Logistic | Residual | 62.1899 | 2.0293 | < 0.001 | [-0.0480, -0.0255] | SVM < Logistic | H5 | < 0.001 | 0.015 | KNN > SVM |
| | | | | z-score | 59.1253 | 3.3361 | < 0.001 | [-0.0352, -0.0126] | SVM < RUSBoost | H7 | < 0.001 | 0.02 | KNN > RUSBoost |
| | | | RUSBoost | Residual | 59.2925 | 4.4720 | 0.016 | [0.0016, 0.0241] | Logistic > RUSBoost | H8 | < 0.001 | 0.025 | SVM < Logistic |
| | | | | z-score | 60.8073 | 5.2156 | < 0.001 | [-0.031, -0.021] | Residual < Z-score | H9 | < 0.001 | 0.03 | SVM < RUSBoost |
| | | B | Naïve Bayes | Residual | 60.8563 | 3.5670 | | | | H10 | 0.002 | 0.035 | Logistic > RUSBoost |
| | | | | z-score | 61.4221 | 2.6536 | | | | H1 | < 0.001 | 0.05 | Residual < Z-score |
| | | | KNN | Residual | 61.6223 | 4.3815 | | | | | | | |
| | | | | z-score | 61.7521 | 2.7153 | | | | | | | |
| | | | SVM | Residual | 51.5807 | 5.4335 | | | | | | | |
| | | | | z-score | 63.4315 | 3.2211 | | | | | | | |
| | | | Logistic | Residual | 63.1784 | 1.8744 | | | | | | | |
| | | | | z-score | 60.2994 | 3.0693 | | | | | | | |
| | | | RUSBoost | Residual | 60.5388 | 4.4808 | | | | | | | |
| | | | | z-score | 58.3177 | 3.7034 | | | | | | | |
| | | C | Naïve Bayes | Residual | 61.5758 | 3.1988 | | | | | | | |
| | | | | z-score | 63.7251 | 2.4656 | | | | | | | |
| | | | KNN | Residual | 60.6363 | 3.0294 | | | | | | | |
| | | | | z-score | 62.8547 | 2.9607 | | | | | | | |
| | | | SVM | Residual | 51.7812 | 3.7352 | | | | | | | |
| | | | | z-score | 63.2867 | 3.3589 | | | | | | | |
| | | | Logistic | Residual | 62.6243 | 2.3485 | | | | | | | |
| | | | | z-score | 60.7915 | 4.8096 | | | | | | | |
| | | | RUSBoost | Residual | 60.4857 | 4.9405 | | | | | | | |
| | | | | z-score | 61.2148 | 4.7537 | | | | | | | |
| | | D | Naïve Bayes | Residual | 63.1794 | 3.4229 | | | | | | | |
| | | | | z-score | 64.0226 | 2.3940 | | | | | | | |
| | | | KNN | Residual | 61.6134 | 3.9225 | | | | | | | |
| | | | | z-score | 61.6134 | 3.9225 | | | | | | | |
| | | | SVM | Residual | 52.3613 | 4.1964 | | | | | | | |
| | | | | z-score | 66.7764 | 2.9571 | | | | | | | |
| | | | Logistic | Residual | 62.3784 | 2.5252 | | | | | | | |
| | | | | z-score | 62.0895 | 5.1080 | | | | | | | |
| | | | RUSBoost | Residual | 61.0774 | 4.7127 | | | | | | | |
| | | | | z-score | 60.6533 | 3.2087 | | | | | | | |
| | AROC | A | Naïve Bayes | Residual | 66.2378 | 2.1406 | 0.001 | [-0.0279, -0.0050] | A < D | H3 | < 0.001 | 0.0083 | A < D |
| | | | | z-score | 68.0236 | 3.1790 | < 0.001 | [-0.0293, -0.0063] | B < D | H5 | < 0.001 | 0.0167 | B < D |
| | | | KNN | Residual | 64.1787 | 3.6558 | < 0.001 | [0.0097, 0.0369] | Naïve Bayes > KNN | H4 | 0.031 | 0.025 | B < C |
| | | | | z-score | 65.1333 | 2.9851 | < 0.001 | [0.0575, 0.0848] | Naïve Bayes > SVM | H1 | < 0.001 | 0.005 | Naïve Bayes > KNN |
| | | | SVM | Residual | 50.1227 | 7.6377 | < 0.001 | [0.0410, 0.0683] | Naïve Bayes > Logistic | H2 | < 0.001 | 0.01 | Naïve Bayes > SVM |
| | | | | z-score | 69.0413 | 3.3664 | < 0.001 | [0.0445, 0.0718] | Naïve Bayes > RUSBoost | H3 | < 0.001 | 0.015 | Naïve Bayes > Logistic |
| | | | Logistic | Residual | 60.6725 | 2.7434 | < 0.001 | [0.0342, 0.0615] | KNN > SVM | H4 | < 0.001 | 0.02 | Naïve Bayes > RUSBoost |
| | | | | z-score | 63.2366 | 3.7168 | < 0.001 | [0.0177, 0.0450] | KNN > Logistic | H5 | < 0.001 | 0.025 | KNN > SVM |
| | | | RUSBoost | Residual | 61.7533 | 4.8717 | < 0.001 | [0.0212, 0.0485] | KNN > RUSBoost | H6 | < 0.001 | 0.03 | KNN > Logistic |
| | | | | z-score | 62.7525 | 4.4215 | 0.008 | [-0.0301, -0.0029] | SVM < Logistic | H7 | < 0.001 | 0.035 | KNN > RUSBoost |
| | | B | Naïve Bayes | Residual | 66.4306 | 3.1301 | < 0.001 | [-0.055, -0.043] | Residual < Z-score | H8 | < 0.001 | 0.04 | SVM < logistic |
| | | | | z-score | 66.6209 | 2.1095 | | | | H9 | 0.009 | 0.045 | SVM < RUSBoost |
| | | | KNN | Residual | 64.3955 | 4.5949 | | | | H1 | < 0.001 | 0.005 | Residual < Z-score |
| | | | | z-score | 65.4467 | 3.8292 | | | | | | | |
| | | | SVM | Residual | 50.5289 | 13.5394 | | | | | | | |
| | | | | z-score | 68.8364 | 2.3262 | | | | | | | |
| | | | Logistic | Residual | 61.5450 | 3.9805 | | | | | | | |
| | | | | z-score | 64.3124 | 3.2277 | | | | | | | |
| | | | RUSBoost | Residual | 61.9905 | 3.8556 | | | | | | | |
| | | | | z-score | 59.6909 | 4.2791 | | | | | | | |
| | | C | | Residual | 67.5523 | 3.1095 | | | | | | | |
| | | | Naïve Bayes | z-score | 68.7035 | 2.3168 | | | | | | | |
| | | | KNN | Residual | 64.3360 | 3.1468 | | | | | | | |
| | | | | z-score | 66.7634 | 3.4886 | | | | | | | |
| | | | SVM | Residual | 53.0719 | 5.9671 | | | | | | | |
| | | | | z-score | 68.7645 | 3.8384 | | | | | | | |
| | | | Logistic | Residual | 60.6240 | 3.7212 | | | | | | | |
| | | | | z-score | 64.4081 | 4.4113 | | | | | | | |
| | | | RUSBoost | Residual | 62.3486 | 4.2655 | | | | | | | |
| | | | | z-score | 62.8287 | 5.1086 | | | | | | | |
| | | D | Naïve Bayes | Residual | 69.2295 | 3.1924 | | | | | | | |
| | | | | z-score | 69.9446 | 2.3887 | | | | | | | |
| | | | KNN | Residual | 65.9151 | 4.0882 | | | | | | | |
| | | | | z-score | 67.9481 | 3.4249 | | | | | | | |
| | | | SVM | Residual | 53.4438 | 6.4185 | | | | | | | |
| | | | | z-score | 72.0124 | 2.3998 | | | | | | | |
| | | | Logistic | Residual | 58.1742 | 3.4906 | | | | | | | |
| | | | | z-score | 66.0674 | 5.1594 | | | | | | | |
| | | | RUSBoost | Residual | 62.4612 | 4.1880 | | | | | | | |
| | | | | z-score | 62.4023 | 3.5042 | | | | | | | |

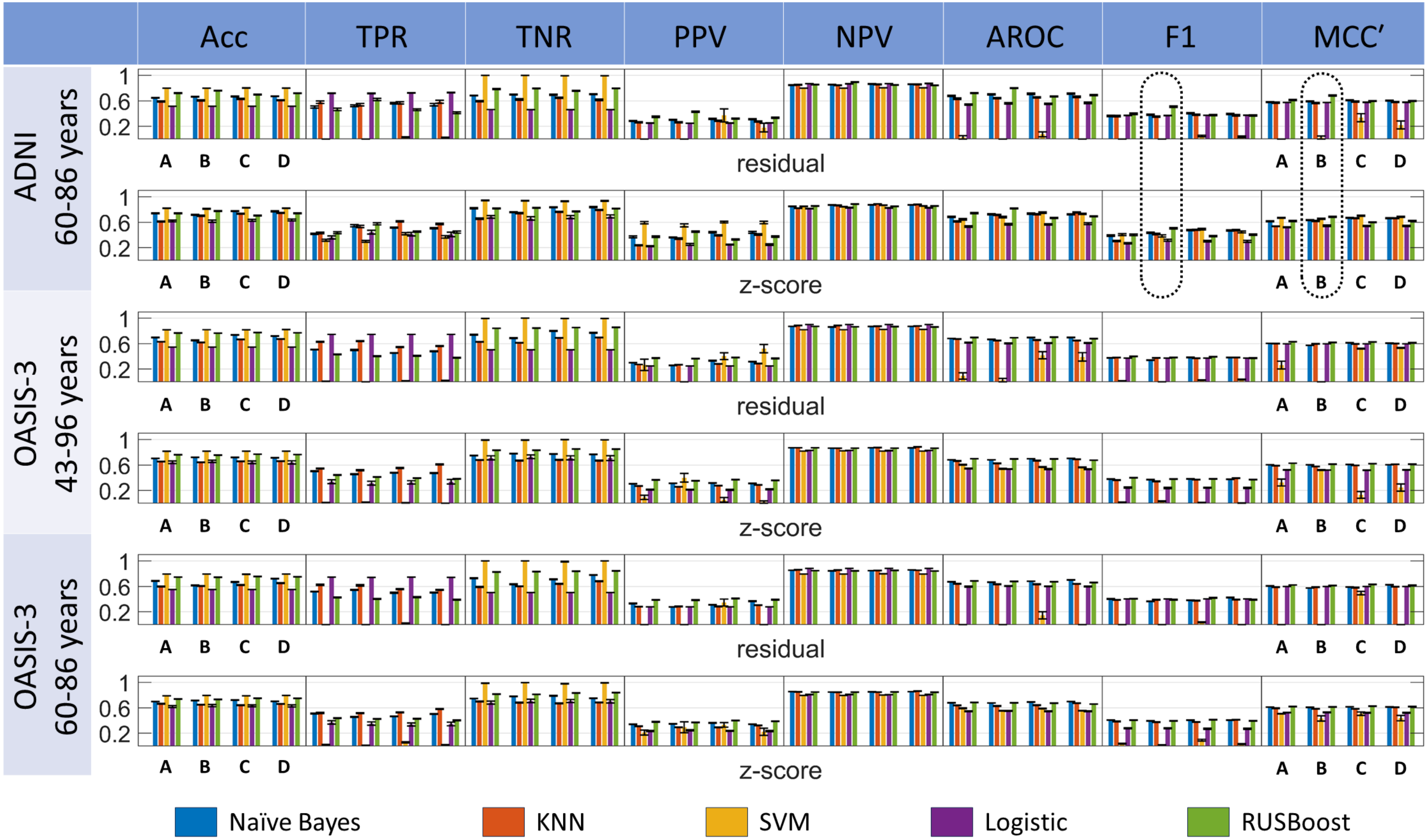

**Supplementary Figure 2:** Comparison among multiple classification pipeline options, involving five classifiers, four feature selection, and two harmonization techniques. Performance is measured directly for ADNI and OASIS-3 imbalanced datasets within a Monte Carlo replication analysis. See **Supplementary Figure 1** for complementary information regarding balanced data analysis and figure caption.

**Supplementary Table 3:** Comparative analysis of classifier models and feature selection algorithms using different normalization techniques (see **Fig. 5** for complementary information) for imbalanced analysis for ADNI dataset

****classifier model pairwise comparison: p value post hoc test (Tukey's HSD)
**** Pairwise comparison: p-value: Adjustment for multiple comparison: Bonferroni
**** NS = Not significant
***** Features selection methods:
A = Average score percentage (MCL app): chi-square, ReliefF, ANOVA & Kruskal Wallis
**B** = ReliefF (MCL app)
**C** = Frequent feature appearances from all feature ranking analysis
**D** = features extracted from statistical analysis (SPSS software): ANOVA, ANCOVA & Kruskal Wallis

| Dataset | Classification | Features selection group | Classifier model | Harmonization | Mean (%) | Standard deviation | Adjustment for multiple comparison: Bonferroni | | | Adjustment for multiple comparison: Benjamini-Hochberg | | | |
|---|---|---|---|---|---|---|---|---|---|---|---|---|---|
| | | | | | | | p-value (**Displaying only statistically significant p-values**) | 95% confidence interval | Pairwise comparison | Hypothesis ID | Raw p-value | adj. p-value | Pairwise comparison |
| ADNI | Accuracy | A | Naïve Bayes | Residual | 64.7521 | 2.3109 | < 0.001 | [-0.0243, -0.0078] | A < B | H3 | < 0.001 | 0.0083 | A < D |
| | | | | z-score | 73.9256 | 3.3967 | < 0.001 | [-0.0265, -0.0099] | A < C | H2 | <0.001 | 0.0167 | A < C |
| | | | KNN | Residual | 59.0909 | 3.7230 | < 0.001 | [-0.0312, -0.0146] | A < D | H1 | < 0.001 | 0.025 | A < B |
| | | | | z-score | 61.0744 | 3.8754 | < 0.001 | [0.0430, 0.0626] | Naïve Bayes > KNN | H5 | 0.033 | 0.0333 | B < D |
| | | | SVM | Residual | 80.1653 | 0 | < 0.001 | [-0.1143, -0.0946] | Naïve Bayes < SVM | H1 | < 0.001 | 0.005 | Naïve Bayes |
| | | | | | | | | | | | | | > KNN |
| | | | | z-score | 81.9008 | 1.2831 | < 0.001 | [0.1272, 0.1469] | Naïve Bayes > Logistic | H2 | < 0.001 | 0.01 | Naïve Bayes < SVM |
| | | | Logistic | Residual | 51.4463 | 0.4546 | < 0.001 | [-0.0364, -0.0167] | Naïve Bayes < RUSBoost | H3 | < 0.001 | 0.015 | Naïve Bayes > Logistic |
| | | | | z-score | 62.0248 | 5.9685 | < 0.001 | [-0.1671, -0.1474] | KNN < SVM | H4 | < 0.001 | 0.02 | Naïve Bayes < - RUSBoost |
| | | | RUSBoost | Residual | 72.1488 | 2.5733 | < 0.001 | [0.0744, 0.0941] | KNN > Logistic | H5 | < 0.001 | 0.025 | KNN < SVM |
| | | | | z-score | 74.0909 | 2.6446 | < 0.001 | [-0.0892, -0.0695] | KNN < RUSBoost | H6 | < 0.001 | 0.03 | KNN > Logistic |
| | | B | Naïve Bayes | Residual | 66.4050 | 2.2317 | < 0.001 | [0.2317, 0.2513] | SVM > Logistic | H7 | < 0.001 | 0.035 | KNN < RUSBoost |
| | | | | z-score | 71.6942 | 2.4703 | < 0.001 | [0.0681, 0.0877] | SVM > RUSBoost | H9 | < 0.001 | 0.04 | SVM > Logistic |
| | | | KNN | Residual | 60.5372 | 3.8566 | < 0.001 | [-0.1734, -0.1538] | Logistic < RUSBoost | H9 | < 0.001 | 0.045 | SVM > RUSBoost |
| | | | | z-score | 70.1240 | 3.8041 | < 0.001 | [-0.069, -0.06] | Residual < Z-score | H10 | < 0.001 | 0.05 | Logistic < RUSBoost |
| | | | SVM | Residual | 80.0826 | 0.3696 | | | | H1 | < 0.001 | 0.05 | Residual < Z-score |
| | | | | z-score | 81.1157 | 2.2478 | | | | | | | |
| | | | Logistic | Residual | 51.3223 | 0.3696 | | | | | | | |
| | | | | z-score | 61.6942 | 8.0720 | | | | | | | |
| | | | RUSBoost | Residual | 76.0331 | 2.3681 | | | | | | | |
| | | | | z-score | 77.6446 | 2.3448 | | | | | | | |
| | | C | Naïve Bayes | Residual | 66.9421 | 3.1384 | | | | | | | |
| | | | | z-score | 77.3554 | 2.3742 | | | | | | | |
| | | | KNN | Residual | 63.0992 | 3.6987 | | | | | | | |
| | | | | z-score | 73.4298 | 3.0487 | | | | | | | |
| | | | SVM | Residual | 80.2893 | 0.9003 | | | | | | | |
| | | | | z-score | 82.9339 | 1.5925 | | | | | | | |
| | | | Logistic | Residual | 51.4050 | 0.6889 | | | | | | | |
| | | | | z-score | 62.8512 | 6.7059 | | | | | | | |
| | | | RUSBoost | Residual | 69.8347 | 2.2988 | | | | | | | |
| | | | | z-score | 70.6612 | 2.9434 | | | | | | | |
| | | D | Naïve Bayes | Residual | 67.1488 | 1.7962 | | | | | | | |
| | | | | z-score | 77.1074 | 3.4661 | | | | | | | |
| | | | KNN | Residual | 60.9091 | 3.6970 | | | | | | | |
| | | | | z-score | 74.8347 | 2.8938 | | | | | | | |
| | | | SVM | Residual | 80.1653 | 0.5363 | | | | | | | |
| | | | | z-score | 82.2314 | 1.8087 | | | | | | | |
| | | | Logistic | Residual | 51.5289 | 0.4852 | | | | | | | |
| | | | | z-score | 63.4298 | 6.5178 | | | | | | | |
| | | | RUSBoost | Residual | 72.0248 | 1.8822 | | | | | | | |
| | | | | z-score | 74.1322 | 2.3847 | | | | | | | |
| | F1 | A | Naïve Bayes | Residual | 36.0377 | 4.4481 | < 0.001 | [-0.0522, -0.0281] | A < B | H2 | < 0.001 | 0.0083 | A < C |
| | | | | z-score | 38.9049 | 4.7583 | < 0.001 | [-0.0568, -0.0328] | A < C | H1 | < 0.001 | 0.0167 | A < B |

| | | | | | | | | | | | | |
|---|---|---|---|---|---|---|---|---|---|---|---|---|
| | | | KNN | Residual | 35.9312 | 4.8959 | < 0.001 | [-0.0501, -0.0261] | A < D | H3 | < 0.001 | 0.025 | A < D |
| | | | | z-score | 30.4679 | 3.5664 | < 0.001 | [0.0068, 0.0354] | Naïve Bayes > KNN | H9 | < 0.001 | 0.005 | SVM < Logistic |
| | | | SVM | Residual | 0 | 0 | < 0.001 | [0.1728, 0.2014] | Naïve Bayes > SVM | H2 | < 0.001 | 0.01 | Naïve Bayes > SVM |
| | | | | z-score | 40.5470 | 6.0897 | < 0.001 | [0.0650, 0.0936] | Naïve Bayes > Logistic | H5 | < 0.001 | 0.015 | KNN > SVM |
| | | | Logistic | Residual | 37.0613 | 0.8350 | < 0.001 | [0.1517, 0.1803] | KNN > SVM | H8 | < 0.001 | 0.02 | SVM < Logistic |
| | | | | z-score | 26.9913 | 4.6160 | < 0.001 | [0.0440, 0.0725] | KNN > Logistic | H10 | < 0.001 | 0.025 | Logistic < RUSBoost |
| | | | RUSBoost | Residual | 39.6844 | 5.3813 | < 0.001 | [-0.0392, -0.0106] | KNN < RUSBoost | H3 | < 0.001 | 0.03 | Logistic > Logistic |
| | | | | z-score | 39.9543 | 5.7453 | < 0.001 | [-0.1221, -0.0935] | SVM < Logistic | H6 | < 0.001 | 0.035 | KNN > Logistic |
| | | B | Naïve Bayes | Residual | 38.1444 | 3.3950 | < 0.001 | [-0.2052, -0.1766] | SVM < RUSBoost | H7 | < 0.001 | 0.04 | KNN < RUSBoost |
| | | | | z-score | 43.2623 | 4.6927 | < 0.001 | [-0.0975, 0.0689] | Logistic < RUSBoost | H1 | < 0.001 | 0.045 | Naïve Bayes |
| | | | | | | | | | | | | | > KNN |
| | | | KNN | Residual | 35.1090 | 4.7882 | < 0.001 | [-0.098, -0.085] | Residual < Z-score | H1 | < 0.001 | 0.05 | Residual < Z-score |
| | | | | z-score | 41.4994 | 5.0098 | | | | | | | |
| | | | SVM | Residual | 0 | 0 | | | | | | | |
| | | | | z-score | 38.4188 | 6.8875 | | | | | | | |
| | | | Logistic | Residual | 36.8673 | 0.6923 | | | | | | | |
| | | | | z-score | 31.3866 | 6.9799 | | | | | | | |
| | | | RUSBoost | Residual | 50.5969 | 5.2048 | | | | | | | |
| | | | | z-score | 50.4421 | 5.0576 | | | | | | | |
| | | C | Naïve Bayes | Residual | 40.3730 | 3.9487 | | | | | | | |
| | | | | z-score | 47.4765 | 3.9164 | | | | | | | |
| | | | KNN | Residual | 37.9349 | 4.4667 | | | | | | | |
| | | | | z-score | 47.9604 | 3.6313 | | | | | | | |
| | | | SVM | Residual | 4.6210 | 5.7673 | | | | | | | |
| | | | | z-score | 49.0687 | 5.9858 | | | | | | | |
| | | | Logistic | Residual | 37.1773 | 0.9580 | | | | | | | |
| | | | | z-score | 30.1444 | 4.7325 | | | | | | | |
| | | | RUSBoost | Residual | 37.5764 | 3.9190 | | | | | | | |
| | | | | z-score | 38.0492 | 3.5583 | | | | | | | |
| | | D | Naïve Bayes | Residual | 39.2905 | 4.3668 | | | | | | | |
| | | | | z-score | 46.9732 | 4.8702 | | | | | | | |
| | | | KNN | Residual | 37.1536 | 4.7976 | | | | | | | |
| | | | | z-score | 47.5438 | 4.3277 | | | | | | | |
| | | | SVM | Residual | 3.3535 | 5.6149 | | | | | | | |
| | | | | z-score | 44.7706 | 6.3947 | | | | | | | |
| | | | Logistic | Residual | 37.3687 | 0.8824 | | | | | | | |
| | | | | z-score | 30.0029 | 6.9896 | | | | | | | |
| | | | RUSBoost | Residual | 36.8080 | 4.9263 | | | | | | | |
| | | | | z-score | 40.4196 | 5.3271 | | | | | | | |
| | MCC′ | A | Naïve Bayes | Residual | 57.7349 | 30945 | < 0.001 | [-0.0465, -0.0097] | A < B | H2 | < 0.001 | 0.0083 | A < C |
| | | | | z-score | 61.3445 | 3.3174 | < 0.001 | [-0.0769, -0.0402] | A < C | H3 | < 0.001 | 0.0167 | A < D |
| | | | KNN | Residual | 56.9601 | 3.9306 | < 0.001 | [-0.0644, -0.0277] | A < D | H4 | < 0.001 | 0.025 | B < C |
| | | | | z-score | 53.5664 | 2.7129 | < 0.001 | [-0.0488, -0.0121] | B < C | H1 | < 0.001 | 0.033 | A < B |
| | | | SVM | Residual | 0 | 0 | < 0.001 | [0.1874, 0.2310] | Naïve Bayes > SVM | H5 | 0.012 | 0.042 | B < D |
| | | | | z-score | 66.7977 | 2.5259 | < 0.001 | [0.0420, 0.0856] | Naïve Bayes > Logistic | H9 | < 0.001 | 0.005 | SVM < RUSBoost |
| | | | Logistic | Residual | 57.4274 | 0.8174 | < 0.001 | [0.1660, 0.2096] | KNN > SVM | H2 | < 0.001 | 0.01 | Naïve Bayes > SVM |
| | | | | z-score | 51.9918 | 3.0235 | < 0.001 | [0.0205, 0.0641] | KNN > Logistic | H5 | < 0.001 | 0.015 | KNN > SVM |
| | | | RUSBoost | Residual | 61.2951 | 3.4053 | 0.004 | [-0.0501, -0.0065] | KNN < RUSBoost | H8 | < 0.001 | 0.02 | SVM < Logistic |
| | | | | z-score | 61.8886 | 3.6280 | < 0.001 | [-0.1672, -0.1236] | SVM < Logistic | H10 | < 0.001 | 0.025 | Logistic < RUSBoost |
| | | B | Naïve Bayes | Residual | 59.3151 | 2.3941 | < 0.001 | [-0.2379, -0.1943] | SVM < RUSBoost | H3 | < 0.001 | 0.03 | Naïve Bayes > Logistic |
| | | | | z-score | 63.2723 | 3.0799 | < 0.001 | [-0.0925, -0.0489] | Logistic < RUSBoost | H6 | < 0.001 | 0.035 | KNN > Logistic |
| | | | KNN | Residual | 56.5625 | 3.5708 | < 0.001 | [-0.129, -0.109] | Residual < Z-score | H7 | < 0.001 | 0.04 | KNN < RUSBoost |
| | | | | z-score | 61.9587 | 3.5055 | | | | H1 | 0.007 | 0.045 | Naïve Bayes > KNN |
| | | | SVM | Residual | 2.3388 | 10.4594 | | | | H1 | < 0.001 | 0.05 | Residual < Z-score |
| | | | | z-score | 65.2231 | 4.1998 | | | | | | | |
| | | | Logistic | Residual | 57.2378 | 0.6779 | | | | | | | |
| | | | | z-score | 54.4625 | 5.0155 | | | | | | | |
| | | | RUSBoost | Residual | 68.3117 | 3.4468 | | | | | | | |
| | | | | z-score | 68.4252 | 3.2375 | | | | | | | |
| | | C | Naïve Bayes | Residual | 60.7965 | 2.9720 | | | | | | | |
| | | | | z-score | 66.6945 | 2.6367 | | | | | | | |
| | | | KNN | Residual | 58.8082 | 3.3697 | | | | | | | |
| | | | | z-score | 66.3830 | 2.5412 | | | | | | | |
| | | | SVM | Residual | 33.5233 | 28.4519 | | | | | | | |
| | | | | z-score | 70.2082 | 3.1579 | | | | | | | |
| | | | Logistic | Residual | 57.5376 | 0.9422 | | | | | | | |
| | | | | z-score | 54.1100 | 3.0002 | | | | | | | |
| | | | RUSBoost | Residual | 59.5593 | 2.6148 | | | | | | | |
| | | | | z-score | 59.9720 | 02.6111 | | | | | | | |
| | | D | Naïve Bayes | Residual | 60.2132 | 2.9667 | | | | | | | |
| | | | | z-score | 66.3907 | 3.4642 | | | | | | | |
| | | | KNN | Residual | 58.1145 | 3.5570 | | | | | | | |
| | | | | z-score | 66.2494 | 2.9748 | | | | | | | |
| | | | SVM | Residual | 21.8907 | 27.6870 | | | | | | | |
| | | | | z-score | 68.4077 | 3.1179 | | | | | | | |
| | | | Logistic | Residual | 57.7281 | 0.8641 | | | | | | | |
| | | | | z-score | 54.2344 | 4.0943 | | | | | | | |
| | | | RUSBoost | Residual | 59.662 | 3.0223 | | | | | | | |
| | | | | z-score | 62.1512 | 3.3414 | | | | | | | |
| | AROC | A | Naïve Bayes | Residual | 67.7181 | 3.4287 | < 0.001 | [-0.0529, -0.0259] | A < B | H1 | < 0.001 | 0.0083 | A < B |
| | | | | z-score | 68.4412 | 3.5404 | < 0.001 | [-0.0421, -0.0152] | A < C | H3 | < 0.001 | 0.0167 | A < D |
| | | | KNN | Residual | 63.4806 | 5.0820 | < 0.001 | [-0.0437, -0.0168] | A < D | H2 | < 0.001 | 0.025 | A < C |
| | | | | z-score | 61.0856 | 4.6171 | < 0.001 | [0.0174, 0.0494] | Naïve Bayes > KNN | H4 | 0.041 | 0.033 | B > C |
| | | | SVM | Residual | 2.5 | 11.1803 | < 0.001 | [0.3292, 0.3612] | Naïve Bayes > SVM | H9 | < 0.001 | 0.005 | SVM < RUSBoost |
| | | | | z-score | 64.4907 | 5.6290 | < 0.001 | [0.1351, 0.1671] | Naïve Bayes > Logistic | H2 | < 0.001 | 0.01 | Naïve Bayes > SVM |
| | | | Logistic | Residual | 53.9708 | 3.9789 | < 0.001 | [0.2958, 0.3278] | KNN > SVM | H5 | < 0.001 | 0.015 | KNN > SVM |
| | | | | z-score | 53.0991 | 5.3586 | < 0.001 | [0.1016, 0.1337] | KNN > Logistic | H8 | < 0.001 | 0.02 | SVM < Logistic |
| | | | RUSBoost | Residual | 72.1981 | 3.5397 | < 0.001 | [-0.0662, -0.0342] | KNN < RUSBoost | H10 | < 0.001 | 0.025 | Logistic < RUSBoost |
| | | | | z-score | 74.3435 | 4.1253 | < 0.001 | [-0.2101, -0.1781] | SVM < Logistic | H3 | < 0.001 | 0.03 | Naïve Bayes > Logistic |
| | | B | Naïve Bayes | Residual | 69.9653 | 3.7476 | < 0.001 | [-0.3780, -0.3460] | SVM < RUSBoost | H6 | < 0.001 | 0.035 | KNN > |
| | | | | | | | | | | | | | Logistic |
| | | | | z-score | 72.3981 | 4.3473 | < 0.001 | [-0.1839, -0.1518] | Logistic < RUSBoost | H7 | < 0.001 | 0.04 | KNN < RUSBoost |
| | | | KNN | Residual | 64.0995 | 4.1817 | < 0.001 | [0.524, 0.535] | Residual < Z-score | H1 | < 0.001 | 0.045 | Naïve Bayes > KNN |
| | | | | z-score | 71.5222 | 3.8474 | | | | H4 | 0.004 | 0.05 | Naïve Bayes < RUSBoost |
| | | | SVM | Residual | 0 | 0 | | | | H1 | < 0.001 | 0.05 | Residual < Z-score |
| | | | | z-score | 68.0750 | 4.7735 | | | | | | | |
| | | | Logistic | Residual | 56.1176 | 4.3962 | | | | | | | |
| | | | | z-score | 56.8838 | 5.7037 | | | | | | | |
| | | | RUSBoost | Residual | 80.1088 | 2.5231 | | | | | | | |
| | | | | z-score | 81.5407 | 2.9199 | | | | | | | |
| | | C | Naïve Bayes | Residual | 71.0606 | 4.6236 | | | | | | | |
| | | | | z-score | 73.4620 | 4.3962 | | | | | | | |
| | | | KNN | Residual | 65.2917 | 4.0369 | | | | | | | |
| | | | | z-score | 72.7023 | 3.7043 | | | | | | | |
| | | | SVM | Residual | 7.3 | 17.8447 | | | | | | | |
| | | | | z-score | 75.0125 | 5.0479 | | | | | | | |
| | | | Logistic | Residual | 55.0972 | 4.4881 | | | | | | | |
| | | | | z-score | 56.4222 | 4.3605 | | | | | | | |
| | | | RUSBoost | Residual | 66.9606 | 3.5426 | | | | | | | |
| | | | | z-score | 66.6745 | 2.9261 | | | | | | | |
| | | D | Naïve Bayes | Residual | 71.3421 | 4.4605 | | | | | | | |
| | | | | z-score | 72.4130 | 3.5032 | | | | | | | |
| | | | KNN | Residual | 66.3704 | 5.0504 | | | | | | | |

| | | | | | | | | | | | | |
|---|---|---|---|---|---|---|---|---|---|---|---|---|
| | | | | z-score | 75.5 | 3.1630 | | | | | | |
| | | | SVM | Residual | 0 | 0 | | | | | | |
| | | | | z-score | 73.2463 | 4.0171 | | | | | | |
| | | | Logistic | Residual | 56.5912 | 5.2079 | | | | | | |
| | | | | z-score | 57.7394 | 5.8033 | | | | | | |
| | | | RUSBoost | Residual | 69.0847 | 3.3470 | | | | | | |
| | | | | z-score | 69.2958 | 3.1485 | | | | | | |

**Supplementary Table 4:** Comparative analysis of classifier models and feature selection algorithms using different normalization techniques (see **Fig. 4** for complementary information) for the imbalanced analysis for OASIS-3 (age 60-86) dataset.

| Dataset | Classification | Features selection group | Classifier model | Harmonization | Mean (%) | Standard deviation | Adjustment for multiple comparison: Bonferroni | | | Adjustment for multiple comparison: Benjamini-Hochberg | | | |
|---|---|---|---|---|---|---|---|---|---|---|---|---|---|
| | | | | | | | p-value (**Displaying only statistically significant p-values**) | 95% confidence interval | Pairwise comparison | Hypothesis ID | Raw p-value | adj. p-value | Pairwise comparison |
| OASIS-3 (age 60-86) | Accuracy | A | Naïve Bayes | Residual | 68.5645 | 2.4772 | < 0.001 | [-0.0193, -0.0060] | A < D | H3 | < 0.001 | 0.0083 | A < D |
| | | | | z-score | 69.8170 | 0.8393 | 0.002 | [-0.0158, -0.0025] | B < C | H4 | < 0.001 | 0.0167 | B < C |
| | | | KNN | Residual | 59.7688 | 2.2880 | < 0.001 | [-0.0237, -0.0105] | B < D | H5 | < 0.001 | 0.025 | B < D |
| | | | | z-score | 66.3295 | 1.7311 | 0.011 | [-0.0146, 0.0013] | C < D | H6 | 0.002 | 0.033 | C < D |
| | | | SVM | Residual | 79.5761 | 0 | < 0.001 | [0.0469, 0.0627] | Naïve Bayes > KNN | H1 | < 0.001 | 0.005 | Naïve Bayes > KNN |
| | | | | z-score | 78.8921 | 0.5173 | < 0.001 | [-0.1098, -0.0941] | Naïve Bayes < SVM | H2 | < 0.001 | 0.01 | Naïve Bayes < SVM |
| | | | Logistic | Residual | 55.1252 | 0.1065 | < 0.001 | [0.0932, 0.1089] | Naïve Bayes > Logistic | H3 | < 0.001 | 0.015 | Naïve Bayes > Logistic |
| | | | | z-score | 61.8015 | 6.7768 | < 0.001 | [-0.0630, -0.0472] | Naïve Bayes < RUSBoost | H4 | < 0.001 | 0.02 | Naïve Bayes < RUSBoost |
| | | | RUSBoost | Residual | 74.5472 | 0.8546 | < 0.001 | [-0.1646, -0.1489] | KNN < SVM | H5 | < 0.001 | 0.025 | KNN < SVM |
| | | | | z-score | 73.8247 | 1.0814 | < 0.001 | [0.0384, 0.0541] | KNN > Logistic | H6 | < 0.001 | 0.03 | KNN > Logistic |
| | | B | Naïve Bayes | Residual | 61.5703 | 2.2044 | < 0.001 | [-0.1178, -0.1020] | KNN < RUSBoost | H7 | < 0.001 | 0.035 | KNN < RUSBoost |
| | | | | z-score | 71.4836 | 1.0664 | < 0.001 | [0.1951, 0.2109] | SVM > Logistic | H8 | < 0.001 | 0.04 | SVM > Logistic |
| | | | KNN | Residual | 60.4721 | 1.7214 | < 0.001 | [0.0390, 0.0547] | SVM > RUSBoost | H9 | < 0.001 | 0.045 | SVM > RUSBoost |
| | | | | z-score | 64.7592 | 1.7221 | < 0.001 | [-0.1640, 0.1483] | Logistic < RUSBoost | H10 | < 0.001 | 0.05 | Logistic < |
| | | | | | | | | | | | | | RUSBoost |
| | | | SVM | Residual | 79.5761 | 0 | < 0.001 | [-0.031, -0.024] | Residual < Z-score | H1 | < 0.001 | 0.05 | Residual < Z-score |
| | | | | z-score | 79.3642 | 0.3746 | | | | | | | |
| | | | Logistic | Residual | 55.1252 | 0.1518 | | | | | | | |
| | | | | z-score | 63.5260 | 7.4813 | | | | | | | |
| | | | RUSBoost | Residual | 74.5568 | 1.1967 | | | | | | | |
| | | | | z-score | 73.3237 | 1.0011 | | | | | | | |
| | | C | Naïve Bayes | Residual | 66.9075 | 2.1717 | | | | | | | |
| | | | | z-score | 72.2929 | 1.4985 | | | | | | | |
| | | | KNN | Residual | 62.2929 | 2.5623 | | | | | | | |
| | | | | z-score | 64.0267 | 1.8333 | | | | | | | |
| | | | SVM | Residual | 79.2389 | 1.1388 | | | | | | | |
| | | | | z-score | 78.9499 | 0.6121 | | | | | | | |
| | | | Logistic | Residual | 55.1734 | 0.1686 | | | | | | | |
| | | | | z-score | 63.2274 | 6.5916 | | | | | | | |
| | | | RUSBoost | Residual | 75.6936 | 1.0066 | | | | | | | |
| | | | | z-score | 75.1156 | 1.0976 | | | | | | | |
| | | D | Naïve Bayes | Residual | 72.2447 | 1.5672 | | | | | | | |
| | | | | z-score | 70.0771 | 1.0406 | | | | | | | |
| | | | KNN | Residual | 65.2216 | 2.0174 | | | | | | | |
| | | | | z-score | 66.2620 | 1.7334 | | | | | | | |
| | | | SVM | Residual | 79.5761 | 0 | | | | | | | |
| | | | | z-score | 79.3545 | 0.3204 | | | | | | | |
| | | | Logistic | Residual | 55.2023 | 1.5938 | | | | | | | |
| | | | | z-score | 62.9480 | 7.3766 | | | | | | | |
| | | | RUSBoost | Residual | 75.1445 | 1.0385 | | | | | | | |
| | | | | z-score | 74.8459 | 0.9654 | | | | | | | |
| | F1 | A | Naïve Bayes | Residual | 40.2892 | 1.8216 | < 0.001 | [0.0045, 0.0171] | A > B | H1 | < 0.001 | 0.0083 | A > B |
| | | | | z-score | 40.7772 | 1.1797 | < 0.001 | [-0.0226, -0.0100] | B < C | H4 | < 0.001 | 0.0167 | B < C |
| | | | KNN | Residual | 38.8089 | 2.3038 | < 0.001 | [-0.0180, -0.0054] | B < D | H5 | < 0.001 | 0.025 | B < D |
| | | | | z-score | 38.5943 | 2.2652 | < 0.001 | [0.0054, 0.0203] | Naïve Bayes > KNN | H2 | 0.025 | 0.033 | A < C |
| | | | SVM | Residual | 0 | 0 | < 0.001 | [0.3668, 0.3817] | Naïve Bayes > SVM | H1 | < 0.001 | | Naïve Bayes > KNN |
| | | | | z-score | 3.4872 | 3.1793 | < 0.001 | [0.0534, 0.0683] | Naïve Bayes > Logistic | H2 | < 0.001 | | Naïve Bayes > SVM |
| | | | Logistic | Residual | 40.3890 | 0.1395 | < 0.001 | [0.3539, 0.3688] | KNN > SVM | H3 | < 0.001 | | Naïve Bayes > Logistic |
| | | | | z-score | 27.7672 | 3.7698 | < 0.001 | [0.0405, 0.0554] | KNN > Logistic | H5 | < 0.001 | | KNN > SVM |
| | | | RUSBoost | Residual | 40.5314 | 1.9544 | < 0.001 | [-0.0229, -0.0079] | KNN < RUSBoost | H6 | < 0.001 | | KNN > Logistic |
| | | | | z-score | 40.5110 | 1.8760 | < 0.001 | [-0.3209, -0.3059] | SVM < Logistic | H7 | < 0.001 | | KNN < RUSBoost |
| | | B | Naïve Bayes | Residual | 36.7397 | 1.2243 | < 0.001 | [-0.3842, -0.3693] | SVM < RUSBoost | H8 | < 0.001 | | SVM < Logistic |
| | | | | z-score | 39.4623 | 1.7538 | < 0.001 | [-0.0708, -0.0559] | Logistic < RUSBoost | H9 | < 0.001 | | SVM < RUSBoost |
| | | | KNN | Residual | 38.8543 | 2.4859 | < 0.001 | [0.014, 0.021] | Residual > Z-score | H10 | < 0.001 | | Logistic < RUSBoost |
| | | | | z-score | 37.3813 | 1.8011 | | | | H1 | < 0.001 | 0.05 | Residual > Z-score |
| | | | SVM | Residual | 0 | 0 | | | | | | | |
| | | | | z-score | 1.2742 | 1.6602 | | | | | | | |
| | | | Logistic | Residual | 40.3279 | 0.2145 | | | | | | | |
| | | | | z-score | 27.6837 | 3.4944 | | | | | | | |
| | | | RUSBoost | Residual | 39.1837 | 2.4782 | | | | | | | |
| | | | | z-score | 39.4522 | 1.8545 | | | | | | | |
| | | C | Naïve Bayes | Residual | 38.1505 | 1.6428 | | | | | | | |
| | | | | z-score | 40.6422 | 1.6719 | | | | | | | |
| | | | KNN | Residual | 37.6445 | 2.7208 | | | | | | | |
| | | | | z-score | 37.4457 | 1.8359 | | | | | | | |
| | | | SVM | Residual | 3.5272 | 3.5586 | | | | | | | |
| | | | | z-score | 8.8575 | 6.3667 | | | | | | | |
| | | | Logistic | Residual | 40.4303 | 0.1320 | | | | | | | |
| | | | | z-score | 26.7457 | 4.0245 | | | | | | | |
| | | | RUSBoost | Residual | 41.9387 | 2.3561 | | | | | | | |
| | | | | z-score | 41.2596 | 2.0285 | | | | | | | |
| | | D | Naïve Bayes | Residual | 42.5352 | 1.7050 | | | | | | | |
| | | | | z-score | 40.7284 | 1.3597 | | | | | | | |
| | | | KNN | Residual | 39.0854 | 1.8219 | | | | | | | |
| | | | | z-score | 41.2112 | 1.7956 | | | | | | | |
| | | | SVM | Residual | 0 | 0 | | | | | | | |
| | | | | z-score | 2.7818 | 3.5328 | | | | | | | |
| | | | Logistic | Residual | 40.4152 | 0.1777 | | | | | | | |
| | | | | z-score | 26.8894 | 4.8358 | | | | | | | |
| | | | RUSBoost | Residual | 38.9842 | 2.6277 | | | | | | | |
| | | | | z-score | 39.4700 | 1.9998 | | | | | | | |
| | MCC′ | A | Naïve Bayes | Residual | 60.6877 | 1.4722 | < 0.001 | [-0.062, -0.035] | A < C | H2 | < 0.001 | 0.0083 | A < C |
| | | | | z-score | 61.1936 | 0.7764 | < 0.001 | [-0.075, -0.048] | B < C | H4 | < 0.001 | 0.0167 | B < C |
| | | | KNN | Residual | 58.7492 | 1.8826 | < 0.001 | [0.038, 0.066] | C > D | H6 | < 0.001 | 0.025 | C > D |
| | | | | z-score | 59.2975 | 1.6312 | < 0.001 | [0.292, 0.325] | Naïve Bayes > SVM | H1 | 0.014 | 0.033 | A > B |
| | | | SVM | Residual | 0 | 0 | < 0.001 | [0.028, 0.060] | Naïve Bayes > Logistic | H2 | < 0.001 | 0.005 | Naibe Bayes > SVM |
| | | | | z-score | 50.6073 | 2.1925 | < 0.001 | [0.276, 0.309] | KNN > SVM | H3 | < 0.001 | 0.01 | Naïve Bayes > Logistic |
| | | | Logistic | Residual | 59.9659 | 0.1309 | < 0.001 | [0.012, 0.044] | KNN > Logistic | H5 | < 0.001 | 0.015 | KNN > SVM |
| | | | | z-score | 52.3186 | 1.9531 | < 0.001 | [-0.047, -0.015] | KNN < RUSBoost | H6 | < 0.001 | 0.02 | KNN > Logistic |
| | | | RUSBoost | Residual | 62.2228 | 1.1789 | < 0.001 | [-0.281, -0.248] | SVM < Logitic | H7 | < 0.001 | 0.025 | KNN < RUSBoost |
| | | | | z-score | 61.9795 | 1.2299 | < 0.001 | [-0.340, -307] | SVM < RUSBoost | H8 | < 0.001 | 0.03 | SVM < Logistic |
| | | B | Naïve Bayes | Residual | 57.4068 | 1.0191 | < 0.001 | [-0.075, -0.043] | Logistic < RUSBoost | H9 | < 0.001 | 0.035 | SVM < RUSBoost |
| | | | | z-score | 60.8060 | 1.0709 | < 0.001 | [-0.064, -0.050] | Residual < Z-score | H10 | < 0.001 | 0.04 | Logistic < RUSBoost |
| | | | KNN | Residual | 58.8325 | 1.9804 | | | | H1 | 0.006 | 0.045 | Nave Bayes > KNN |
| | | | | z-score | 58.2795 | 1.3272 | | | | H4 | 0.010 | 0.05 | Naïve Bayes < RUSBoost |
| | | | SVM | Residual | 0 | 0 | | | | H1 | < 0.01 | 0.05 | Residual < Z-score |
| | | | | z-score | 43.2712 | 18.8271 | | | | | | | |
| | | | Logistic | Residual | 59.9076 | 0.2012 | | | | | | | |
| | | | | z-score | 52.8537 | 2.1486 | | | | | | | |
| | | | RUSBoost | Residual | 61.5705 | 1.5664 | | | | | | | |
| | | | | z-score | 61.2843 | 1.2171 | | | | | | | |
| | | C | Naïve Bayes | Residual | 59.1189 | 1.2407 | | | | | | | |
| | | | | z-score | 61.6382 | 1.1850 | | | | | | | |
| | | | KNN | Residual | 58.1054 | 2.1207 | | | | | | | |
| | | | | z-score | 58.2028 | 1.4118 | | | | | | | |
| | | | SVM | Residual | 49.4829 | 11.9440 | | | | | | | |
| | | | | z-score | 50.8442 | 12.2602 | | | | | | | |
| | | | Logistic | Residual | 60.0040 | 0.1223 | | | | | | | |
| | | | | z-score | 52.1995 | 2.0986 | | | | | | | |
| | | | RUSBoost | Residual | 63.3101 | 1.4303 | | | | | | | |
| | | | | z-score | 62.7716 | 1.3253 | | | | | | | |
| | | D | Naïve Bayes | Residual | 62.6722 | 1.2459 | | | | | | | |
| | | | | z-score | 61.2064 | 0.9460 | | | | | | | |
| | | | KNN | Residual | 59.4432 | 1.4282 | | | | | | | |
| | | | | z-score | 60.9945 | 1.3456 | | | | | | | |
| | | | SVM | Residual | 0 | 0 | | | | | | | |
| | | | | z-score | 43.6616 | 19.0091 | | | | | | | |
| | | | Logistic | Residual | 59.9890 | 0.1662 | | | | | | | |
| | | | | z-score | 52.2091 | 2.4415 | | | | | | | |
| | | | RUSBoost | Residual | 61.6997 | 1.5805 | | | | | | | |
| | | | | z-score | 61.8194 | 1.1930 | | | | | | | |
| | AROC | A | Naïve Bayes | Residual | 67.2232 | 1.4855 | 0.009 | [-0.0255, -0.0026] | A < C | H4 | < 0.001 | 0.0083 | B < C |
| | | | | z-score | 67.9178 | 1.0937 | < 0.001 | [-0.0320, -0.0092] | B < C | H2 | 0.002 | 0.0167 | A < C |
| | | | KNN | Residual | 64.1155 | 1.9691 | 0.008 | [0.0027, 0.0256] | C > D | H6 | 0.002 | 0.025 | C > D |
| | | | | z-score | 63.7735 | 2.2731 | < 0.001 | [0.0288, 0.0560] | Naïve Bayes > KNN | H1 | < 0.001 | 0,005 | Naïve Bayes > KNN |
| | | | SVM | Residual | 0 | 0 | < 0.001 | [0.3656, 0.3927] | Naïve Bayes > SVM | H2 | < 0.001 | 0,01 | Naïve Bayes > SVM |
| | | | | z-score | 59.2248 | 2.8784 | < 0.001 | [0.0958, 0.1230] | Naïve Bayes > Logistic | H3 | < 0.001 | 0,015 | Naïve Bayes > Logistic |
| | | | Logistic | Residual | 59.6277 | 3.4603 | < 0.001 | [0.3232, 0.3503] | KNN > SVM | H5 | < 0.001 | 0,02 | KNN > SVM |
| | | | | z-score | 54.5287 | 2.3967 | < 0.001 | [0.0534, 0.0805] | KNN > Logistic | H6 | < 0.001 | 0,025 | KNN > Logistic |
| | | | RUSBoost | Residual | 68.6673 | 1.0967 | < 0.001 | [-0.0466, -0.0194] | KNN < RUSBoost | H7 | < 0.001 | 0,03 | KNN < RUSBoost |
| | | | | z-score | 68.3061 | 1.1908 | < 0.001 | [-0.2833, -0.2562] | SVM < Logistic | H8 | < 0.001 | 0,035 | SVM < Logistic |
| | | B | Naïve Bayes | Residual | 66.5713 | 0.8977 | < 0.001 | [-0.3833, -0.3562] | SVM < RUSBoost | H9 | < 0.001 | 0,04 | SVM < RUSBoost |
| | | | | z-score | 67.5055 | 1.0227 | < 0.001 | [-0.1135, -0.0864] | Logistic < RUSBoost | H10 | < 0.001 | 0,045 | Logistic < RUSBoost |
| | | | KNN | Residual | 63.3681 | 1.9988 | < 0.001 | [-0.105, -0.092] | Residual < Z-score | H1 | < 0.001 | 0.05 | Residual < Z-score |
| | | | | z-score | 62.8953 | 1.8717 | | | | | | | |
| | | | SVM | Residual | 0 | 0 | | | | | | | |
| | | | | z-score | 55.3609 | 2.4932 | | | | | | | |
| | | | Logistic | Residual | 60.6272 | 2.5508 | | | | | | | |
| | | | | z-score | 55.0623 | 2.7527 | | | | | | | |
| | | | RUSBoost | Residual | 67.7937 | 1.1148 | | | | | | | |
| | | | | z-score | 67.6416 | 1.3682 | | | | | | | |
| | | C | Naïve Bayes | Residual | 68.1437 | 1.0794 | | | | | | | |
| | | | | z-score | 69.1354 | 0.9408 | | | | | | | |
| | | | KNN | Residual | 63.3636 | 2.2262 | | | | | | | |
| | | | | z-score | 64.0967 | 1.9409 | | | | | | | |
| | | | SVM | Residual | 14.4219 | 25.6353 | | | | | | | |
| | | | | z-score | 58.7988 | 2.0479 | | | | | | | |
| | | | Logistic | Residual | 60.0521 | 2.3341 | | | | | | | |
| | | | | z-score | 54.6324 | 2.7014 | | | | | | | |
| | | | RUSBoost | Residual | 67.4480 | 1.0431 | | | | | | | |
| | | | | z-score | 67.3300 | 1.0749 | | | | | | | |
| | | D | Naïve Bayes | Residual | 70.3262 | 0.9983 | | | | | | | |
| | | | | z-score | 69.8087 | 0.9064 | | | | | | | |
| | | | KNN | Residual | 63.9925 | 1.6694 | | | | | | | |
| | | | | z-score | 67.1069 | 1.3086 | | | | | | | |
| | | | SVM | Residual | 0 | 0 | | | | | | | |
| | | | | z-score | 55.5158 | 1.5998 | | | | | | | |
| | | | Logistic | Residual | 60.0638 | 2.5941 | | | | | | | |
| | | | | z-score | 54.5454 | 2.8761 | | | | | | | |
| | | | RUSBoost | Residual | 66.0498 | 1.4774 | | | | | | | |
| | | | | z-score | 65.8849 | 1.1820 | | | | | | | |

**Supplementary Table 5:** Analysis of Variance for F1 score of balanced data analysis: measured for cross-validation measurements.

## Analysis of Variance

| Source | Sum Sq. | d.f. | Mean Sq. | F | Prob>F |
|---|---|---|---|---|---|
| Features | 0.0225 | 3 | 0.00751 | 0.4 | 0.7544 |
| Classifiers | 2.8335 | 4 | 0.70837 | 37.54 | 0 |
| DataFile | 0.3808 | 2 | 0.1904 | 10.09 | 0 |
| Harmonization | 0.6898 | 1 | 0.68985 | 36.56 | 0 |
| Features:Classifiers | 1.3179 | 12 | 0.10983 | 5.82 | 0 |
| Features:DataFile | 0.0702 | 6 | 0.01169 | 0.62 | 0.7148 |
| Features:Harmonization | 0.1454 | 3 | 0.04847 | 2.57 | 0.0527 |
| Classifiers:DataFile | 19.3356 | 8 | 2.41695 | 128.09 | 0 |
| Classifiers:Harmonization | 5.084 | 4 | 1.271 | 67.36 | 0 |
| DataFile:Harmonization | 0.0579 | 2 | 0.02893 | 1.53 | 0.2161 |
| Features:Classifiers:DataFile | 1.733 | 24 | 0.07221 | 3.83 | 0 |
| Features:Classifiers:Harmonization | 0.7343 | 12 | 0.06119 | 3.24 | 0.0001 |
| Features:DataFile:Harmonization | 0.1092 | 6 | 0.01821 | 0.96 | 0.4474 |
| Classifiers:DataFile:Harmonization | 0.2013 | 8 | 0.02516 | 1.33 | 0.2218 |
| Features:Classifiers:DataFile:Harmonization | 1.0659 | 24 | 0.04441 | 2.35 | 0.0002 |
| Error | 43.0205 | 2280 | 0.01887 | | |
| Total | 76.8019 | 2399 | | | |

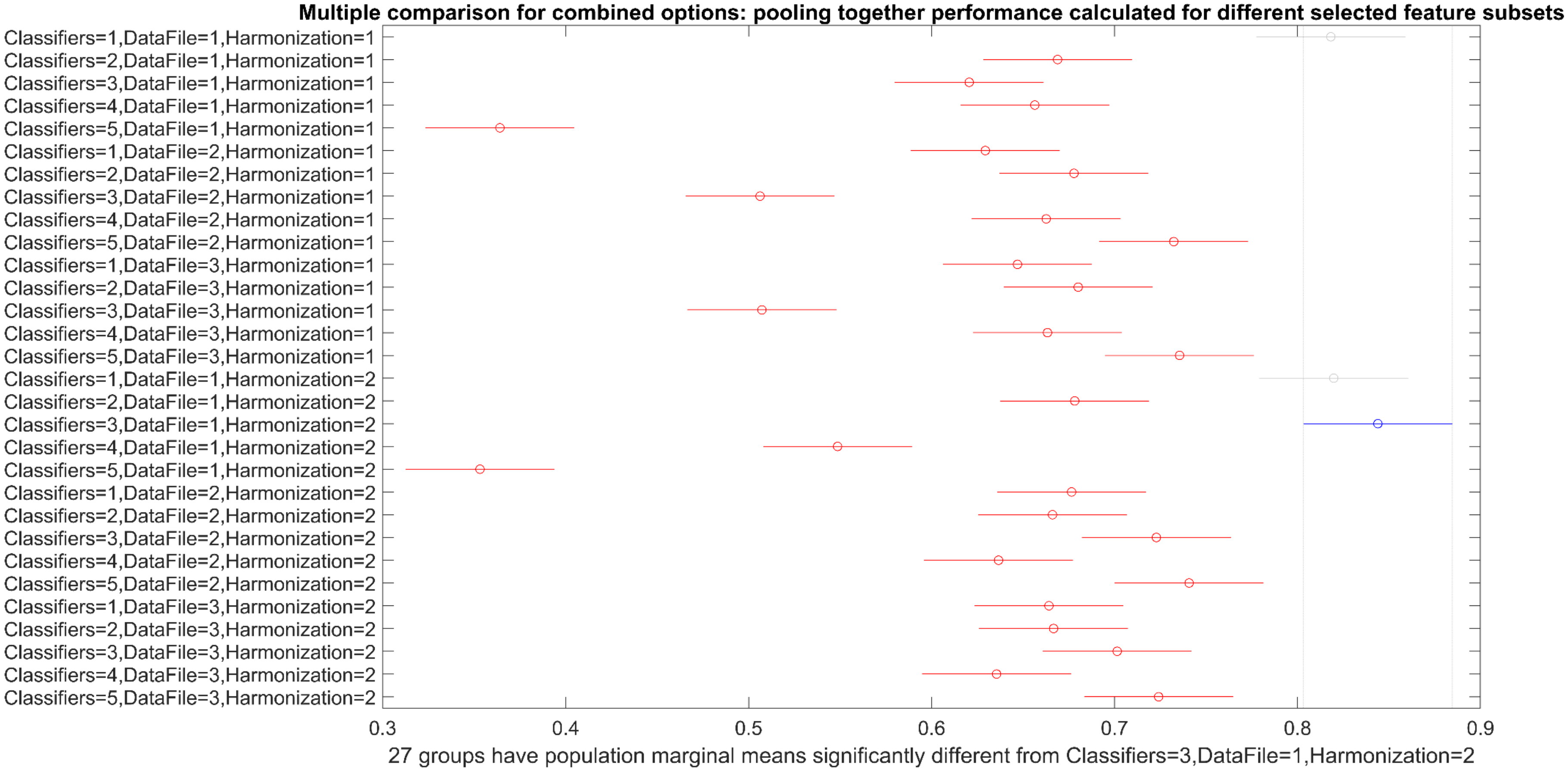

**Supplementary Figure 3:** Multiple comparisons for F1 score of balanced data analysis: measured for cross-validation (CV) measurements. Five classification models in this order: 1) Naïve Bayes, 2) KNN, 3) SVM, 4) LR, and 5) RUSBoost. Three datasets in this order: 1) ADNI, 2) OASIS-3, 3) OASIS-3 restricted to same age range as in ADNI dataset. Two harmonization procedures in this order: 1) residual and 2) z-score. The y-axis shows the label for each results entry, plotted per row, corresponding to the combination among classifier, data type, and harmonization options. The performance measure for the different feature selection subsets were pooled together in this and accompanying analyses. Best results for F1 score achieved by SVM applied for z-score harmonization of ADNI data (blue bar), which was significantly different from all the other combinations except for Naïve Bayes, which also showed "best" results for both residuals and z-score harmonization of ADNI data (grey bars).

**Supplementary Table 6:** Analysis of Variance for MCC score for balanced data analysis: measured for cross-validation measurements.

## Analysis of Variance

| Source | Sum Sq. | d.f. | Mean Sq. | F | Prob>F |
|---|---|---|---|---|---|
| Features | 0.0195 | 3 | 0.00651 | 0.32 | 0.8135 |
| Classifiers | 7.4044 | 4 | 1.85109 | 90 | 0 |
| DataFile | 1.1936 | 2 | 0.59681 | 29.02 | 0 |
| Harmonization | 1.0844 | 1 | 1.08444 | 52.73 | 0 |
| Features:Classifiers | 0.948 | 12 | 0.079 | 3.84 | 0 |
| Features:DataFile | 0.0936 | 6 | 0.01561 | 0.76 | 0.6024 |
| Features:Harmonization | 0.1747 | 3 | 0.05824 | 2.83 | 0.0371 |
| Classifiers:DataFile | 19.7818 | 8 | 2.47273 | 120.22 | 0 |
| Classifiers:Harmonization | 6.0071 | 4 | 1.50176 | 73.02 | 0 |
| DataFile:Harmonization | 0.2974 | 2 | 0.14871 | 7.23 | 0.0007 |
| Features:Classifiers:DataFile | 1.3252 | 24 | 0.05522 | 2.68 | 0 |
| Features:Classifiers:Harmonization | 0.6832 | 12 | 0.05693 | 2.77 | 0.001 |
| Features:DataFile:Harmonization | 0.0987 | 6 | 0.01645 | 0.8 | 0.5698 |
| Classifiers:DataFile:Harmonization | 0.9919 | 8 | 0.12398 | 6.03 | 0 |
| Features:Classifiers:DataFile:Harmonization | 0.9486 | 24 | 0.03952 | 1.92 | 0.0046 |
| Error | 46.8942 | 2280 | 0.02057 | | |
| Total | 87.9464 | 2399 | | | |

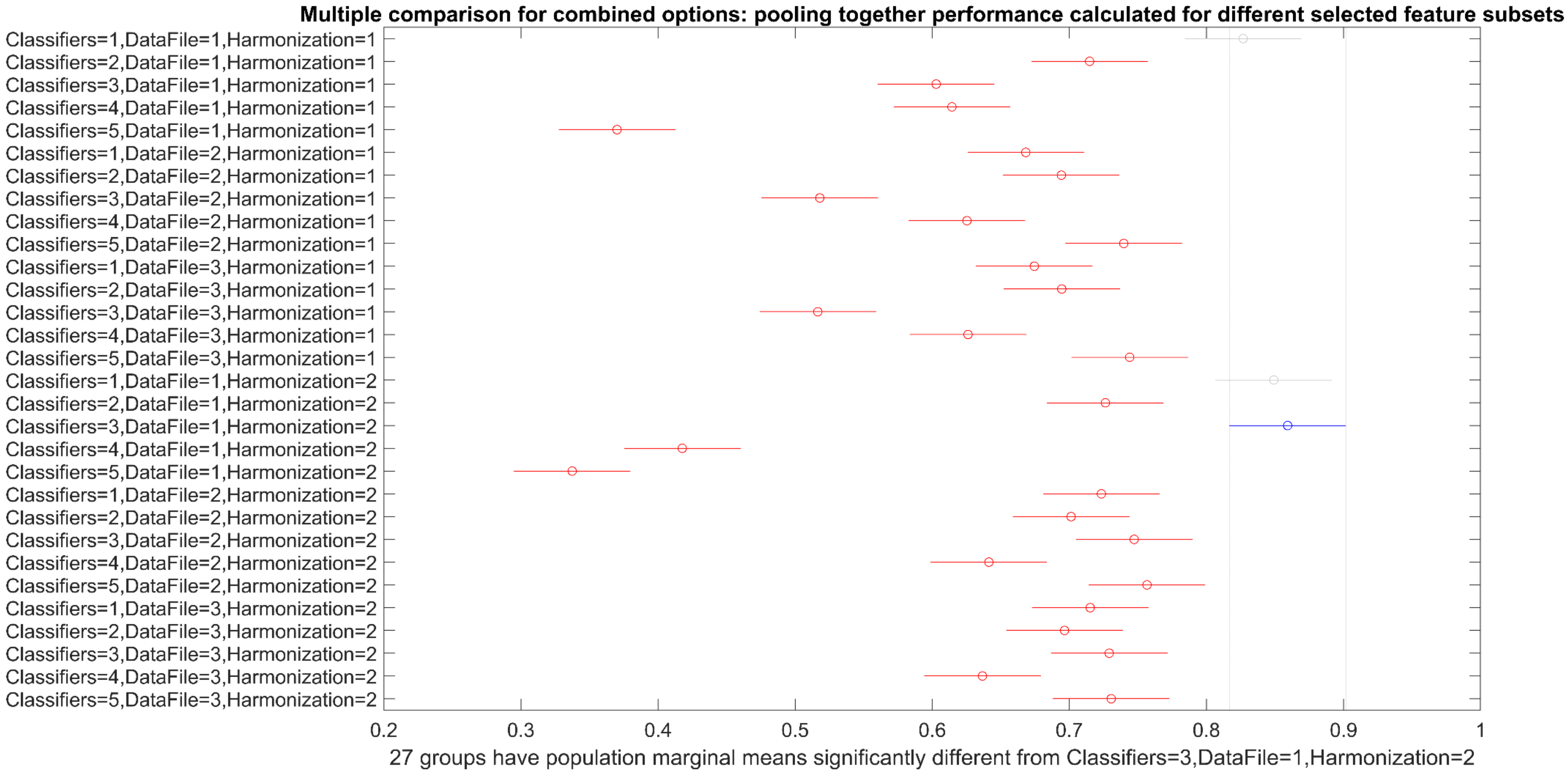

**Supplementary Figure 4:** Multiple comparisons for MCC score of balanced data analysis: measured for cross-validation (CV) measurements. Best results for MCC score achieved by SVM applied for z-score harmonization of ADNI data (blue bar), which was significantly different to the other combinations except for Naïve Bayes, which also showed "best" results for both residuals and z-score harmonization of ADNI data (grey bars). See **Supplementary Figure 3** caption for better understanding of representation.

**Supplementary Table 7:** Analysis of Variance for F1 score of imbalanced data analysis: measured for cross-validation measurements.

## Analysis of Variance

| Source | Sum Sq. | d.f. | Mean Sq. | F | Prob>F |
|---|---|---|---|---|---|
| Features | 1.72 | 3 | 0.5733 | 39.46 | 0 |
| Classifiers | 120.341 | 4 | 30.0854 | 2070.85 | 0 |
| DataFile | 2.331 | 2 | 1.1653 | 80.21 | 0 |
| Harmonization | 2.132 | 1 | 2.1322 | 146.76 | 0 |
| Features:Classifiers | 3.497 | 12 | 0.2915 | 20.06 | 0 |
| Features:DataFile | 0.327 | 6 | 0.0545 | 3.75 | 0.001 |
| Features:Harmonization | 0.62 | 3 | 0.2068 | 14.23 | 0 |
| Classifiers:DataFile | 2.411 | 8 | 0.3014 | 20.74 | 0 |
| Classifiers:Harmonization | 54.967 | 4 | 13.7419 | 945.89 | 0 |
| DataFile:Harmonization | 0.525 | 2 | 0.2626 | 18.07 | 0 |
| Features:Classifiers:DataFile | 0.71 | 24 | 0.0296 | 2.04 | 0.0021 |
| Features:Classifiers:Harmonization | 2.295 | 12 | 0.1912 | 13.16 | 0 |
| Features:DataFile:Harmonization | 0.25 | 6 | 0.0416 | 2.86 | 0.0088 |
| Classifiers:DataFile:Harmonization | 1.027 | 8 | 0.1284 | 8.84 | 0 |
| Features:Classifiers:DataFile:Harmonization | 0.522 | 24 | 0.0217 | 1.5 | 0.0574 |
| Error | 33.124 | 2280 | 0.0145 | | |
| Total | 226.8 | 2399 | | | |

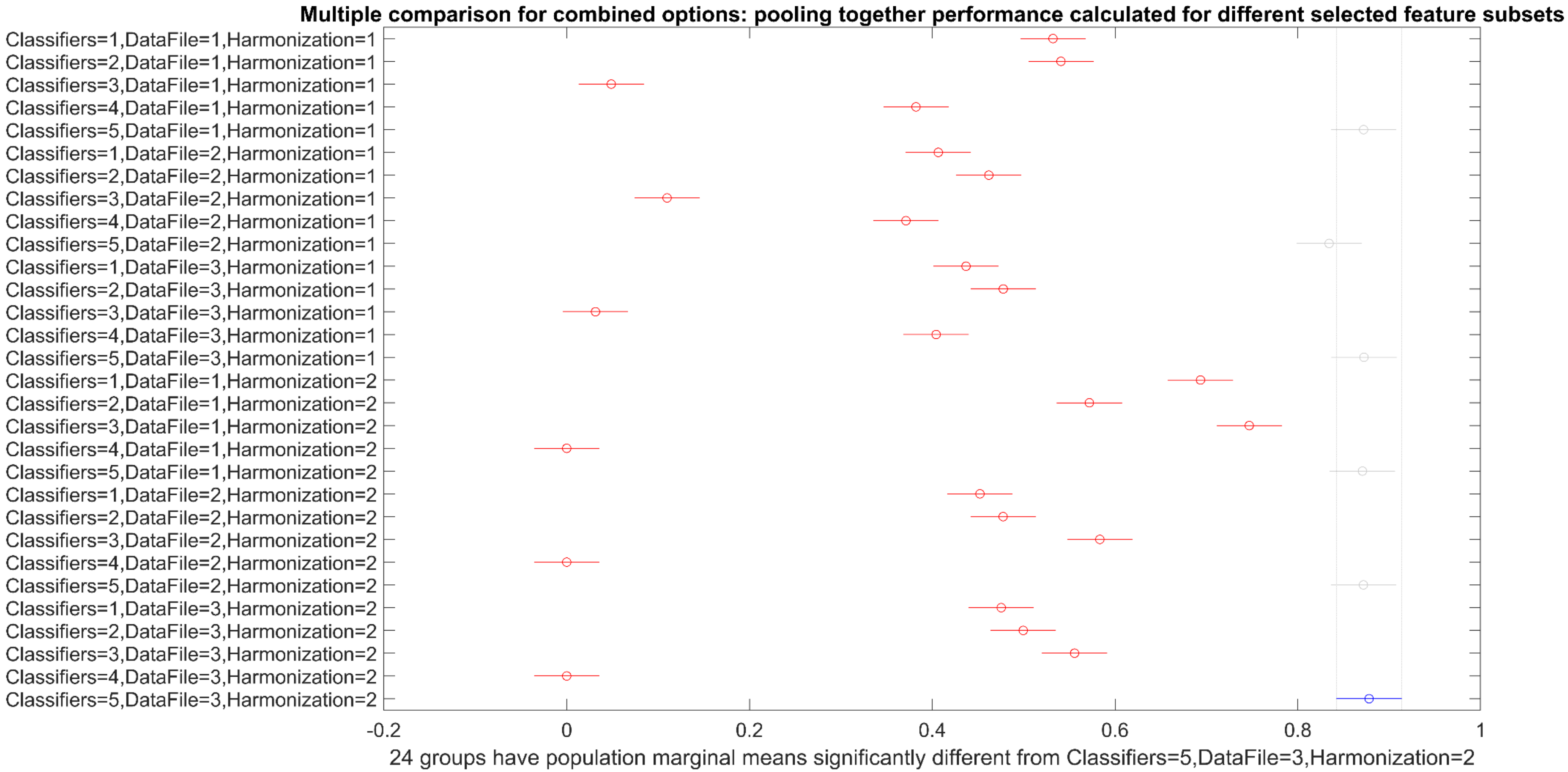

**Supplementary Figure 5:** Multiple comparisons for F1 score of imbalanced data analysis: measured for cross-validation (CV) measurements. Best results for F1 score achieved by RUSBoost for all the six combinations between data type and harmonization approach (blue and grey bars almost vertically arranged). See **Supplementary Figure 3** caption for better understanding of representation.

**Supplementary Table 8:** Analysis of Variance for MCC score of imbalanced data analysis: measured for cross-validation measurements.

## Analysis of Variance

| Source | Sum Sq. | d.f. | Mean Sq. | F | Prob>F |
|---|---|---|---|---|---|
| Features | 0.779 | 3 | 0.2597 | 15.5 | 0 |
| Classifiers | 117.797 | 4 | 29.4493 | 1757.74 | 0 |
| DataFile | 1.304 | 2 | 0.6521 | 38.92 | 0 |
| Harmonization | 0.089 | 1 | 0.0886 | 5.29 | 0.0216 |
| Features:Classifiers | 2.112 | 12 | 0.176 | 10.51 | 0 |
| Features:DataFile | 0.197 | 6 | 0.0328 | 1.96 | 0.068 |
| Features:Harmonization | 0.994 | 3 | 0.3313 | 19.78 | 0 |
| Classifiers:DataFile | 1.896 | 8 | 0.2371 | 14.15 | 0 |
| Classifiers:Harmonization | 82.452 | 4 | 20.6129 | 1230.32 | 0 |
| DataFile:Harmonization | 0.985 | 2 | 0.4927 | 29.41 | 0 |
| Features:Classifiers:DataFile | 0.754 | 24 | 0.0314 | 1.87 | 0.0062 |
| Features:Classifiers:Harmonization | 3.78 | 12 | 0.315 | 18.8 | 0 |
| Features:DataFile:Harmonization | 0.431 | 6 | 0.0719 | 4.29 | 0.0003 |
| Classifiers:DataFile:Harmonization | 2.727 | 8 | 0.3408 | 20.34 | 0 |
| Features:Classifiers:DataFile:Harmonization | 1.533 | 24 | 0.0639 | 3.81 | 0 |
| Error | 38.199 | 2280 | 0.0168 | | |
| Total | 256.03 | 2399 | | | |

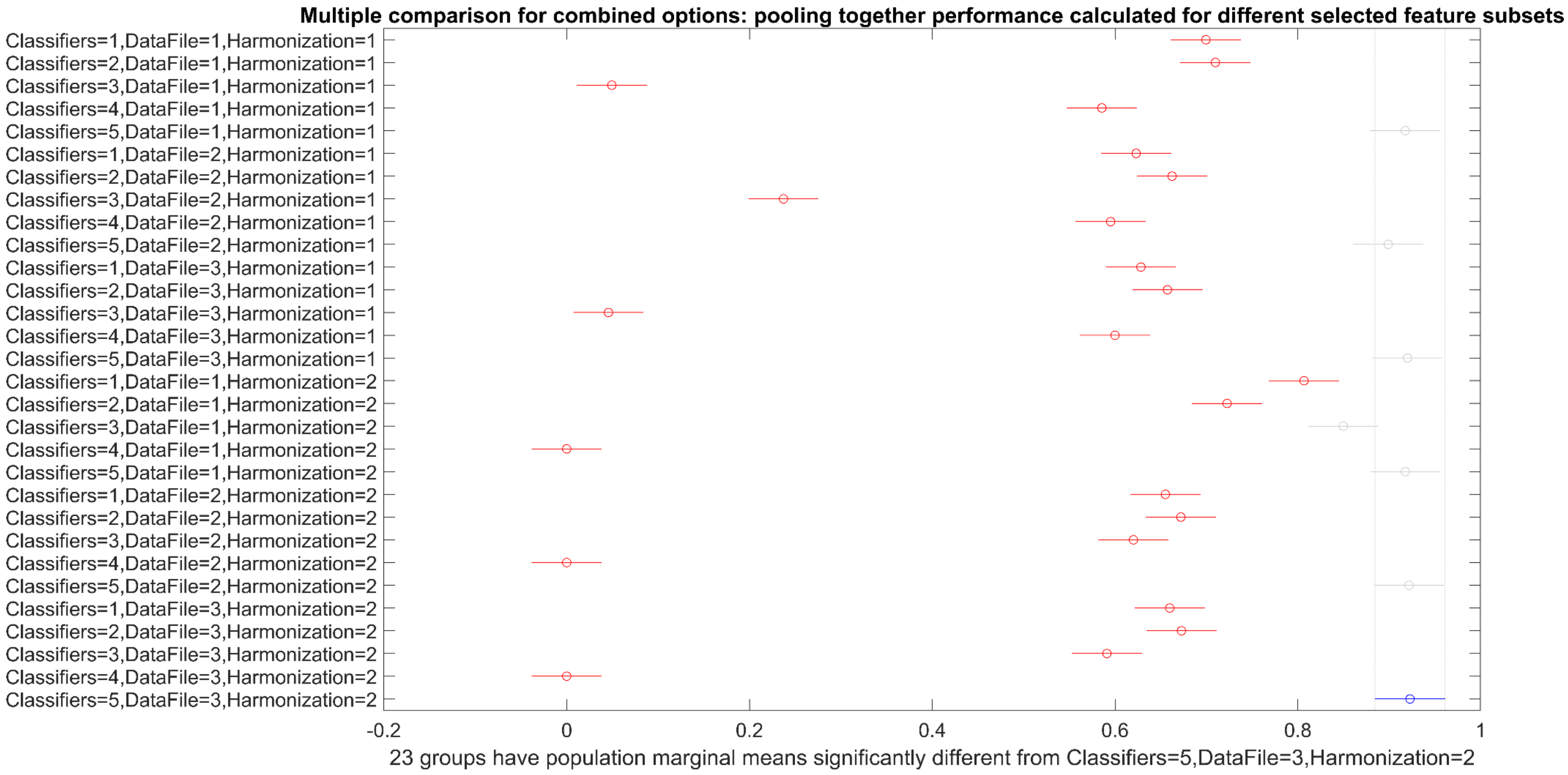

**Supplementary Figure 6:** Multiple comparisons for MCC score of imbalanced data analysis: measured for cross-validation (CV) measurements. Best results for MCC score achieved by RUSBoost for all the six combinations between data type and harmonization approach (blue and grey bars almost vertically arranged). See **Supplementary Figure 3** caption for better understanding of representation.

**Supplementary Table 9:** Analysis of Variance for F1 score of balanced data analysis: measured for nested-CV (holdout) measurements.

## Analysis of Variance

| Source | Sum Sq. | d.f. | Mean Sq. | F | Prob>F |
|---|---|---|---|---|---|
| Features | 0.0322 | 3 | 0.01074 | 2.61 | 0.0496 |
| Classifiers | 1.2708 | 4 | 0.31771 | 77.34 | 0 |
| DataFile | 0.142 | 2 | 0.07101 | 17.28 | 0 |
| Harmonization | 0.0408 | 1 | 0.04075 | 9.92 | 0.0017 |
| Features:Classifiers | 0.0508 | 12 | 0.00424 | 1.03 | 0.4165 |
| Features:DataFile | 0.0721 | 6 | 0.01202 | 2.93 | 0.0076 |
| Features:Harmonization | 0.0288 | 3 | 0.00959 | 2.33 | 0.0721 |
| Classifiers:DataFile | 1.0093 | 8 | 0.12616 | 30.71 | 0 |
| Classifiers:Harmonization | 1.7741 | 4 | 0.44353 | 107.96 | 0 |
| DataFile:Harmonization | 0.0018 | 2 | 0.00089 | 0.22 | 0.8049 |
| Features:Classifiers:DataFile | 0.1483 | 24 | 0.00618 | 1.5 | 0.0551 |
| Features:Classifiers:Harmonization | 0.0825 | 12 | 0.00687 | 1.67 | 0.0665 |
| Features:DataFile:Harmonization | 0.0589 | 6 | 0.00982 | 2.39 | 0.0264 |
| Classifiers:DataFile:Harmonization | 0.1135 | 8 | 0.01419 | 3.45 | 0.0006 |
| Features:Classifiers:DataFile:Harmonization | 0.1077 | 24 | 0.00449 | 1.09 | 0.3434 |
| Error | 9.3665 | 2280 | 0.00411 | | |
| Total | 14.3001 | 2399 | | | |

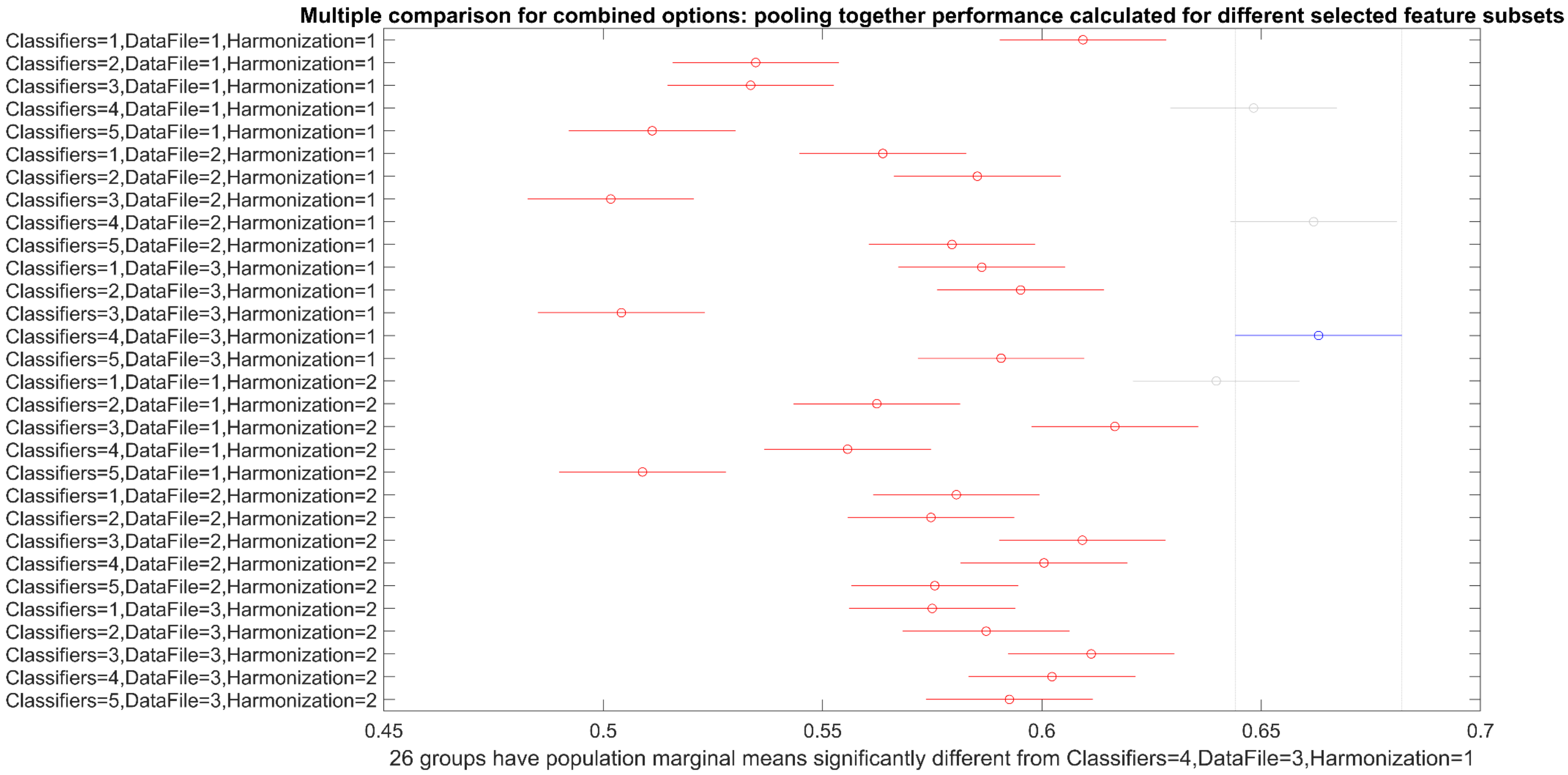

**Supplementary Figure 7:** Multiple comparisons for F1 score of balanced data analysis: measured for nested-CV (holdout) measurements. Best results for F1 score achieved by LR for all the three datasets using the residual harmonization approach (blue bar and upper grey bars), which is significantly different from all the other outcomes except from results achieved by Naïve Bayes for ADNI dataset and z-score harmonization (lower grey bar). Notice the dramatic difference between this score results (check the range of x-axis ticks) and the analogous for the CV analysis shown in **Supplementary Figure 3**, revealing the overfitting in traditional CV analysis, which is controlled by the nested-CV approach as shown with the present outcome. See **Supplementary Figure 3** caption for better understanding of representation.

**Supplementary Table 10:** Analysis of Variance for MCC score of balanced data analysis: measured for nested CV (holdout) measurements.

## Analysis of Variance

| Source | Sum Sq. | d.f. | Mean Sq. | F | Prob>F |
|---|---|---|---|---|---|
| Features | 0.03659 | 3 | 0.0122 | 4.36 | 0.0045 |
| Classifiers | 0.71671 | 4 | 0.17918 | 64.03 | 0 |
| DataFile | 0.02605 | 2 | 0.01302 | 4.65 | 0.0096 |
| Harmonization | 0.49182 | 1 | 0.49182 | 175.75 | 0 |
| Features:Classifiers | 0.03177 | 12 | 0.00265 | 0.95 | 0.4991 |
| Features:DataFile | 0.02748 | 6 | 0.00458 | 1.64 | 0.133 |
| Features:Harmonization | 0.00878 | 3 | 0.00293 | 1.05 | 0.3714 |
| Classifiers:DataFile | 0.36239 | 8 | 0.0453 | 16.19 | 0 |
| Classifiers:Harmonization | 1.45617 | 4 | 0.36404 | 130.09 | 0 |
| DataFile:Harmonization | 0.00184 | 2 | 0.00092 | 0.33 | 0.7195 |
| Features:Classifiers:DataFile | 0.09569 | 24 | 0.00399 | 1.42 | 0.0828 |
| Features:Classifiers:Harmonization | 0.05668 | 12 | 0.00472 | 1.69 | 0.0632 |
| Features:DataFile:Harmonization | 0.00394 | 6 | 0.00066 | 0.23 | 0.9654 |
| Classifiers:DataFile:Harmonization | 0.06136 | 8 | 0.00767 | 2.74 | 0.0052 |
| Features:Classifiers:DataFile:Harmonization | 0.0333 | 24 | 0.00139 | 0.5 | 0.9807 |
| Error | 6.38023 | 2280 | 0.0028 | | |
| Total | 9.79082 | 2399 | | | |

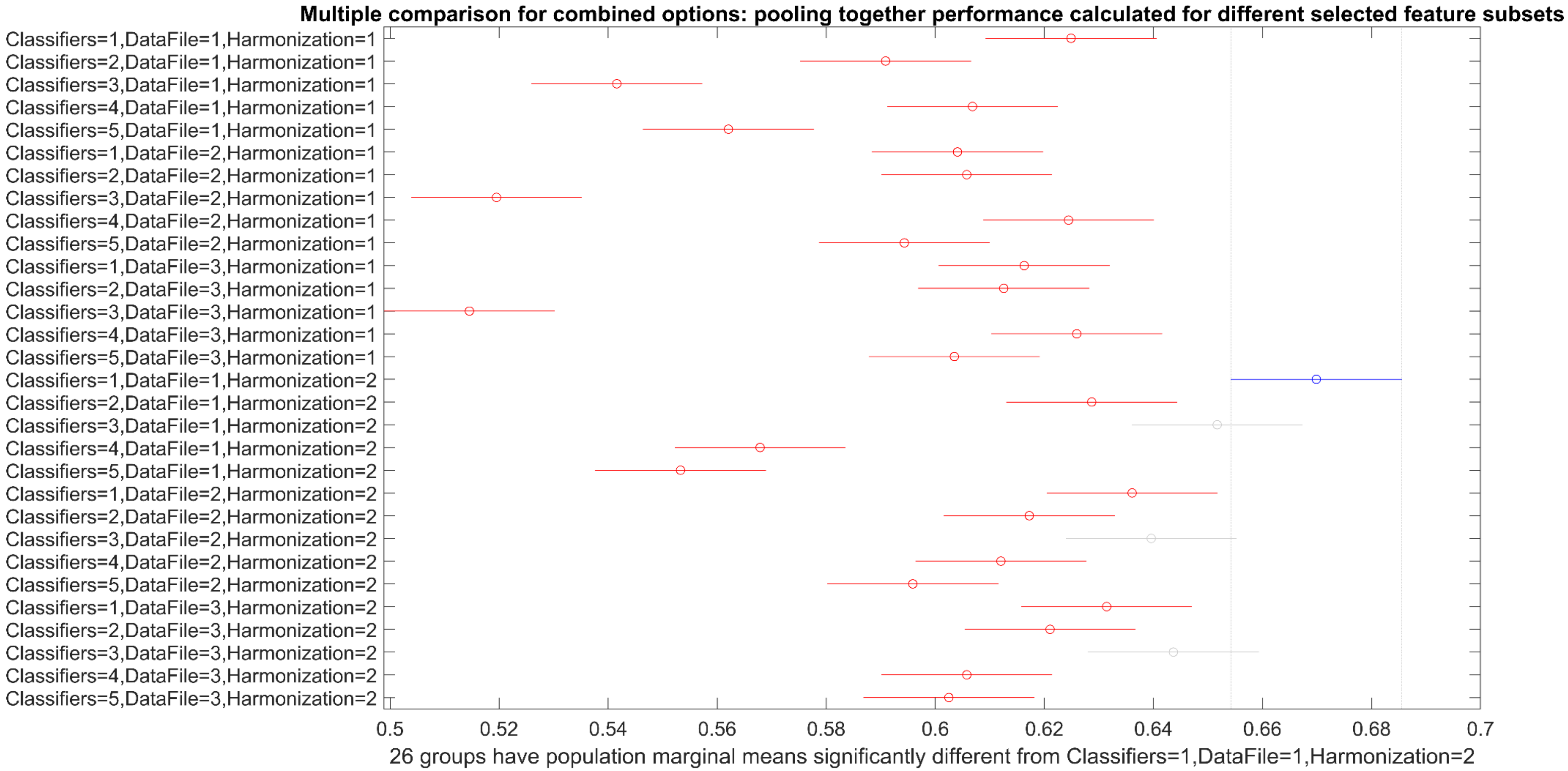

**Supplementary Figure 8:** Multiple comparisons for MCC score of balanced data analysis: measured for nested-CV (holdout) measurements. Best results for MCC score achieved by Naïve Bayes for ADNI dataset and z-score harmonization (blue bar), which is significantly different from all the other outcomes except from results achieved by SVM for all the three datasets and the z-score harmonization approach (grey bars). Notice the dramatic difference between this score results (check the range of x-axis ticks) and the analogous for the CV analysis shown in **Supplementary Figure 4**, revealing the overfitting in traditional CV analysis, which is controlled by the nested-CV approach as shown with the present outcome. See **Supplementary Figure 3** caption for better understanding of representation.

**Supplementary Table 11:** Analysis of Variance for F1 score of imbalanced data analysis: measured for nested CV (holdout) measurements.

## Analysis of Variance

| Source | Sum Sq. | d.f. | Mean Sq. | F | Prob>F |
|---|---|---|---|---|---|
| Features | 0.092 | 3 | 0.03068 | 29.04 | 0 |
| Classifiers | 33.4944 | 4 | 8.3736 | 7925.5 | 0 |
| DataFile | 1.8628 | 2 | 0.93142 | 881.58 | 0 |
| Harmonization | 0.1365 | 1 | 0.13654 | 129.23 | 0 |
| Features:Classifiers | 0.3514 | 12 | 0.02928 | 27.71 | 0 |
| Features:DataFile | 0.2078 | 6 | 0.03464 | 32.78 | 0 |
| Features:Harmonization | 0.0294 | 3 | 0.0098 | 9.27 | 0 |
| Classifiers:DataFile | 3.0345 | 8 | 0.37932 | 359.02 | 0 |
| Classifiers:Harmonization | 4.0383 | 4 | 1.00957 | 955.54 | 0 |
| DataFile:Harmonization | 1.7682 | 2 | 0.88409 | 836.78 | 0 |
| Features:Classifiers:DataFile | 0.4592 | 24 | 0.01913 | 18.11 | 0 |
| Features:Classifiers:Harmonization | 0.0865 | 12 | 0.00721 | 6.82 | 0 |
| Features:DataFile:Harmonization | 0.0746 | 6 | 0.01244 | 11.77 | 0 |
| Classifiers:DataFile:Harmonization | 2.7372 | 8 | 0.34215 | 323.84 | 0 |
| Features:Classifiers:DataFile:Harmonization | 0.0993 | 24 | 0.00414 | 3.92 | 0 |
| Error | 2.4089 | 2280 | 0.00106 | | |
| Total | 50.881 | 2399 | | | |

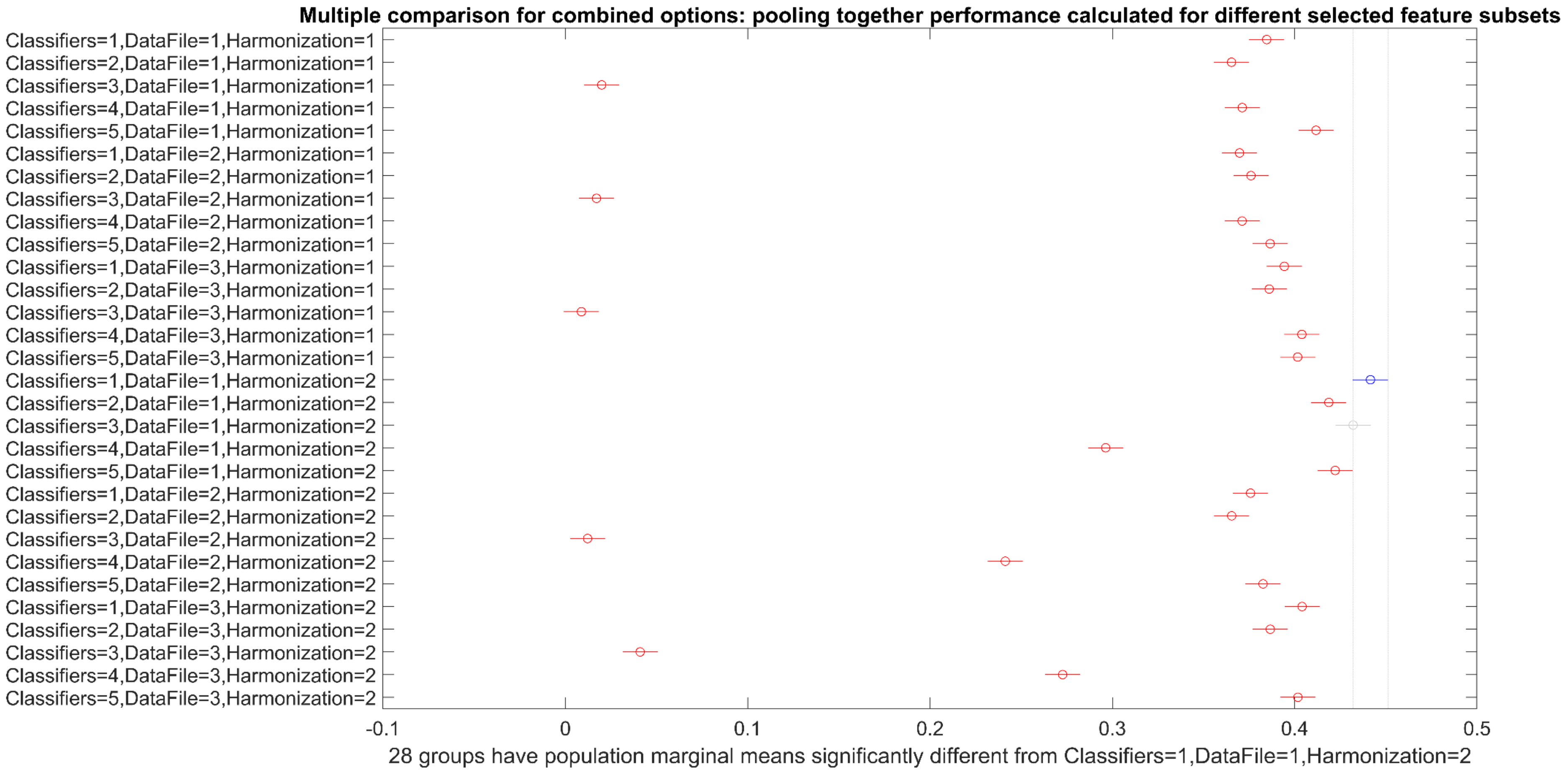

**Supplementary Figure 9:** Multiple comparisons for F1 score of imbalanced data analysis: measured for nested CV (holdout) measurements. Best results for F1 score achieved by Naïve Bayes for ADNI dataset and z-score harmonization (blue bar), which is significantly different from all the other outcomes except from results achieved by SVM for ADNI dataset and the z-score harmonization (grey bar). Notice the dramatic difference between this score results (check the range of x-axis ticks) and the analogous for the CV analysis shown in **Supplementary Figure 5**, revealing the overfitting in traditional CV analysis, which is controlled by the nested CV approach as shown with the present outcome. See **Supplementary Figure 3** caption for better understanding of representation.

**Supplementary Table 12:** Analysis of Variance for MCC score of imbalanced data analysis: measured for nested CV (holdout) measurements.

## Analysis of Variance

| Source | Sum Sq. | d.f. | Mean Sq. | F | Prob>F |
|---|---|---|---|---|---|
| Features | 0.5949 | 3 | 0.19831 | 42.12 | 0 |
| Classifiers | 26.0127 | 4 | 6.50318 | 1381.21 | 0 |
| DataFile | 0.3457 | 2 | 0.17286 | 36.71 | 0 |
| Harmonization | 1.615 | 1 | 1.61501 | 343.01 | 0 |
| Features:Classifiers | 1.8135 | 12 | 0.15112 | 32.1 | 0 |
| Features:DataFile | 0.3614 | 6 | 0.06024 | 12.79 | 0 |
| Features:Harmonization | 0.9067 | 3 | 0.30222 | 64.19 | 0 |
| Classifiers:DataFile | 0.8642 | 8 | 0.10803 | 22.94 | 0 |
| Classifiers:Harmonization | 8.6766 | 4 | 2.16915 | 460.71 | 0 |
| DataFile:Harmonization | 1.974 | 2 | 0.98702 | 209.63 | 0 |
| Features:Classifiers:DataFile | 1.2599 | 24 | 0.0525 | 11.15 | 0 |
| Features:Classifiers:Harmonization | 4.0931 | 12 | 0.34109 | 72.44 | 0 |
| Features:DataFile:Harmonization | 0.4804 | 6 | 0.08007 | 17.01 | 0 |
| Classifiers:DataFile:Harmonization | 4.6077 | 8 | 0.57596 | 122.33 | 0 |
| Features:Classifiers:DataFile:Harmonization | 1.863 | 24 | 0.07762 | 16.49 | 0 |
| Error | 10.735 | 2280 | 0.00471 | | |
| Total | 66.2039 | 2399 | | | |

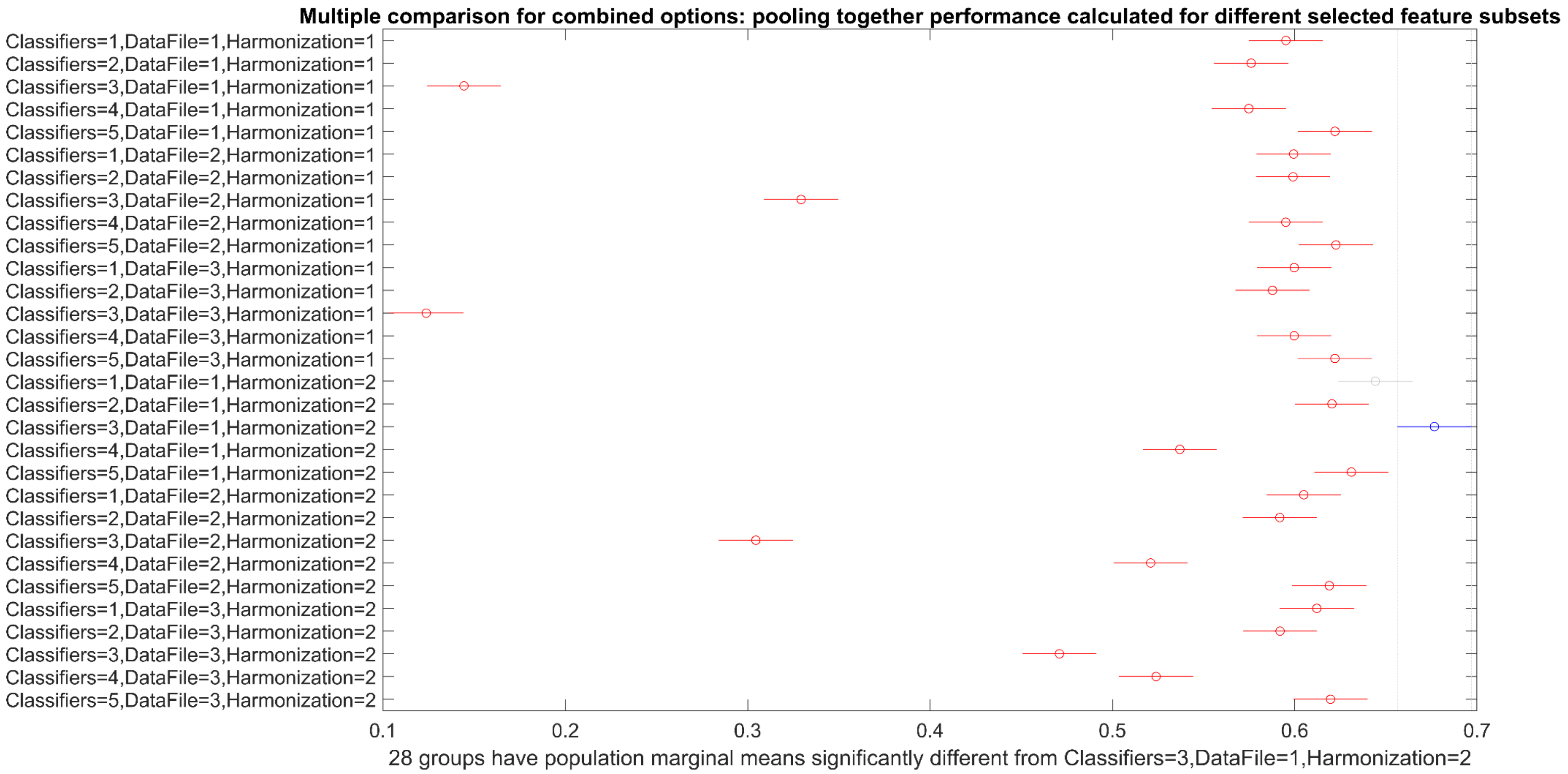

**Supplementary Figure 10:** Multiple comparisons for MCC score of imbalanced data analysis: measured for nested CV (holdout) measurements. Best results for MCC score achieved by SVM for ADNI dataset and z-score harmonization (blue bar), which is significantly different from all the other outcomes except from results achieved by Naïve Bayes for ADNI dataset and the z-score harmonization (grey bar). Notice the dramatic difference between this score results (check the range of x-axis ticks) and the analogous for the CV analysis shown in **Supplementary Figure 6**, revealing the overfitting in traditional CV analysis, which is controlled by the nested CV approach as shown with the present outcome. See **Supplementary Figure 3** caption for better understanding of representation.

**Supplementary Table 13:** Comparison of classification performance for balanced data analysis for different classification methods and feature selection criteria (see **Figure 5** caption for more information).

| Features | Models | Dataset | Acc (Mean (SD)) (%) | AROC (Mean (SD)) (%) | F1 (Mean (SD)) (%) | MCC (Mean (SD)) (%) | Acc (Mean (SD)) (%) | AROC (Mean (SD)) (%) | F1 (Mean (SD)) (%) | MCC′ (Mean (SD)) (%) |
|---|---|---|---|---|---|---|---|---|---|---|
| | | | residual harmonization | | | | z-score harmonization | | | |
| **A) Average score percentage (MCL app): chi-square, ReliefF, ANOVA & Kruskal Wallis (20 features)** | | | | | | | | | | |
| **brainSegVolNotVent<br>entorhinal<br>inf-Lat-Vent<br>lateral-Ventricle parahippocampal<br>hippocampus<br>temporalpole<br>fusiform<br>lateralorbitofrontal<br>insula<br>precentral<br>isthmuscingulate<br>parsopercularis<br>posteriorcingulate<br>accumbens-area<br>rostralmiddlefrontal<br>middletemporal<br>rostralanteriorcingulate<br>amygdala<br>parsorbitalis** | **Naïve Bayes** | **ADNI (age: 60-86)** | 59.69 (7.77) | 69.60 (10.78) | 57.43 (9.87) | 59.74 (7.85) | 63.75 (5.91) | 73.15 (6.31) | 60.25 (8.05) | 64.30 (6.21) |
| | | **OASIS-3 (age: 43-96)** | 59.83 (3.08) | 65.57 (4.01) | 57.53 (3.70) | 59.94 (3.10) | 62.93 (2.34) | 67.89 (2.46) | 58.36 (3.69) | 63.28 (2.36) |
| | | **OASIS-3 (age: 60-86)** | 60.73 (3.13) | 66.24 (2.14) | 58.51 (3.96) | 60.90 (3.22) | 63.00 (3.29) | 68.02 (3.18) | 57.91 (5.18) | 63.40 (3.30) |
| | **KNN** | **ADNI (age: 60-86)** | 57.29 (5.72) | 62.48 (6.34) | 51.50 (7.64) | 57.58 (5.79) | 59.79 (6.12) | 67.31 (7.40) | 52.61 (9.81) | 60.49 (6.56) |
| | | **OASIS-3 (age: 43-96)** | 59.81 (3.29) | 65.52 (3.09) | 59.33 (4.23) | 59.83 (3.29) | 61.11 (2.68) | 64.90 (3.06) | 56.93 (4.21) | 61.43 (2.82) |
| | | **OASIS-3 (age: 60-86)** | 61.06 (3.00) | 64.18 (3.66) | 59.46 (4.44) | 61.15 (2.99) | 60.57 (2.34) | 65.13 (2.99) | 57.84 (4.07) | 60.73 (2.36) |
| | **SVM** | **ADNI (age: 60-86)** | 56.04 (7.40) | 53.33 (20.67) | 54.86 (8.18) | 56.06 (7.41) | 65.83 (8.87) | 73.71 (8.56) | 62.22 (10.20) | 66.22 (9.04) |
| | | **OASIS-3 (age: 43-96)** | 52.48 (4.44) | 50.36 (12.97) | 50.94 (5.38) | 52.49 (4.46) | 63.12 (3.89) | 68.46 (3.48) | 59.75 (5.14) | 63.33 (3.91) |
| | | **OASIS-3 (age: 60-86)** | 50.09 (5.94) | 50.12 (7.64) | 49.51 (5.79) | 50.10 (5.95) | 63.84 (3.24) | 69.04 (3.37) | 61.51 (4.24) | 63.97 (3.23) |
| | **Logistic** | **ADNI (age: 60-86)** | 60.83 (3.42) | 57.92 (8.03) | 64.91 (2.07) | 61.12 (3.33) | 57.19 (9.47) | 61.69 (10.58) | 56.46 (9.36) | 57.22 (9.50) |
| | | **OASIS-3 (age: 43-96)** | 61.67 (1.67) | 58.87 (3.62) | 65.92 (1.02) | 62.04 (1.63) | 60.15 (4.52) | 65.30 (5.16) | 59.08 (4.29) | 60.18 (4.54) |
| | | **OASIS-3 (age: 60-86)** | 61.82 (2.08) | 60.67 (2.74) | 66.06 (1.29) | 62.19 (2.03) | 59.10 (3.33) | 63.24 (3.72) | 58.48 (3.13) | 59.13 (3.34) |
| | **RUSBoost** | **ADNI (age: 60-86)** | 54.69 (9.46) | 49.56 (15.28) | 47.14 (13.79) | 54.73 (9.80) | 56.46 (10.40) | 52.25 (10.00) | 52.88 (12.58) | 56.56 (10.59) |
| | | **OASIS-3 (age: 43-96)** | 58.95 (4.53) | 61.49 (4.75) | 58.13 (7.09) | 59.13 (4.62) | 59.91 (4.16) | 61.94 (4.38) | 56.79 (5.62) | 60.14 (4.23) |
| | | **OASIS-3 (age: 60-86)** | 59.22 (4.46) | 61.75 (4.87) | 58.29 (6.22) | 59.29 (4.47) | 60.57 (5.02) | 62.75 (4.42) | 60.44 (7.31) | 60.81 (5.22) |
| **B) ReliefF (19 features)** | | | | | | | | | | |
| **entorhinal fusiform Inf-Lat-Vent Temporalpole Posteriorcingulate Isthmuscingulate hippocampus parahippocampal parsopercularis Lateral-Ventricle transversetemporal precentral insula frontalpole superiortemporal amygdala lateralorbitofrontal lingual BrainSegVolNotVent** | **Naïve Bayes** | **ADNI (age: 60-86)** | 63.54 (9.28) | 71.10 (10.80) | 62.04 (10.54) | 63.60 (9.31) | 69.17 (6.54) | 77.73 (7.08) | 69.21 (7.90) | 69.28 (6.56) |
| | | **OASIS-3 (age: 43-96)** | 59.25 (3.31) | 65.48 (2.95) | 56.19 (3.70) | 59.47 (3.51) | 62.65 (2.44) | 68.41 (2.700) | 57.05 (5.50) | 63.13 (2.26) |
| | | **OASIS-3 (age: 60-86)** | 60.71 (3.45) | 66.43 (3.13) | 60.34 (4.29) | 60.86 (3.57) | 60.99 (2.57) | 66.62 (2.11) | 55.05 (3.73) | 61.42 (2.65) |
| | **KNN** | **ADNI (age: 60-86)** | 59.17 (9.99) | 62.67 (10.26) | 53.32 (13.56) | 59.45 (10.06) | 62.71 (5.61) | 70.06 (7.91) | 57.99 (6.96) | 63.11 (5.73) |
| | | **OASIS-3 (age: 43-96)** | 59.27 (3.15) | 62.86 (4.05) | 57.14 (5.41) | 59.37 (3.14) | 59.53 (1.56) | 63.66 (2.23) | 53.44 (4.54) | 60.01 (1.79) |
| | | **OASIS-3 (age: 60-86)** | 61.58 (4.38) | 64.40 (4.59) | 61.65 (4.99) | 61.62 (4.38) | 61.39 (2.71) | 65.45 (3.83) | 56.91 (5.46) | 61.75 (2.72) |
| | **SVM** | **ADNI (age: 60-86)** | 53.85 (6.49) | 45.04 (24.83) | 52.92 (7.11) | 53.88 (6.56) | 65.73 (7.66) | 70.67 (8.50) | 64.37 (7.90) | 65.89 (7.71) |
| | | **OASIS-3 (age: 43-96)** | 51.97 (4.58) | 50.03 (13.39) | 50.90 (4.60) | 51.97 (4.59) | 62.78 (2.12) | 67.72 (2.56) | 59.94 (3.15) | 62.93 (2.11) |
| | | **OASIS-3 (age: 60-86)** | 51.58 (5.41) | 50.53 (13.54) | 50.47 (6.14) | 51.58 (5.43) | 63.14 (3.21) | 68.84 (2.33) | 59.19 (4.40) | 63.43 (3.22) |
| | **Logistic** | **ADNI (age: 60-86)** | 60.41 (4.11) | 55.83 (7.98) | 64.81 (2.59) | 60.70 (4.11) | 55.21 (7.20) | 59.63 (9.66) | 53.66 (8.26) | 55.27 (7.27) |
| | | **OASIS-3 (age: 43-96)** | 61.73 (2.47) | 59.03 (3.17) | 65.99 (1.48) | 62.10 (2.39) | 60.94 (3.46) | 65.43 (4.01) | 59.35 (3.69) | 61.00 (3.49) |
| | | **OASIS-3 (age: 60-86)** | 62.83 (1.93) | 61.55 (3.98) | 66.68 (1.21) | 63.18 (1.87) | 60.26 (3.06) | 64.31 (3.23) | 59.71 (3.29) | 60.30 (3.07) |
| | **RUSBoost** | **ADNI (age: 60-86)** | 57.19 (7.81) | 55.52 (12.06) | 53.15 (10.19) | 57.39 (8.04) | 53.33 (9.83) | 48.88 (14.29) | 48.46 (11.21) | 53.48 (10.23) |
| | | **OASIS-3 (age: 43-96)** | 59.81 (3.17) | 61.77 (4.02) | 57.88 (7.02) | 60.07 (3.18) | 59.58 (3.90) | 61.19 (4.37) | 56.22 (6.16) | 59.58 (4.67) |
| | | **OASIS-3 (age: 60-86)** | 60.40 (4.49) | 61.99 (3.86) | 59.16 (7.03) | 60.54 (4.48) | 58.16 (3.63) | 59.69 (4.28) | 57.76 (6.80) | 58.32 (3.70) |
| **C) Frequent feature appearances from all feature ranking analysis (7 features)** | | | | | | | | | | |
| **BrainSegVolNotVent Inf-Lat-Vent entorhinal lateralorbitofrontal Lateral-Ventricle parahippocampal hippocampus** | **Naïve Bayes** | **ADNI (age: 60-86)** | 62.40 (8.51) | 69.10 (11.24) | 61.57 (9.44) | 62.44 (8.56) | 67.40 (5.16) | 74.08 (6.35) | 63.90 (6.26) | 67.83 (5.31) |
| | | **OASIS-3 (age: 43-96)** | 59.74 (3.92) | 67.30 (3.76) | 54.70 (3.81) | 60.10 (4.20) | 63.91 (3.13) | 70.14 (2.67) | 58.52 (5.51) | 64.40 (2.99) |
| | | **OASIS-3 (age: 60-86)** | 61.13 (2.90) | 67.55 (3.11) | 56.67 (3.92) | 61.58 (3.20) | 63.23 (2.59) | 68.70 (2.32) | 57.65 (5.00) | 63.73 (2.47) |
| | **KNN** | **ADNI (age: 60-86)** | 58.23 (7.14) | 63.77 (7.81) | 53.02 (9.90) | 58.47 (7.24) | 63.96 (6.05) | 70.85 (6.90) | 58.00 (9.50) | 64.68 (6.31) |
| | | **OASIS-3 (age: 43-96)** | 61.67 (3.11) | 65.36 (3.30) | 58.93 (4.01) | 61.86 (3.20) | 62.84 (2.33) | 67.18 (2.59) | 59.60 (2.92) | 63.04 (2.37) |
| | | **OASIS-3 (age: 60-86)** | 60.45 (2.91) | 64.34 (3.15) | 57.13 (3.28) | 60.64 (3.03) | 62.59 (2.83) | 66.76 (3.49) | 59.07 (3.42) | 62.85 (2.96) |
| | **SVM** | **ADNI (age: 60-86)** | 56.04 (7.85) | 57.92 (8.88) | 55.75 (7.29) | 56.06 (7.96) | 64.27 (7.16) | 72.33 (6.54) | 59.99 (8.12) | 64.69 (7.39) |
| | | **OASIS-3 (age: 43-96)** | 53.16 (4.60) | 54.82 (6.75) | 51.16 (5.74) | 53.18 (4.65) | 64.10 (3.22) | 70.34 (3.24) | 61.12 (4.34) | 64.30 (3.21) |
| | | **OASIS-3 (age: 60-86)** | 51.77 (3.71) | 53.07 (5.97) | 50.28 (4.07) | 51.78 (3.74) | 63.09 (3.39) | 68.76 (3.84) | 59.85 (4.79) | 63.59 (3.36) |
| | **Logistic** | **ADNI (age: 60-86)** | 60.42 (4.73) | 57.48 (9.00) | 64.99 (2.35) | 60.70 (4.66) | 58.96 (6.27) | 61.75 (6.58) | 57.44 (6.47) | 59.07 (6.32) |
| | | **OASIS-3 (age: 43-96)** | 62.56 (2.19) | 60.53 (3.27) | 66.50 (1.31) | 62.91 (2.13) | 61.13 (3.24) | 64.79 (4.76) | 60.43 (2.75) | 61.16 (3.25) |
| | | **OASIS-3 (age: 60-86)** | 62.26 (2.41) | 60.62 (3.72) | 66.34 (1.51) | 62.62 (2.35) | 60.78 (4.81) | 64.41 (4.41) | 60.89 (4.51) | 60.79 (4.81) |
| | **RUSBoost** | **ADNI (age: 60-86)** | 56.98 (9.15) | 56.54 (13.69) | 52.37 (12.10) | 57.16 (9.60) | 55.00 (7.45) | 52.13 (13.40) | 50.10 (9.26) | 55.13 (7.71) |
| | | **OASIS-3 (age: 43-96)** | 59.74 (3.87) | 62.22 (3.73) | 58.47 (5.16) | 59.81 (3.88) | 60.13 (4.38) | 62.25 (4.10) | 59.45 (5.65) | 60.21 (4.39) |
| | | **OASIS-3 (age: 60-86)** | 60.26 (4.82) | 62.35 (4.27) | 58.83 (6.67) | 60.49 (4.94) | 61.08 (4.68) | 62.83 (5.11) | 59.71 (5.89) | 61.21 (4.75) |
| **D) SPSS statistical analysis' extracted features (8 features)** | | | | | | | | | | |
| **Entorhinal<br>Inf-Lat-Vent<br>Hippocampus<br>Lateral-Ventricle<br>lateralorbitofrontal<br>BrainSegVolNotVent<br>Accumbens-area<br>middle temporal** | **Naïve Bayes** | **ADNI (age: 60-86)** | 64.06 (7.40) | 69.56 (7.38) | 62.70 (7.38) | 64.17 (7.47) | 66.15 (4.37) | 73.54 (5.29) | 62.54 (4.90) | 66.53 (4.49) |
| | | **OASIS-3 (age: 43-96)** | 61.65 (3.42) | 67.67 (2.47) | 57.07 (3.58) | 62.11 (3.76) | 63.23 (2.08) | 70.11 (1.62) | 58.27 (3.26) | 63.63 (2.08) |
| | | **OASIS-3 (age: 60-86)** | 62.81 (3.16) | 69.23 (3.19) | 59.00 (2.91) | 63.18 (3.42) | 63.61 (2.27) | 69.94 (2.39) | 59.38 (3.44) | 64.02 (2.39) |
| | **KNN** | **ADNI (age: 60-86)** | 60.52 (8.93) | 64.63 (10.86) | 56.06 (10.97) | 60.85 (9.19) | 62.40 (7.10) | 68.44 (8.93) | 56.34 (8.76) | 63.19 (7.64) |
| | | **OASIS-3 (age: 43-96)** | 61.11 (2.82) | 65.07 (2.74) | 58.70 (3.98) | 61.24 (2.87) | 62.24 (2.64) | 67.37 (3.01) | 59.91 (3.75) | 62.41 (2.70) |
| | | **OASIS-3 (age: 60-86)** | 61.51 (3.89) | 65.92 (4.09) | 59.82 (4.05) | 61.61 (3.92) | 62.97 (2.77) | 67.95 (3.42) | 61.09 (2.65) | 63.08 (2.85) |
| | **SVM** | **ADNI (age: 60-86)** | 50.63 (8.00) | 54.42 (11.26) | 49.91 (8.77) | 50.63 (8.05) | 63.65 (6.71) | 70.50 (9.06) | 60.07 (8.16) | 63.87 (6.69) |
| | | **OASIS-3 (age: 43-96)** | 50.15 (4.10) | 50.38 (5.61) | 47.69 (4.20) | 50.16 (4.14) | 65.13 (2.81) | 70.79 (2.50) | 62.88 (2.90) | 65.28 (2.87) |
| | | **OASIS-3 (age: 60-86)** | 52.36 (4.18) | 53.44 (6.42) | 51.40 (5.66) | 52.36 (4.20) | 66.58 (2.91) | 72.01 (2.40) | 63.94 (3.41) | 66.78 (2.96) |
| | **Logistic** | **ADNI (age: 60-86)** | 59.90 (4.32) | 57.83 (8.36) | 64.59 (2.60) | 60.20 (4.28) | 55.52 (8.23) | 60.65 (10.37) | 54.72 (7.80) | 55.59 (8.29) |
| | | **OASIS-3 (age: 43-96)** | 62.37 (2.13) | 59.08 (2.99) | 66.37 (1.27) | 62.72 (2.06) | 62.44 (3.40) | 66.12 (3.59) | 61.33 (3.78) | 62.47 (3.40) |
| | | **OASIS-3 (age: 60-86)** | 62.03 (2.59) | 58.17 (3.49) | 66.14 (1.59) | 62.38 (2.53) | 62.08 (5.10) | 66.07 (5.16) | 61.84 (5.04) | 62.09 (5.11) |
| | **RUSBoost** | **ADNI (age: 60-86)** | 55.31 (10.01) | 53.44 (13.73) | 51.80 (11.19) | 55.53 (10.33) | 56.04 (8.60) | 53.17 (12.20) | 52.14 (12.73) | 56.13 (8.78) |
| | | **OASIS-3 (age: 43-96)** | 58.61 (3.97) | 60.62 (4.23) | 57.31 (6.98) | 58.72 (4.02) | 58.31 (2.86) | 61.07 (3.46) | 57.77 (3.72) | 58.42 (2.94) |
| | | **OASIS-3 (age: 60-86)** | 60.80 (4.60) | 62.46 (4.19) | 60.00 (6.98) | 61.08 (4.71) | 60.47 (3.17) | 62.40 (3.50) | 59.12 (6.36) | 60.65 (3.21) |

**Supplementary Table 14:** Comparison of classification performance for imbalanced data analysis for different classification methods and feature selection criteria (see Figure 5 caption for more information).

| Features | Models | Dataset | Acc (Mean (SD)) (%) | AROC (Mean (SD)) (%) | F1 (Mean (SD)) (%) | MCC (Mean (SD)) (%) | Acc (Mean (SD)) (%) | AROC (Mean (SD)) (%) | F1 (Mean (SD)) (%) | MCC' (Mean (SD)) (%) |
|---|---|---|---|---|---|---|---|---|---|---|
| | | | residual harmonization | | | | z-score harmonization | | | |
| **A) Average score percentage (MCL app): chi-square, ReliefF, ANOVA & Kruskal Wallis** | | | | | | | | | | |
| **brainSegVolNotVent entorhinal inf-Lat-Vent lateral-Ventricle parahippocampal hippocampus temporalpole fusiform lateralorbitofrontal insula precentral isthmuscingulate parsopercularis posteriorcingulate accumbens-area rostralmiddlefrontal middletemporal rostralanteriorcingulate amygdala parsorbitalis** | **Naïve Bayes** | **ADNI (age: 60-86)** | 64.75 (2.31) | 67.72 (3.43) | 36.04 (4.45) | 57.73 (3.09) | 73.93 (3.37) | 68.44 (3.54) | 38.90 (4.76) | 61.34 (3.32) |
| | | **OASIS-3 (age: 43-96)** | 69.81 (1.73) | 68.03 (0.98) | 37.61 (1.05) | 60.28 (0.81) | 70.34 (0.78) | 68.18 (0.85) | 37.81 (1.38) | 60.50 (0.88) |
| | | **OASIS-3 (age: 60-86)** | 68.56 (2.48) | 67.22 (1.49) | 40.29 (1.82) | 60.69 (1.47) | 69.82 (0.84) | 67.92 (1.09) | 40.78 (1.18) | 61.19 (0.78) |
| | **KNN** | **ADNI (age: 60-86)** | 59.09 (3.73) | 63.48 (5.08) | 35.93 (4.90) | 56.96 (3.93) | 61.07 (3.88) | 61.09 (4.62) | 30.47 (3.57) | 53.57 (2.71) |
| | | **OASIS-3 (age: 43-96)** | 62.82 (1.20) | 67.38 (1.28) | 37.88 (1.30) | 60.04 (1.00) | 65.42 (1.90) | 66.33 (1.71) | 36.13 (1.86) | 58.87 (1.40) |
| | | **OASIS-3 (age: 60-86)** | 59.77 (2.29) | 64.12 (1.97) | 38.81 (2.30) | 58.75 (1.88) | 66.33 (1.73) | 63.77 (2.27) | 38.59 (2.27) | 59.30 (1.63) |
| | **SVM** | **ADNI (age: 60-86)** | 80.17 (0) | 2.50 (11.18) | 0 (0) | 0 (0) | 81.90 (1.28) | 64.49 (5.63) | 40.55 (6.09) | 66.80 (2.53) |
| | | **OASIS-3 (age: 43-96)** | 81.95 (0.22) | 9.10 (22.24) | 1.07 (1.69) | 26.18 (26.93) | 81.60 (0.44) | 60.45 (3.42) | 1.10 (2.07) | 32.32 (24.39) |
| | | **OASIS-3 (age: 60-86)** | 79.58 (0) | 0 (0) | 0 (0) | 0 (0) | 78.89 (0.52) | 59.22 (2.88) | 3.49 (3.18) | 50.61 (2.19) |
| | **Logistic** | **ADNI (age: 60-86)** | 51.45 (0.45) | 53.97 (3.98) | 37.06 (0.83) | 57.43 (0.82) | 62.02 (5.97) | 53.10 (5.36) | 26.99 (4.62) | 51.99 (3.02) |
| | | **OASIS-3 (age: 43-96)** | 54.63 (0.09) | 61.45 (2.79) | 37.11 (0.04) | 59.51 (0.04) | 64.44 (9.19) | 54.38 (2.20) | 24.62 (4.13) | 52.22 (1.80) |
| | | **OASIS-3 (age: 60-86)** | 55.13 (0.11) | 59.63 (3.46) | 40.39 (0.14) | 59.97 (0.13) | 61.80 (6.78) | 54.53 (2.40) | 27.77 (3.77) | 52.32 (1.95) |
| | **RUSBoost** | **ADNI (age: 60-86)** | 72.15 (2.57) | 72.20 (3.54) | 39.68 (5.38) | 61.30 (3.41) | 74.09 (2.64) | 74.34 (4.13) | 39.95 (5.74) | 61.89 (3.63) |
| | | **OASIS-3 (age: 43-96)** | 76.72 (0.81) | 69.73 (1.00) | 39.93 (1.96) | 62.84 (1.21) | 76.15 (0.69) | 69.99 (0.98) | 40.01 (1.64) | 62.74 (1.01) |
| | | **OASIS-3 (age: 60-86)** | 74.55 (0.85) | 68.67 (1.10) | 40.53 (1.95) | 62.22 (1.18) | 73.82 (1.08) | 68.31 (1.19) | 40.51 (1.88) | 62.10 (1.20) |
| **B) ReliefF** | | | | | | | | | | |
| **entorhinal fusiform Inf-Lat-Vent Temporalpole Posteriorcingulate Isthmuscingulate hippocampus parahippocampal parsopercularis Lateral-Ventricle transversetemporal precentral insula frontalpole superiortemporal amygdala lateralorbitofrontal lingual BrainSegVolNotVent** | **Naïve Bayes** | **ADNI (age: 60-86)** | 66.40 (2.23) | 69.97 (3.75) | 38.14 (3.39) | 59.32 (2.39) | 71.69 (2.47) | 72.40 (4.35) | 43.26 (4.69) | 63.27 (3.08) |
| | | **OASIS-3 (age: 43-96)** | 65.21 (1.85) | 66.80 (1.09) | 34.03 (1.55) | 57.46 (1.13) | 72.13 (0.89) | 68.13 (0.78) | 37.00 (1.40) | 60.26 (0.82) |
| | | **OASIS-3 (age: 60-86)** | 61.57 (2.20) | 66.57 (0.90) | 36.74 (1.22) | 57.41 (1.02) | 71.48 (1.07) | 67.51 (1.02) | 39.46 (1.75) | 60.81 (1.07) |
| | **KNN** | **ADNI (age: 60-86)** | 60.54 (3.86) | 64.10 (4.18) | 35.11 (4.79) | 56.56 (3.57) | 70.12 (3.80) | 71.52 (3.85) | 41.50 (5.01) | 61.96 (3.51) |
| | | **OASIS-3 (age: 43-96)** | 61.77 (1.54) | 64.74 (1.61) | 37.57 (1.61) | 59.77 (1.30) | 64.21 (1.63) | 62.66 (1.79) | 34.21 (1.75) | 57.49 (1.13) |
| | | **OASIS-3 (age: 60-86)** | 60.47 (1.72) | 63.37 (2.00) | 38.85 (2.49) | 58.83 (1.98) | 64.76 (1.72) | 62.90 (1.87) | 37.38 (1.80) | 58.28 (1.33) |
| | **SVM** | **ADNI (age: 60-86)** | 80.08 (0.37) | 0 (0) | 0 (0) | 2.34 (10.46) | 81.12 (2.25) | 68.08 (4.77) | 38.42 (6.89) | 65.22 (4.20) |
| | | **OASIS-3 (age: 43-96)** | 82.00 (0.00) | 2.50 (11.18) | 0 (0) | 0 (0) | 81.75 (0.42) | 53.89 (4.01) | 2.90 (2.29) | 52.08 (2.52) |
| | | **OASIS-3 (age: 60-86)** | 79.58 (0) | 0 (0) | 0 (0) | 0 (0) | 79.36 (0.37) | 55.36 (2.49) | 1.27 (1.66) | 43.27 (18.83) |
| | **Logistic** | **ADNI (age: 60-86)** | 51.32 (0.37) | 56.12 (4.40) | 36.87 (0.69) | 57.24 (0.68) | 61.69 (8.07) | 56.88 (5.70) | 31.39 (6.98) | 54.46 (5.02) |
| | | **OASIS-3 (age: 43-96)** | 54.65 (0.10) | 60.38 (2.35) | 37.12 (0.05) | 59.52 (0.04) | 65.64 (8.31) | 53.53 (2.06) | 23.72 (4.45) | 52.12 (1.27) |
| | | **OASIS-3 (age: 60-86)** | 55.13 (0.15) | 60.63 (2.55) | 40.33 (0.21) | 59.91 (0.20) | 63.53 (7.48) | 55.06 (2.75) | 27.68 (3.49) | 52.85 (2.15) |
| | **RUSBoost** | **ADNI (age: 60-86)** | 76.03 (2.37) | 80.11 (2.52) | 50.60 (5.20) | 68.31 (3.45) | 77.64 (2.34) | 81.54 (2.92) | 50.44 (5.06) | 68.43 (3.24) |
| | | **OASIS-3 (age: 43-96)** | 76.52 (0.76) | 69.25 (0.88) | 38.17 (1.89) | 61.90 (1.13) | 75.60 (1.00) | 69.52 (1.08) | 37.85 (1.47) | 61.46 (1.01) |
| | | **OASIS-3 (age: 60-86)** | 74.56 (1.20) | 67.79 (1.11) | 39.18 (2.48) | 61.57 (1.57) | 73.32 (1.00) | 67.64 (1.37) | 39.45 (1.85) | 61.28 (1.22) |
| **C) Frequent feature appearances from all feature ranking analysis** | | | | | | | | | | |
| **BrainSegVolNotVent<br>Inf-Lat-Vent<br>entorhinal<br>lateralorbitofrontal<br>Lateral-Ventricle<br>parahippocampal<br>hippocampus** | **Naïve Bayes** | **ADNI (age: 60-86)** | 66.94 (3.14) | 71.06 (4.62) | 40.37 (3.95) | 60.80 (2.97) | 77.36 (2.37) | 73.46 (4.40) | 47.48 (3.92) | 66.69 (2.64) |
| | | **OASIS-3 (age: 43-96)** | 73.63 (1.80) | 69.41 (1.41) | 38.27 (2.03) | 61.24 (1.47) | 72.06 (1.05) | 69.80 (0.75) | 38.08 (1.25) | 60.87 (0.82) |
| | | **OASIS-3 (age: 60-86)** | 66.91 (2.17) | 68.14 (1.08) | 38.15 (1.64) | 59.12 (1.24) | 72.29 (1.50) | 69.14 (0.94) | 40.64 (1.67) | 61.64 (1.19) |
| | **KNN** | **ADNI (age: 60-86)** | 63.10 (3.70) | 65.29 (4.04) | 37.93 (4.47) | 58.81 (3.37) | 73.43 (3.05) | 72.70 (3.70) | 47.96 (3.63) | 66.38 (2.54) |
| | | **OASIS-3 (age: 43-96)** | 66.52 (1.48) | 65.62 (1.96) | 37.00 (1.72) | 59.57 (1.26) | 65.73 (1.59) | 66.90 (1.75) | 36.72 (1.59) | 59.32 (1.15) |
| | | **OASIS-3 (age: 60-86)** | 62.29 (2.56) | 63.36 (2.23) | 37.64 (2.72) | 58.11 (2.12) | 64.03 (1.83) | 64.10 (1.94) | 37.45 (1.84) | 58.20 (1.41) |
| | **SVM** | **ADNI (age: 60-86)** | 80.29 (0.90) | 7.30 (17.84) | 4.62 (5.77) | 33.52 (28.45) | 82.93 (1.59) | 75.01 (5.05) | 49.07 (5.99) | 70.21 (3.16) |
| | | **OASIS-3 (age: 43-96)** | 81.91 (0.17) | 41.68 (24.84) | 2.39 (1.53) | 52.13 (1.81) | 81.97 (1.07) | 56.71 (2.22) | 0.79 (2.43) | 12.87 (22.94) |
| | | **OASIS-3 (age: 60-86)** | 79.24 (1.14) | 14.42 (2.56) | 3.53 (3.56) | 49.48 (11.94) | 78.95 (0.61) | 58.80 (2.05) | 8.86 (6.37) | 50.84 (12.26) |
| | **Logistic** | **ADNI (age: 60-86)** | 51.40 (0.69) | 55.10 (4.49) | 37.18 (0.96) | 57.54 (0.94) | 62.85 (6.71) | 56.42 (4.36) | 30.14 (4.73) | 54.11 (3.00) |
| | | **OASIS-3 (age: 43-96)** | 54.62 (0.06) | 60.72 (2.74) | 37.10 (0.03) | 59.50 (0.03) | 64.34 (9.25) | 53.51 (3.08) | 24.14 (3.39) | 51.84 (2.14) |
| | | **OASIS-3 (age: 60-86)** | 55.17 (0.17) | 60.05 (2.33) | 40.43 (0.13) | 60.00 (0.12) | 63.23 (6.59) | 54.63 (2.70) | 26.75 (4.02) | 52.20 (2.10) |
| | **RUSBoost** | **ADNI (age: 60-86)** | 69.83 (2.30) | 66.96 (3.54) | 37.58 (3.92) | 59.56 (2.61) | 70.66 (2.94) | 66.67 (2.93) | 38.05 (3.56) | 59.97 (2.61) |
| | | **OASIS-3 (age: 43-96)** | 77.36 (0.55) | 70.10 (0.92) | 39.33 (2.05) | 62.74 (1.17) | 76.99 (0.69) | 69.51 (1.27) | 38.11 (1.34) | 62.01 (0.84) |
| | | **OASIS-3 (age: 60-86)** | 75.69 (1.01) | 67.45 (1.04) | 41.94 (2.36) | 63.31 (1.43) | 75.12 (1.10) | 67.33 (1.07) | 41.26 (2.03) | 62.77 (1.33) |
| **D) features extracted from statistical analysis (SPSS software): ANOVA, ANCOVA & Kruskal Wallis** | | | | | | | | | | |
| **Entorhinal<br>Inf-Lat-Vent<br>Hippocampus<br>Lateral-Ventricle<br>lateralorbitofrontal<br>BrainSegVolNotVent<br>Accumbens-area<br>middle temporal** | **Naïve Bayes** | **ADNI (age: 60-86)** | 67.15 (1.80) | 71.34 (4.46) | 39.29 (4.37) | 60.21 (2.97) | 77.11 (3.47) | 72.41 (3.50) | 46.97 (4.87) | 66.39 (3.46) |
| | | **OASIS-3 (age: 43-96)** | 71.81 (1.52) | 69.79 (1.89) | 37.99 (1.49) | 60.78 (1.05) | 71.53 (0.96) | 70.43 (0.83) | 37.43 (1.16) | 60.40 (0.76) |
| | | **OASIS-3 (age: 60-86)** | 72.24 (1.57) | 70.33 (1.00) | 42.54 (1.71) | 62.67 (1.25) | 70.08 (1.04) | 69.81 (0.91) | 40.73 (1.36) | 61.21 (0.95) |
| | **KNN** | **ADNI (age: 60-86)** | 60.91 (3.70) | 66.37 (5.05) | 37.15 (4.80) | 58.11 (3.56) | 74.83 (2.89) | 75.50 (3.16) | 47.54 (4.33) | 66.25 (2.97) |
| | | **OASIS-3 (age: 43-96)** | 67.02 (1.22) | 64.98 (1.96) | 37.94 (1.59) | 60.25 (1.14) | 65.85 (1.58) | 69.20 (1.54) | 39.10 (1.70) | 61.04 (1.23) |
| | | **OASIS-3 (age: 60-86)** | 65.22 (2.02) | 63.99 (1.67) | 39.09 (1.82) | 59.44 (1.43) | 66.26 (1.73) | 67.11 (1.31) | 41.21 (1.80) | 60.99 (1.35) |
| | **SVM** | **ADNI (age: 60-86)** | 80.17 (0.54) | 0 (0) | 3.35 (5.61) | 21.89 (27.69) | 82.23 (1.81) | 73.25 (4.02) | 44.77 (6.39) | 68.41 (3.12) |
| | | **OASIS-3 (age: 43-96)** | 82.04 (0.28) | 39.15 (29.6) | 3.37 (2.19) | 53.40 (2.68) | 81.86 (0.18) | 56.32 (1.90) | 0.08 (0.37) | 24.52 (25.16) |
| | | **OASIS-3 (age: 60-86)** | 79.58 (0) | 0 (0) | 0 (0) | 0 (0) | 79.35 (0.32) | 55.52 (1.60) | 2.78 (3.53) | 43.66 (19.01) |
| | **Logistic** | **ADNI (age: 60-86)** | 51.53 (0.49) | 56.59 (5.21) | 37.37 (0.88) | 57.73 (0.86) | 63.43 (6.52) | 57.74 (5.80) | 30.00 (6.99) | 54.23 (4.09) |
| | | **OASIS-3 (age: 43-96)** | 54.64 (0.08) | 60.83 (2.88) | 37.11 (0.04) | 59.51 (0.03) | 64.19 (11.00) | 53.40 (2.29) | 24.02 (4.25) | 52.21 (1.78) |
| | | **OASIS-3 (age: 60-86)** | 55.20 (0.16) | 60.06 (2.59) | 40.42 (0.18) | 59.99 (0.17) | 62.95 (7.38) | 54.55 (2.88) | 26.89 (4.84) | 52.21 (2.44) |
| | **RUSBoost** | **ADNI (age: 60-86)** | 72.02 (1.88) | 69.08 (3.35) | 36.81 (4.93) | 59.66 (3.02) | 74.13 (2.38) | 69.30 (3.15) | 40.42 (5.33) | 62.15 (3.34) |
| | | **OASIS-3 (age: 43-96)** | 77.03 (0.99) | 67.80 (0.74) | 37.17 (1.95) | 61.57 (1.25) | 76.58 (0.73) | 67.57 (0.61) | 37.08 (1.33) | 61.37 (0.83) |
| | | **OASIS-3 (age: 60-86)** | 75.14 (1.04) | 66.05 (1.48) | 38.98 (2.63) | 61.70 (1.58) | 74.85 (0.97) | 65.88 (1.18) | 39.47 (2.00) | 61.82 (1.19) |